\documentclass[a4paper,11pt,twoside]{book}

\RequirePackage[colorlinks,citecolor=teal,urlcolor=blue,linkcolor=blue]{hyperref}

\usepackage{subfigure}
\usepackage[english]{babel}
\usepackage{amsmath,amssymb,amsbsy,amsthm}
\usepackage{IEEEtrantools}
\usepackage[dvipsnames]{xcolor}
\usepackage{pifont} 
\usepackage{siunitx}
\usepackage{hyperref}
\usepackage{empheq}
\usepackage{cite}

\newcommand{\EQ}{EQ}
\newcommand{\ST}{ST}
\newcommand{\FLAT}{FL}
\newcommand{\p}{\partial}
\newcommand{\SU}{\mathrm{SU}}
\newcommand{\su}{\mathfrak{su}}
\newcommand{\dg}{\dagger}
\newcommand{\one}{\mathbf{1}}
\newcommand{\Tr}{\mathrm{Tr}}
\newcommand{\mg}[2]{\vcenter{\hbox{\includegraphics[scale=#1]{#2}}}}

\newcommand{\LGCNN}{{L-CNN}}
\newcommand{\LConv}{L-Conv}
\newcommand{\LBL}{{L-Bilin}}
\newcommand{\LAct}{{L-Act}}
\newcommand{\LExp}{{L-Exp}}
\newcommand{\Plaq}{Plaq}
\newcommand{\Poly}{Poly}
\newcommand{\Trace}{Trace}
\newcommand{\LCB}{L-CB}

\newcommand{\Conv}[3]{\mbox{{\LCB}(#1, #2, #3)}}
\newcommand{\GTr}{\mbox{{Trace}}}
\newcommand{\Lin}[2]{\mbox{{Linear}(#1, #2)}}
\newcommand{\CCD}[3]{C2D(#1, #2, #3)}

\renewcommand{\Re}{\mathrm{Re}}
\renewcommand{\Im}{\mathrm{Im}}

\newcommand{\nnum}[1]{\num[exponent-product = {\times}]{#1}}
\newcommand{\bnum}[1]{\pmb{\num[exponent-product = {\times}]{#1}}}

\DeclareMathOperator*{\argmin}{arg\,min}

\title{Equivariant neural networks in lattice field theories}
\author{Matteo Favoni}
\date{\today}

\begin{document}
\begin{titlepage}
	\vspace{-3cm}
	\begin{center}
	\includegraphics[scale=0.3]{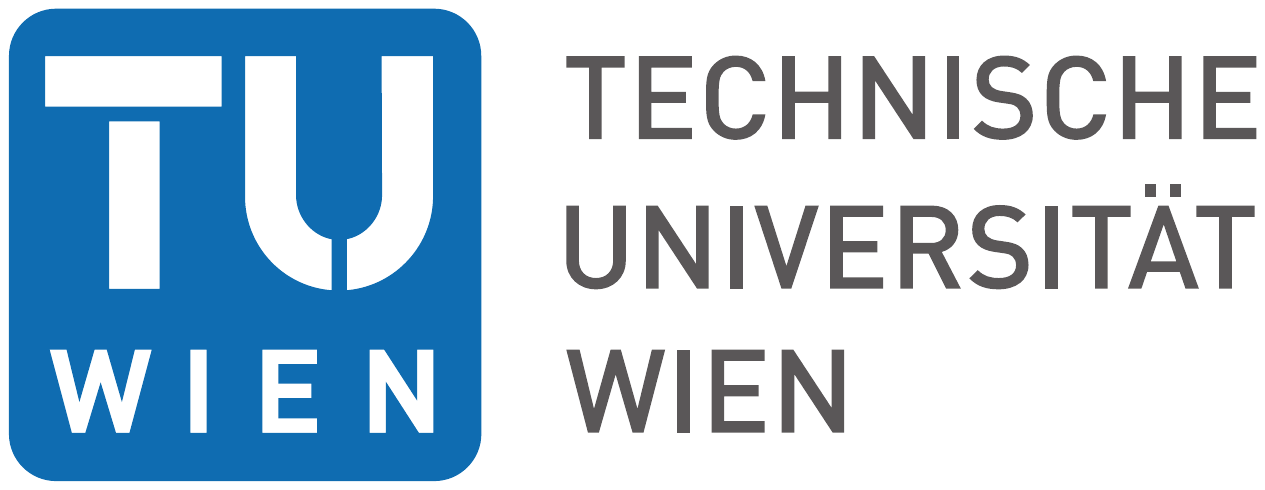}\\[0.5cm]
	\end{center}	
	\vspace{-2.5cm}
	
	\begin{center}
	{\LARGE DISSERTATION\\[1.0cm]}
	{\LARGE\textbf{Symmetry-preserving neural networks in lattice field theories}\\[1.0cm]}
	\end{center}
	
	\begin{center}    
		{\normalsize ausgef\"uhrt zum Zwecke der Erlangung des akademischen Grades eines Doktors der Technischen Wissenschaften unter der Leitung von}\\[0.5cm]
	\end{center}
	\begin{center}    
		{\normalsize 	Priv.Doz.\ Dr.techn.\ Andreas Ipp\\
                [0.5cm]}
	\end{center}
	\begin{center}    
		{\normalsize und mit der Unterstützung von}\\[0.5cm]
	\end{center}
	\begin{center}    
		{\normalsize 	Dr.techn.\ David I.\ M{\"u}ller\\
                [0.5cm]}
	\end{center}
	\begin{center}    
		{\normalsize eingereicht an der Technischen Universit{\"a}t Wien,\\
			Fakult{\"a}t f{\"u}r Physik} \\[0.5cm]     
	\end{center} 
	\begin{center}    
		{\normalsize von\\
			Dott.~Mag.\ Matteo Favoni\\ [0.5cm]
        }
	\end{center} 
	\begin{center} 
	\noindent\begin{tabular}{ll}
		\makebox[2cm]{Wien, am 16.12.2024}\hspace*{4cm} & \makebox[4cm]{\hrulefill}  \\[2.0cm]
	\end{tabular}

	\begin{tabular}{p{5cm}@{}p{5cm}@{}p{5cm}@{}}
		\makebox[4cm]{\hrulefill} & \makebox[4cm]{\hrulefill} & \makebox[4cm]{\hrulefill} \\
		Dr.\ Andreas Ipp    &  Prof.\ Tilo Wettig & Prof.\ Kai Zhou    \\
		(Betreuer)                & (Gutachter)               & (Gutachter)                 \\
	\end{tabular}
	\end{center} 
		
\end{titlepage}
\thispagestyle{empty}
\newpage{}


\frontmatter
\chapter*{Abstract}

This thesis deals with neural networks that respect symmetries and presents the advantages in applying them to lattice field theory problems. The concept of equivariance is explained, together with the reason why such a property is crucial for the network to preserve the desired symmetry. The benefits of choosing equivariant networks are first illustrated for translational symmetry on a complex scalar field toy model. The discussion is then extended to gauge theories, for which Lattice Gauge Equivariant Convolutional Neural Networks (L-CNNs) are specifically designed ad hoc. Regressions of physical observables such as Wilson loops are successfully solved by L-CNNs, whereas traditional architectures which are not gauge symmetric perform significantly worse. 
Finally, we introduce the technique of neural gradient flow, which is an ordinary differential equation solved by neural networks, and propose it as a method to generate lattice gauge configurations.

\addcontentsline{toc}{section}{Abstract}

\newpage{}
\newpage{}


\chapter*{Zusammenfassung}

Diese Arbeit befasst sich mit neuronalen Netzen, die Symmetrien respektieren, und stellt die Vorteile ihrer Anwendung auf gitterfeldtheoretische Probleme dar. Das Konzept der Äquivarianz wird erklärt, zusammen mit der Begründung, warum eine solche Eigenschaft für das Netz entscheidend ist, um die gewünschte Symmetrie zu erhalten. Die Vorteile der Wahl äquivarianter Netze werden zunächst für die Translationssymmetrie eines komplexen Skalarfeld-Spielzeugmodells erläutert. Die Diskussion wird dann auf Eichtheorien ausgedehnt, für die Lattice Gauge Equivariant Convolutional Neural Networks (L-CNNs) speziell entworfen werden. Regressionen von physikalischen Observablen wie Wilson-Schleifen werden von L-CNNs erfolgreich gelöst, während traditionelle Architekturen, die nicht eichsymmetrisch sind, deutlich schlechter abschneiden. Schließlich stellen wir die Technik des neuronalen Gradientenflusses vor, bei der es sich um eine gewöhnliche Differentialgleichung handelt, die von neuronalen Netzen gelöst wird, und schlagen sie als eine Methode  zur Generierung von Gittereichkonfigurationen vor.

\addcontentsline{toc}{section}{Zusammenfassung}


\chapter*{Preface}

\addcontentsline{toc}{section}{Preface}

The techniques and results presented in this thesis are largely based on the following published articles:

\begin{itemize}
	\item  S.~Bulusu, M.~Favoni, A.~Ipp D.~M\"uller and D.~Schuh,
	\href{https://doi.org/10.1103/PhysRevD.104.074504}{Phys.\ Lett.\ D {\bf 104}, 074504 (2021)}
	[\href{https://arxiv.org/abs/2103.14686}{2103.14686}] \cite{Bulusu:2021rqz}
	\item M.~Favoni, A.~Ipp, D.~M\"uller and D.~Schuh, \href{https://doi.org/10.1103/PhysRevLett.128.032003}{Phys.\ Rev.\ Lett. {\bf 128}, 032003 (2022)}
	[\href{https://arxiv.org/abs/2012.12901}{2012.12901}] \cite{Favoni:2020reg}
	\item   M.~Favoni, A.~Ipp and D.~M\"uller, \href{https://doi.org/10.1051/epjconf/202227409001}{EPJ Web of Conferences {\bf 274}, 09001 (2022)}
	[\href{https://arxiv.org/abs/2212.00832}{2212.00832}] \cite{Favoni:2022mcg}
	
\end{itemize} 
\newpage{}
\newpage{}


\chapter*{Acknowledgements}

\addcontentsline{toc}{section}{Acknowledgements}

First of all, I want to thank my supervisor Andreas Ipp for giving me the opportunity to work on this project. I also want to thank him along with my co-supervisor David Müller for their guidance over the years, especially for managing the pandemic situation seamlessly and making it less burdensome. My colleague Daniel Schuh has been another key presence along this journey; I thank him for his friendship and numerous exchanges on our doctoral projects and on a wide variety of topics. I also thank Srinath Bulusu, whose collaboration was very helpful at the beginning of this project. I want to express my deep gratitude to Alexander Rothkopf for hosting me at the University of Stavanger during my stay abroad. I also thank his collaborators Rasmus Larsen, Daniel Alvestad and Gaurang Parkar for useful discussions.
Finally, I wholeheartedly thank my family, my mother Donata Danna, my father Dario Favoni, and my partner Letizia Savino for their constant support and for always believing in me.

\newpage{}
\newpage{}

\tableofcontents

\mainmatter
\numberwithin{equation}{section}

\chapter{Introduction} \label{chap:intro}

In the past century, quantum field theory has established itself as the appropriate mathematical framework to describe the fundamental constituents of nature. The Standard Model of particles \cite{Weinberg:1968, Salam:1968rm} is the most successful theory based on such a framework, in that it reproduces all the 
experimental results found at particle colliders within a $5 \sigma$ confidence level. Tensions between this theory and experiments do exist, for example for the value of the anomalous magnetic dipole moment \cite{Aguillard:2023mmm, Aoyama:2020ynm}, and, even more importantly, some crucial questions are left unaddressed, such as neutrino oscillations and dark matter. Precise predictions are paramount to reveal deviations from experiments and to identify promising candidate theories to explain physics beyond the Standard Model. Perturbative calculations are possible, but other approaches have been developed. One of the most prominent is lattice field theory, a discretization of quantum field theory on a fictitious grid. Its great appeal lies in the possibility of treating non-perturbative problems, arising for example in the low-energy regime of quantum chromodynamics (QCD), and, moreover, it enables to run simulations on a computer. These are typically performed with Monte Carlo methods \cite{Metropolis:1949}, which are very effective but run into issues such as critical slowing down \cite{Sokal:1997} and the sign problem \cite{Aarts_2016, Gattringer:2016kco}.

The scientific community has witnessed a deep surge of interest in artificial intelligence in the last decade, with a large variety of applications in many fields, such as image recognition \cite{Russakovsky:2015}, natural language processing \cite{OpenAI:2023}, games like chess \cite{Silver:2018}, protein folding \cite{Jumper:2021} and cancer detection \cite{Tran:2021}. A very successful subfield of artificial intelligence is machine learning, which is an umbrella term for programs capable of gradually improving their performance on a given task without explicit instructions. This resembles the human way of learning from experience, and the brain structure has indeed inspired the main machine learning algorithm in use nowadays: artificial neural networks. The elementary unit of such models is an artificial neuron (also called unit, node or simply neuron), and is linked to other neurons via edges, mimicking how biological neurons are connected to each other via synapses in the brain. Artificial neurons receive input signals, which are typically real numbers, multiply each of them by a weight, and output the sum of the result. Usually, neurons are followed by an activation function which introduces some non-linearity \cite{Dubey:2022}.

The perceptron \cite{Rosenblatt:1957, Rosenblatt:1958}, an architecture comprising of just one neuron and an activation function, was proposed in 1957 for binary classification and represents one of the very first machine learning models to be implemented. In a multi-layer perceptron (MLP), also referred to as dense network, neurons are interconnected and organized in layers, with a visible layer representing the input that is then processed by a sequence of hidden layers leading to the output. The larger the number of hidden layers, the deeper the network is said to be, hence the name deep learning \cite{LeCun:2015}. Stacking many layers one after the other allows the network to identify relevant pieces of information and patterns, starting from simple concepts and combining them to achieve higher levels of abstraction. MLPs are feed-forward neural networks, that is, the flow of information only goes from the input to the output. Therefore, no backward connections between neurons are present.

One of the strategies to deal with data is supervised learning, in which a network is provided with a dataset containing some input and the corresponding output data, also called labels. The parameters of the network (i.e.~the aforementioned weights multiplying the input of each neuron) are usually randomly initialized and, consequently, the network's outputs are in general very distant from the labels. The distance is encoded in a loss function. It is common practice to split the dataset into training, validation and test set. The training samples are used to minimize the loss function by updating the network's weights using some form of gradient descent \cite{Ruder:2016}. The validation set helps to avoid a phenomenon called overfitting \cite{Ying:2019}, which causes the network to memorize the samples without truly extracting useful information. The test set reports on the ability of the network to generalize to unseen data. Given this procedure, a neural network serves as the best approximating function of the true underlying map between the input and the corresponding labels present in the dataset. If the true function is continuous, it is always possible to approximate it with arbitrary accuracy for a sufficiently wide or deep dense network with at least one hidden layer and an activation function, as guaranteed by various versions of the universal approximation theorem \cite{Cybenko:1989, Lu:2017a}.

Neural networks have recently garnered the attention of the physics community, with successful applications in many fields. In lattice field theory, a pioneering work in phase classification was conducted in \cite{Carrasquilla:2017}, followed by applications to $\phi^4$ scalar field theory, lattice QCD, Ising, XY, Potts and Yukawa models, where neural networks were applied to classical \cite{Zhou:2018ill} and topological \cite{Wang:2020hji} phase transitions, to universality classes \cite{Bachtis:2020ajb}, and to infer order parameters \cite{Bluecher:2020mjt}, action parameters \cite{Shanahan:2018vcv} and thermodynamic properties \cite{Nicoli:2021njz}. Several efforts have been made to alleviate the sign problem with the aid of neural-network-based algorithms \cite{Alexandru:2020wrj, Alvestad:2022abf, Alvestad:2023jgl, Kanwar:2023otc}. Fruitful applications have been reported in the context of the renormalization group \cite{Hu:2019nea, Bachtis:2021eww} and also vice versa renormalization group transformations have been used to interpret machine learning models \cite{Bachtis:2020fly}. A thriving field is represented by generative models, in particular normalizing flows \cite{Pawlowski:2018qxs}, although another option is represented by diffusion models \cite{Wang:2023sry}. A strong relationship between stochastic normalizing flows \cite{Wu:2020} and Jarzynski's equality was found \cite{Caselle:2022acb}, which helped to mitigate topological freezing with out-of-equilibrium simulations \cite{Bonanno:2024udh} and sampling the lattice Nambu-Goto string \cite{Caselle:2023mvh}.

Since Noether published her theorem \cite{Noether:1918} more than a hundred years ago, the role of symmetries in physics has become pivotal, especially in field theories. For instance, the Standard Model describes the three fundamental interactions between elementary particles as gauge theories. Even though symmetry-agnostic neural networks have been proven to be a powerful tool also in lattice field theory, constraining them to respect the underlying symmetries can facilitate the learning process. This acknowledgement had an impact also in computer vision. Images are typically characterized by global translational equivariance, which induces convolutions \cite{Jaehne:1997}. These constitute the basis of convolutional neural networks (CNNs). A convolution is defined as the sum of the element-wise product of a kernel (or filter) with an equally-sized patch of the image. This operation is repeated over the whole input, such that the output is a new image rather than a single output. Since the parameters of the kernel do not change while being applied at each pixel, the convolution has the property of weight sharing, which is essential for translational symmetry. Also, a compact kernel enables the network to scan the image in search of local features. Many kernels with different weights can be applied and their outputs are organized in separate channels, and several convolutional layers can be stacked, usually interposing activation functions and pooling layers. These are usually employed for downsampling. Forms of the universal approximation theorem exist also for CNNs \cite{Yarotsky:2022, Zhou:2020ucn}. The main ingredients of CNNs were introduced with the Neocognitron \cite{Fukushima:1980} in 1979, and more recently the advantages of CNNs became apparent. For instance, they played a major role in the ImageNet Large Scale Visual Recognition Challenge (ILSVRC)~\cite{Russakovsky:2015}, a competition held yearly from 2010 to 2017, with AlexNet~\cite{Krizhevsky:2017} becoming the first CNN to win it in 2012. Despite the success, most of the CNN architectures used in ILSVRC do not preserve symmetry under translations. For example, a flattening layer completely breaks the symmetry. This can be avoided by substituting it with a global average pooling (GAP) \cite{Lin:2013gap}, which is a feature of ResNet~\cite{He:2015del}, the winning CNN architecture of ILSVRC in 2015. In lattice field theory too, CNNs that are not translationally invariant have been employed even though translational symmetry was a property of the system under consideration \cite{Zhou:2018ill,  Bachtis:2020ajb, Bluecher:2020mjt, Bachtis:2020fly, Wetzel:2017ooo, Padavala:2021, Wang:2021}. In Chapter~\ref{chap:translations}, we will extensively discuss the symmetry properties of CNNs and show the benefits of applying a symmetry-respecting network in the context of a complex scalar field theory.

Available a priori knowledge about specific problems has prompted researchers to design ad-hoc neural networks, as in the case of physics-informed neural networks \cite{Karniadakis:2021}, Hamiltonian neural networks \cite{Greydanus:2019} and Lagrangian neural networks \cite{Cranmer:2020, Muller:2023}. The commitment to incorporate symmetries stands out, and has given rise to a very active field, with a focus on global symmetries, with group equivariant convolutional networks \cite{Cohen:2016, Cohen:2016a, Worrall:2017, Worrall:2018, Ecker:2018, Veeling:2018, Lafarge:2020, Pang:2020, Scaife:2021, Kondor:2018a, Cheng:2019xrt, Esteves:2020a, Rath:2020a, Gerken:2021sla, Celledoni:2021, Aronsson:2022, Zhdanov:2023, Hossain:2023, Edixhoven:2023} and on local ones, with gauge equivariant convolutional networks \cite{Cohen:2019xsh, Luo:2020stn, Finzi:2020, Favoni:2020reg, Nagai:2021bhh, Aronsson:2023rli, Lehner:2023bba, Lehner:2023prf, Holland:2024muu} and equivariant transformers \cite{Nagai:2023fxt, Tomiya:2023jdy}. Chapter~\ref{chap:lcnns} is devoted to the design of a neural network able to preserve gauge symmetry on the lattice.

In the context of configuration sampling, a particularly active field is represented by gauge equivariant normalizing flows \cite{Kanwar:2020xzo, Nicoli:2021njz, Boyda:2020hsi, Albergo:2021bna, Abbott:2022, Nicoli:2023qsl}. In Chapter~\ref{chap:ngf}, we propose a possible alternative method to generate gauge link configurations, similar to the one proposed in \cite{Bacchio:2022vje}.

Finally, we summarize our findings in the Conclusion chapter~\ref{chap:conclusions} and present further details in the Appendix.

\newpage
\chapter{Translationally-symmetric neural networks in scalar field theory} \label{chap:translations}
\chaptermark{Translational symmetry}

The numerous successful applications of neural networks in scientific contexts represent an empirical demonstration that they are a powerful and reliable tool. This is also supported from a theoretical standpoint, in that neural networks can learn to approximate any function. In practice, though, there can be many obstacles preventing a neural network from an effective learning procedure. For instance, a limited amount of samples in the dataset, the training phase requiring too many iterations, the optimization procedure getting stuck in unsatisfactory local minima of the loss function. An approach that aims at facilitating learning is to take into account the symmetries of the problem under examination and to design the network in a way that preserves these symmetries. In physics, and in particular in field theories, systems are often invariant under translations. An example is a complex scalar field theory, which is the focus of this Chapter. We first describe such a theory and discretize it on the lattice. Then we explain how a non-zero chemical potential gives rise to a sign problem and how it can be circumvented with the use of a duality transformation. After that, individual layers of CNNs are presented and their translational symmetry properties are discussed. Three architecture types are proposed and their performance is evaluated on three different tasks. For each task, we show how the incorporation of translational symmetry proves to be beneficial for the network's performance.

\section{Dual formulation of a complex scalar field} \label{sec:scalar}

The Lagrangian density of a complex scalar field $\phi (x^ \mu)$ subject to the potential $V(\phi ^* \phi)$ can be written as
\begin{equation}
    \mathcal{L} (\p_{\mu} \phi, \p_{\mu} \phi^*, \phi, \phi ^*) = \p_{\mu} \phi^* (x) \p^{\mu} \phi (x) - V(\phi^* (x) \phi (x)),
\end{equation}
with $\mu \in \{0, 1, ..., d-1\}$ running over the $d$ spacetime indices and the convention used for the metric is $\eta_{\mu\nu} = \mathrm{diag} (+, -, \dots, -)$.
The action
\begin{equation}
    S = \int \! \mathrm{d}^d x \, \mathcal{L}
    \label{Scalar:generic_action}
\end{equation}
is then invariant under translations in time and space $x ^{\prime \, \mu} = x ^\mu + a ^\mu$. The field transforms as
\begin{equation}
    \phi' (x ^{\prime \, \mu}) = \phi (x ^\mu),
    \label{Scalar:field_transf}
\end{equation}
while derivatives remain unchanged:
\begin{equation}
    \p' _\mu = \frac{\p}{\p x ^{\prime \, \mu}} = \frac{\p x ^\nu}{\p x ^{\prime \, \mu}} \frac{\p}{\p x ^\nu} = \frac{\p (x ^{\prime \, \nu} - a ^\nu)}{\p x ^{\prime \, \mu}} \frac{\p}{\p x ^\nu} = \delta _\mu ^\nu \frac{\p}{\p x ^\nu} = \frac{\p}{\p x ^\mu} = \p _\mu.
    \label{Scalar:ders_inv}
\end{equation}
In the action of the field $\phi '$
\begin{equation}
    S' = \int \! \mathrm{d}^d x' \, \left( \p'_\mu \phi ^{\prime \, *} (x') \, \p^{\prime \, \mu} \phi' (x') - V(\phi ^{\prime \, *} (x') \, \phi' (x')) \right),
\end{equation}
the integration measure is not modified by a translation, nor are the kinetic term and the interacting term, as one can see from Eqs.~\eqref{Scalar:field_transf} and \eqref{Scalar:ders_inv}. This proves that the action is invariant under translations. As a remark, if the Lagrangian density explicitly depends on the position $x ^\mu$ the action is manifestly not translationally invariant.

As stated in Noether's first theorem~\cite{Noether:1918}, finite continuous symmetries of the action are associated with the existence of conserved quantities, which in the case of invariance under spacetime translations are represented by the energy-momentum tensor.

The action~\eqref{Scalar:generic_action} also exhibits a global $U(1)$ symmetry under the transformation 
\mbox{$\phi \rightarrow e^{i \alpha} \phi$}, which implies the presence of another conserved charge. In this context, a chemical potential can be introduced with a modification of the time derivative given by $D_0 = \p _0 - i \mu$~\cite{Brauner:2020zov}. The action then becomes
\begin{equation}
    S = \int \! \mathrm{d}x_0\, \mathrm{d}^{d-1} \mspace{-2mu} x \mspace{2mu} \left( \lvert D_0 \phi \rvert^2 - \lvert \p_i \phi \rvert^2 - V(\lvert \phi \rvert) \right), \label{Scalar:action}
\end{equation}
where $i \in \{1, 2, \dots, d-1\}$ runs over the space indices. In the following, we will focus on a complex scalar field with quartic interaction subject to the potential
\begin{equation}
    V(\lvert \phi \rvert) = m^2 \lvert \phi \rvert^2 + \lambda \lvert \phi \rvert^4,
\end{equation}
where $m$ is the mass and $\lambda$ the coupling constant.
The action takes the form
\begin{equation}
    S  = \mspace{-5mu} \int \mspace{-4mu} \mathrm{d}x_0 \, \mathrm{d}^{D-1} \mspace{-2mu} x \mspace{2mu} \left(  \lvert D_0 \phi \rvert^2 - \lvert \partial_i \phi \rvert^2  - m^2 \lvert \phi \rvert^2 - \lambda \lvert \phi \rvert^4 \right) \mspace{-2mu}.
    \label{Scalar:quartic_action}
\end{equation}
It is possible to introduce an imaginary time coordinate by performing the Wick rotation
\begin{equation}
x_0 = i x_d,\quad x_d \in \mathbb{R}.
\end{equation}
We explicitly write the effect of this transformation on the time covariant derivative:
\begin{equation}
    \lvert D_0 \phi \rvert^2 = D_0 \phi \,(D_0 \phi) ^\dg = (\p_0 - i \mu) \phi \, (\p_0 + i \mu) \phi^* = - (\p_d + \mu) \phi \, (\p_d - \mu) \phi^*,
\end{equation}
where we used the fact that
$\p_0 = -i\p_d$. We can define the Euclidean action $S_E$ as follows:
\begin{equation}
    S_E = i S \mspace{-1mu} = \mspace{-5mu} \int \mspace{-4mu} \mathrm{d}^d x \mspace{-2mu} \left( \lvert \p_i \phi \rvert^2  \mspace{-2mu} + (\p_d + \mu) \phi \, (\p_d - \mu) \phi^* \mspace{-2mu} + m^2 \lvert \phi \rvert^2 \mspace{-2mu} + \lambda \lvert \phi \rvert^4 \right) \mspace{-2mu}.
    \label{Scalar:cont_euclidean_action}
\end{equation}

We now proceed to the discretization of the theory. The points that define the lattice are located at the positions $x^\mu = a n^\mu$, with $n^\mu \in \mathbb{R}$ and $a = a^i \; \forall i \in \{1, \dots, d\}$, thus assuming the lattice to be equally spaced in every dimension. The integration over spacetime is replaced by a sum over all lattice sites,
\begin{equation}
    \int \mathrm{d}^d x \rightarrow a^d \displaystyle \sum_x.
\end{equation}
Derivatives cannot be taken on the lattice because of its discontinuous nature, therefore a natural substitute is the difference quotient
\begin{equation}
    \p_\mu \phi(x) \rightarrow \displaystyle \frac{\phi_{x+a\hat{\mu}} - \phi_x}{a},
    \label{Scalar:diff_quotient}
\end{equation}
which, in the $a \rightarrow 0$ limit, coincides with the definition of the derivative. The notation $\hat{\mu}$ refers to the unit vector pointing in the $\mu$~direction. It is possible to reformulate the two terms involving the chemical potential in Eq.~\eqref{Scalar:cont_euclidean_action} as follows:
\begin{equation}
    \displaystyle \pm \mu = -\p_d \left( \mathrm{e} ^{\mp \mu x_d} \right) \mathrm{e} ^{\pm \mu x_d} \rightarrow \frac{\mathrm{e} ^{\mp \mu (x_d + a)} - \mathrm{e} ^{\mp \mu x_d}}{a} \, \mathrm{e} ^{\pm \mu x_d} = \frac{\mathrm{e} ^{\mp \mu a} - 1}{a},
\end{equation}
thus yielding
\begin{align}
    (\p_d + \mu) \, \phi &\rightarrow  \displaystyle \frac{\phi_{x+a\hat{d}} - \mathrm{e} ^{-\mu a} \phi_x}{a} \\
    (\p_d - \mu) \, \phi^* &\rightarrow \displaystyle \frac{\phi^*_{x+a\hat{d}} - \mathrm{e} ^{\mu a} \phi^*_x}{a}.
\end{align}
The discretized version of the Euclidean action can be reworked in the following way:
\begin{equation}
\begin{split}
    S_E &\rightarrow a^d \sum_x \left[ \sum_{\nu=1}^{d-1} \left( \frac{\phi_{x+a\hat{\nu}} - \phi_x}{a} \frac{\phi^*_{x+a\hat{\nu}} - \phi^*_x}{a} \right) \right. \\
    & \left. \mathrel{\phantom{= a^d \sum ( \sum (}}+ \frac{\phi_{x+a\hat{d}} - \mathrm{e} ^{-\mu a} \phi_x}{a} \frac{\phi^*_{x+a\hat{d}} - \mathrm{e} ^{\mu a} \phi^*_x}{a} +m^2 \lvert \phi_x \rvert^2 + \lambda \lvert \phi_x \rvert^4 \right] = \\
    &= a^{d-2} \sum_x \left[ \left(2d + (am)^2 \right) \lvert \phi_x \rvert^2 
    + a^2 \lambda \lvert \phi_x \rvert^4 \phantom{\sum_{\nu = 1}^2} \right. \\
    & \left. \mathrel{\phantom{= a^{d-2} \sum_x}}- \sum_{\nu = 1}^2 \left( e^{a \mu \mspace{2mu} \delta_{\nu, 2}} \phi_x^* \, \phi_{x + \hat{\nu}} + e^{-a \mu \mspace{2mu} \delta_{\nu, 2}} \phi_x^* \, \phi_{x - \hat{\nu}} \right) \right],
    \label{Scalar:general_discr_euclidean_action}
\end{split}
\end{equation}
where $\delta_{\nu,2}$ denotes the Kronecker delta and we made use of the observation that
\begin{equation}
    \sum_x \sum_{\nu=1}^d \phi_{x+a\hat{\nu}} \, \phi^*_{x+a\hat{\nu}} = \sum_x \sum_{\nu=1}^d \phi_x \, \phi^*_x = d \sum_x \lvert \phi_x \rvert^2.
\end{equation}
Dimensionless physical parameters are introduced by
\begin{align}
    &m' = am, \nonumber \\
    &\eta = 2d + m^{\prime \, 2}, \nonumber \\
    &\mu' = a \mu, \nonumber \\
    &\lambda' = a^2 \lambda.
    \label{Scalar:dimless_params}
\end{align}
We restrict our interest to a $1+1$ dimensional system with extension $L$ in the space direction and the inverse of the temperature $1/T$ in the imaginary time coordinate. The lattice is equipped with periodic boundary conditions and we set $a=1$. These assumptions lead to the simplification of Eq.~\eqref{Scalar:general_discr_euclidean_action} given below:
\begin{equation}
    S_\text{lat} = \sum_x \left[ \eta \lvert \phi_x \rvert^2 + \lambda \lvert \phi_x \rvert^4 - \sum_{\nu = 1}^2 \left( e^{\mu \mspace{2mu} \delta_{\nu, 2}} \phi_x^* \, \phi_{x + \hat{\nu}} + e^{- \mu \mspace{2mu} \delta_{\nu, 2}} \phi_x^* \, \phi_{x - \hat{\nu}} \right) \right].
    \label{Scalar:lat_action}
\end{equation}

The statistical-mechanical interpretation of the path integral allows to write the partition function
\begin{equation}
    Z = \int \! \mathcal{D}\phi \, e^{-S_\text{lat}}
\end{equation}
and treat the term $e^{-S_\text{lat}}$ as a probability distribution, which enables Monte Carlo sampling. This is possible only if the action $S_\text{lat}$ is real, which is due to the fact that complex weights cannot be interpreted in a probabilistic manner, ruling out Monte Carlo simulations as a tool to study the physics of the system. One often refers to systems exhibiting this type of behavior as suffering a sign problem. As apparent in~\eqref{Scalar:lat_action}, non-zero chemical potentials give rise to complex terms. Many approaches have been tried in order to get rid of the sign problem or at least alleviate it~\cite{Aarts_2016, Gattringer:2016kco}, recently involving also machine learning techniques~\cite{Alexandru:2020wrj, Alvestad:2022abf, Alvestad:2023jgl, Kanwar:2023otc}. For the theory under examination, it can be completely removed by means of a duality transformation, that replaces the complex field $\phi$ with two integer fields $k_{x, \nu}$ and $l_{x, \nu}$, with one component per each direction $\nu=1, 2$ and the index $x\in\{1,2,\ldots,N\}$ is a label for the lattice site. This dual formulation is also called flux representation and the details of the derivation can be found in~\cite{Gattringer:2013df}. 
The partition function obtained with these new fields is
\begin{IEEEeqnarray}{rCl}
    \nonumber Z &=& \sum_{ \{ k, l \} } \left( \prod_{x, \nu} \frac{1}{(\lvert k_{x, \nu} \rvert + l_{x, \nu})! l_{x, \nu}!} \right) \left( \prod_x e^{\mu k_{x,2}} W(f_x) \right) \\
    && \> \times \left( \prod_x \delta \left( \sum_\nu (k_{x, \nu} - k_{x-\hat{\nu}, \nu}) \right) \right), \IEEEeqnarraynumspace \label{Scalar:flux_representation}
\end{IEEEeqnarray}
where the summation is intended to run over all possible values of $k_{x, \nu}$ and $l_{x, \nu}$ at all lattice sites:
\begin{align}
    \sum_{ \{ k, l \} } &= \prod_{x, \nu} \mspace{4mu} \sum_{k_{x, \nu} = - \infty}^\infty \mspace{8mu} \sum_{l_{x, \nu} = 0}^\infty \nonumber \\
    &=\sum_{k_{1,1}=-\infty}^\infty  \mspace{8mu} \sum_{l_{1,1}=0}^\infty \mspace{8mu} \sum_{k_{1,2}=-\infty}^\infty \mspace{4mu} \sum_{l_{1,2}=0}^\infty \cdots \sum_{k_{N,2}=-\infty}^\infty \mspace{4mu} \sum_{l_{N,2}=0}^\infty.
\end{align}
As can be seen in the sum, the field $k$ takes values in $\mathbb{Z}$, while $l$ is a non-negative integer. The third term in Eq.~\eqref{Scalar:flux_representation} imposes that the flux conservation
\begin{equation}
    \sum_{\nu} \left( k_{x, \nu} - k_{x - \hat{\nu}, \nu} \right) = 0 \label{Scalar:flux_conservation}
\end{equation}
has to be respected, otherwise the Kronecker delta yields zero, meaning that configurations of $k$ and $l$ fields for which Eq.~\eqref{Scalar:flux_conservation} is not satisfied do not contribute to the partition function and can therefore be considered unphysical.
The function $W(f_x)$ has the expression
\begin{equation}
    W(f_x)=\int_0^\infty \mathrm{d}t\, t^{f_x+1}\mathrm{e}^{-\eta t^2-\lambda t^4}, \label{Scalar:W}
\end{equation}
and its non-negative integer-valued argument is given by
\begin{equation}
    f_x=\sum_\nu[|k_{x,\nu}|+|k_{x-\hat{\nu},\nu}|+2(l_{x,\nu}+l_{x-\hat{\nu},\nu})].
    \label{Scalar:f_x}
\end{equation}
This quantity has the following useful properties:
\begin{align}
    \frac{\p W}{\p \eta} &= -W(f_x + 2),
    \label{Scalar:der_1} \\
    \frac{\p W}{\p \lambda} &= -W(f_x + 4)
    \label{Scalar:der_2}.
\end{align}

As in statistical mechanics, taking derivatives of the partition function with respect to physical parameters gives us observables that can be computed. For the system under examination, by making use of Eqs.~\eqref{Scalar:der_1} and~\eqref{Scalar:der_2} it is possible to compute their ensemble averages with
\begin{IEEEeqnarray}{rCcCl}
    \langle n \rangle &=& \frac{T}{V} \frac{\p \ln Z}{\p \mu} &=& \frac{1}{N_x N_t} \left \langle \sum_x k_{x, 2} \right \rangle,
    \label{Scalar:n} \\
    \langle \lvert \phi \rvert^2 \rangle &=& - \frac{T}{V} \frac{\p \ln Z}{\p \eta} &=& \frac{1}{N_x N_t} \left \langle \sum_x \frac{W(f_x + 2)}{W(f_x)} \right \rangle,
    \label{Scalar:phi} \\
    \langle |\phi|^4 \rangle &=& -\frac{T}{V}\frac{\p\ln Z}{\p \lambda} &=& \frac{1}{N_x N_t} \left \langle \sum_x\frac{W(f_x+4)}{W(f_x)} \right \rangle,
    \label{Scalar:phi4}
    \IEEEeqnarraynumspace
\end{IEEEeqnarray}
where $N_x$ is the number of lattice sites in the space dimension and $N_t$ refers to the extension in time. In the upcoming studies, the first two quantities will be used, 
namely the particle number density~$n$ and the squared absolute value of the field $\lvert \phi \rvert^2$.

For our purpose, it is crucial to notice that the observables in Eqs.~\eqref{Scalar:n}, \eqref{Scalar:phi} and \eqref{Scalar:phi4} are invariant under spacetime translations, thus the dual formulation inherits the same symmetry that was present at the level of the action in \eqref{Scalar:generic_action}. The global $U(1)$ symmetry mentioned before is instead reflected in the flux conservation~\eqref{Scalar:flux_conservation}. Here, our focus of interest is translational invariance and in the next section we are going to explain how it is implemented into neural networks.

\section{Symmetry properties of neural network layers} \label{sec:layers}
\sectionmark{Symmetry properties of NN layers}

The physical system discussed so far is an ideal situation where neural networks can be applied. We can associate with a certain configuration $\{k_{x, \mu}, l_{x, \mu}\}$ the values of  $n$ and $|\phi| ^ 2$ as given by Eqs.~\eqref{Scalar:n} and~\eqref{Scalar:phi}, which resembles a typical regression task in image processing, such as age prediction from two-dimensional pictures of faces~\cite{Angulu2018}. There are two main differences worth to be noted, though: first, an image is characterized by one real value per pixel in the grayscale case or three in an RGB image, while here the lattice is defined by four integer values per site; second, the configurations that will be used are equipped with periodic boundary conditions, a property that pictures normally do not possess and is necessary for the system to by translationally invariant.

Neural networks are an excellent tool to approximate any functions, and the universal approximation theorem~\cite{Cybenko:1989, Lu:2017a}
supports this with a solid mathematical foundation. Given the ability of these networks to learn relevant features, it may be expected that also symmetries can be identified as an important aspect of a problem, but in general they cannot be learned exactly, only approximately. The core of this work is the following: if a network respects a certain symmetry by construction, it is expected to perform better than a network that has a similar structure but internally breaks the symmetry. We want to verify this statement focusing first on the aforementioned translational symmetry of the complex scalar field presented so far, and extend the discussion to gauge symmetry in the next chapter.
The type of network that will be used here is a CNN. Requiring that such networks respect translational symmetry, formally corresponds to impose that their output is invariant under translations. This translates into a more general concept when discussing individual layers, and as a matter of fact the sufficient condition for a network to be invariant under translations is that its layers are equivariant under translations, which gives the network a higher degree of expressivity. Equivariance of a layer $\Phi$ can be defined generally for any symmetry group $\mathcal{G}$ as
\begin{equation}
        \Phi(L_{\! g} \, x) = L^{\mspace{2mu} \prime}_{\! g} \, \Phi(x),
        \label{Layers:equivariance}
\end{equation}
where $L_{\! g}$ is a group transformation and $g$ is a group element. For situations in which $L_g = L'_g$, equivariance corresponds to the commutative property of the layer with a group transformation. Moreover, invariance is the special case for which $L'_g = 1$.

A network that respects invariance under a global symmetry group is called Group Convolutional Neural Network (G-CNN), as defined in~\cite{Cohen:2016} and on which the following discussion is based. If the group $\mathcal{G}$ is the translation group $\mathbb{T}$, the G-CNN structure coincides with one of a conventional CNN.

We will now analyze individual layers and their symmetry properties one by one.

\subsection{Convolutional layers}

The definition of a convolution on a two-dimensional lattice is given by \cite{Cohen:2016}
\begin{equation}
	[f \star \psi](x) = \sum_{y \in \mathbb{Z}^2} f(y) \, \psi(y - x) = \sum_{y \in \mathbb{Z}^2} f(x + y) \, \psi(y), \label{Layers:cross_correlation}
\end{equation}
where the feature map $f$ and the kernel (or filter) $\psi$ are real-valued functions:
\begin{align}
f : \mathbb{Z}^2 &\rightarrow \mathbb{R}, \\
\psi : \mathbb{Z}^2 &\rightarrow \mathbb{R}.
\end{align}
The definition~\eqref{Layers:cross_correlation} is typically used in the machine learning community, and may differ in other fields, e.g.~in mathematics, where the argument of $\psi$ has the opposite sign. Since the kernel's weights are optimized independently of the orientation of the kernel, the two definitions are equivalent from the point of view of the learning process. We also note that in Eq.~\eqref{Layers:cross_correlation} a bias term $b$ can be added, but it does not affect the discussion about layer symmetry.

Usually, the kernel is localized, therefore there is only a small subset of $\mathbb Z^2$ that is non-zero, but for simplicity we keep the sum running over the whole $\mathbb Z^2$ space. The domain of the feature map $f$ is a finite subset $F \subset \mathbb Z^2$ equipped with periodic boundary conditions. We avoid writing them explicitly by assuming that the feature map repeats periodically outside $F$. A convenient way of realizing such boundary conditions is by customizing an appropriate padding. In \textit{PyTorch}~\cite{Paszke:2019}, it is already implemented and is called circular padding. It is worth noticing that periodicity guarantees that the output of the convolution Eq.~\eqref{Layers:cross_correlation} has the same size as the feature map $f$.

We define a translation of the feature map as
\begin{align}
[L_t f](x) = f(x - t),
\end{align} 
where $t$ is an element of the translation group  $\mathbb{T}$, which can be identified with an element of $\mathbb{Z}^2$. Equivariance under translations can be proved for the convolution with the following sequence of equalities:
\begin{align}
	[L_t f \star \psi](x) &= \sum_{y \in \mathbb{Z}^2} f(y - t) \psi(y - x) \nonumber \\
	&= \sum_{y' \in \mathbb{Z}^2} f(y') \psi(y' - (x - t)) \nonumber \\
	&=	[f \star \psi](x - t) \nonumber \\
	&=	[L_t [f \star \psi]](x).
\end{align}

What has been implied in the definition in Eq.~\eqref{Layers:cross_correlation} is that the convolution has a stride~$s$ of one. The stride of a convolution is the distance between the points where the kernel is applied.

More generally,
\begin{equation}
	[f \star \psi]_s(x) =  \sum_{y \in \mathbb{Z}^2} f(y) \psi(y - sx) \label{Layers:cross_correlation_strides}
\end{equation}
can be used as a definition for convolutions with any stride $s \geq 1$, which coincides with the previous one in Eq.~\eqref{Layers:cross_correlation} if \mbox{$s=1$}. In the case \mbox{$s \geq 2$}, we will call the convolutions strided, and we highlight two relevant features: the size of the convolution output is smaller than the feature map, and translational equivariance is broken in general. The latter can be shown by considering a translation \mbox{$t \in \mathbb{T}$} with its components $t_1$ and $t_2$ satisfying both \mbox{$|t_1| < s$} and \mbox{$|t_2| < s$}. For instance, if we choose \mbox{$t= (1, 0)$}, its action on the feature map results in
\begin{align}
	[L_{t} f \star \psi]_s (x) &= \sum_{y \in \mathbb{Z}^2} f(y - t) \psi(y - s x) \nonumber \\
 &= \sum_{y' \in \mathbb{Z}^2} f(y') \psi(y' - s x + t) \nonumber \\
  &= \sum_{y' \in \mathbb{Z}^2} f(y') \psi(y' - s (x - t / s)).
\end{align}
If equivariance was respected, it would be possible to rewrite the formula above in terms of a shifted position $x' = x - t/s \in \mathbb{Z}^2$, but this is impossible since $t/s \notin \mathbb{Z}^2$. However, strided convolutions are equivariant if we restrict ourselves to the subgroup $\mathbb T_s \subset \mathbb T$, whose elements are translations which are multiples of $s$. In such a situation, $t_1/s \in \mathbb{Z}$ and $t_2/s \in \mathbb{Z}$, and we can write the following:
\begin{align}
	[L_{t} f \star \psi]_s (x) &= \sum_{y \in \mathbb{Z}^2} f(y - t) \psi(y - s x) \nonumber \\
	&= \sum_{y' \in \mathbb{Z}^2} f(y') \psi(y' - s (x - t / s)) \nonumber \\
		&= \sum_{y' \in \mathbb{Z}^2} f(y') \psi(y' - s x') \nonumber \\
				&=[L_{t'} [f \star \psi]_s](x),
\end{align}
with $t' = t / s \in \mathbb T$. To summarize, a convolutional layer with a given stride is equivariant only under translations which are multiples of that stride. It is possible to achieve equivariance under any translation only in the case~\mbox{$s = 1$}. The generalization to more than one feature map, which corresponds to having multiple channels, can be accommodated easily. It is worth noting that a strided convolution is equivalent to a convolution with~\mbox{$s = 1$} followed by a subsampling step, meaning that the input size gets reduced.

\subsection{Spatial pooling layers}

Typically, spatial pooling layers are employed as subsamplers, since the most common choice is a stride \mbox{$s \ge 2$}. Here, however, we will break down spatial pooling into two distinct steps: a pooling step and a subsampling step. For the former, we can start considering average pooling. It can be viewed as a special case of convolution for which the weights of the kernel $\psi$ are all set to the same value, i.e.~the reciprocal of the number of points in the kernel,~\mbox{$1/N_{\psi}$}. This does not affect the equivariance property of the layer, which is instead broken by the subsequent subsampling, except for the subset of translations multiple of the stride, as described previously. 

This does not only apply to average pooling, but remains true for any kind of spatial pooling $P$, as can be shown by examining its action on the feature map $f$:
\begin{equation}
    P f(x) = \mspace{-3mu} \underset{y \in U_x}{P} \mspace{-3mu} f(y),
\end{equation}
where $U_x$ are subsets of the feature map domain $F$ and the index $x$ denotes how the kernel moves through the feature map. If we now apply a translation to the feature map, then
\begin{align}
    P L_t f(x) &= \mspace{-3mu} \underset{y \in U_x}{P} \mspace{-3mu} f(y - t) \nonumber \\
    &= \mspace{-10mu} \underset{y' \in U_{x - t}}{P} \mspace{-10mu} f(y') \nonumber \\
    &= L_t P f(x),
\end{align}
which proves that equivariance as defined in Eq.~\eqref{Layers:equivariance} is preserved. Therefore, as it was for convolutions, also for spatial pooling layers we pinpointed that the symmetry-breaking aspect is the stride, not the layer structure itself.

It is important to emphasize that spatial pooling layers with \mbox{$s = 1$} do respect translational equivariance and can be included in an architecture if one wishes this property to be incorporated, even though they do not perform subsampling as typically expected.

\subsubsection{Flattening and global pooling}

A widely spread practice in visual computing is to flatten the output of the convolutional part of the network and process it further with a fully connected linear part. A translation of the input is reflected into a permutation at the level of the flattening layer output, while the weights in the dense layers do not change their position, hence leading in general to a different final output and breaking translational symmetry. This can be avoided by means of a global pooling layer between the last layer in the convolutional part and the first linear layer, an example of which is the GAP, first introduced in~\cite{Lin:2013gap}. There, a feature map was created for each class, and the average of each feature map was directly fed into a softmax layer. While this method effectively preserves translational symmetry, a more general approach is the insertion of dense layers between global pooling and the softmax operation in order to provide the network with a higher level of expressivity.

\section{Architecture types} \label{sec:archs}

\begin{figure}
    \centering
    \hfill
    \subfigure[~Equivariant architecture (EQ)]{\includegraphics[width=0.48\textwidth]{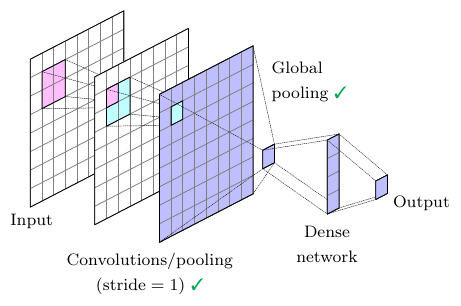}}
    \hfill
    \subfigure[~Strided architecture (ST)]{\includegraphics[width=0.48\textwidth]{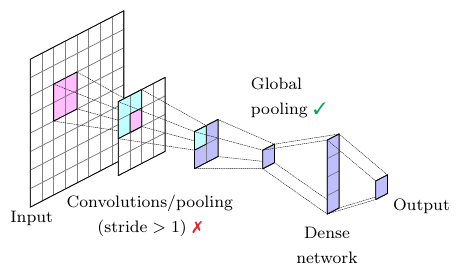}}
    \hfill
    \subfigure[~Flattening architecture (FL)]{\includegraphics[width=0.48\textwidth]{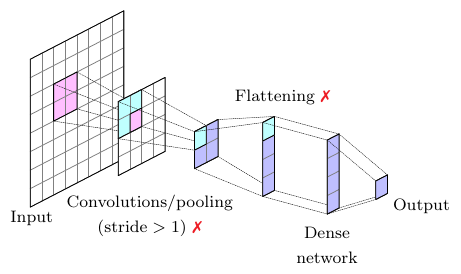}}
    \caption{
        The three different architecture types chosen for the comparison. Operations preserving translational symmetry are indicated by a check mark (\textcolor{Green}{\ding{51}}), while those violating it are marked with a cross (\textcolor{Red}{\ding{55}}). A convolutional or pooling layer with a stride greater than one or a flattening layer are responsible for breaking the symmetry. A global pooling layer allows for the application of the same network to different lattice sizes. The number of channels in each layer (not shown) does not impact the translational symmetry properties. Image from \cite{Bulusu:2021rqz}.}
    \label{fig:architectures}
\end{figure}

In order to study how significant symmetry is for the performance of neural networks, we make a comparison between architectures which respect it and others which break it. Here, we consider three types of architectures. The first one is translationally equivariant, and will be labeled as \EQ{}. The other two do not preserve translational symmetry: a strided architecture, indicated by \ST{}, contains spatial pooling layers with a stride larger than one, and the \FLAT{} architecture, which uses a flattening step and spatial pooling layers with stride larger than one. After the global pooling or flattening step, it is possible to append a dense network without affecting the symmetry properties of the architecture. The three CNN types are shown in Fig.~\ref{fig:architectures}.

The \FLAT{} type is vastly employed on image data, but is affected by a drawback which in lattice field theory contexts is highly relevant: this kind of architecture is limited to the size of the configurations used for training, while the \EQ{} and \ST{} types can be applied to different sizes.

Lattice studies are typically conducted as approximations of field theories, where spacetime is continuous and infinite. It is important to find a balance between lattice size and computation time. This is because on one side lattice calculations need to yield results as close as possible to the physical case, while on the other larger lattices require longer simulations. For this reason, an advantageous strategy is to train on rather small configurations first and then use the optimized model on larger configurations, since training is the more time-consuming part. If the application on different lattice sizes is not possible, transfer learning \cite{Bozinovski:2020} can be of assistance, though at the cost of additional training, as is the case for \FLAT{} architectures.

Let us now examine the architectures in more detail. Input configurations are fed to a convolutional layer, whose stride is one for all architecture types. The kernel size can be odd or even, and the use of circular padding ensures that the output maintains the same size as the input and that translational equivariance is respected also at the boundaries. Next, an activation function such as ReLU is used to introduce non-linearity. This does not undermine equivariance, since activation functions act pointwise. This pair of layers constitutes the main core of a CNN and is consecutively repeated multiple times, meaning that the network becomes deeper and thus becomes more expressive. Non-equivariant architectures also feature the presence of a spatial pooling layer with stride larger than one between at least one of the activation function layers and the following convolution. In the \EQ{} and the \ST{} types, the convolutional part is concluded by a global pooling, while \FLAT{} architectures end in a flattening layer. Global pooling removes position dependence, so any translational symmetry still existent is inherited by the output, while this is not the case for flattening. Afterwards, an MLP can be used to process the result even further before leading to the output.

There are a few more comments to be made about the architectures. It is possible to include longer-range correlations by the introduction of dilated convolutions~\cite{Yu:2016jsd}. If their stride is one, equivariance is preserved, as can be shown considering that dilated convolutions are equivalent to standard convolutions with a larger kernel in which some of the weights are set to zero.

As mentioned previously, spatial pooling layers with a stride of one do not break translational symmetry, but do not perform any subsampling either. If one is interested in subsampling respecting symmetries, coset pooling~\cite{Cohen:2016a} meets the requirements. The problems we are tackling in this chapter only involve local quantities, while coset pooling is a non-local operation, hence it is not expected to be an effective choice here.

In the tasks that the CNNs will be given, different types of observables have to be predicted. Depending on whether such observables are extensive or intensive, there can be a fitting choice of global pooling that allows the network to generalize better across different lattice sizes. In a 2D problem, extensive variables scale with the area, whereas intensive ones remain constant under area changes. This means that in a regression task on an extensive quantity, a global sum pooling is the most suitable option, while a global average pooling factors out the area dependence, which is ideal for an intensive observable. Classification tasks are connected to the prediction of decision boundaries between different categories, therefore the previous discussion is not applicable and in general it is not trivial to identify a priori an optimal global pooling layer.

A strategy that has been used in visual computing to overcome the limitations of a network that breaks a certain symmetry is data augmentation. There, a specific input is passed to the network multiple times after having undergone different group transformations. This is intended to teach the network the invariance under such transformations, effectively learning the underlying symmetry present in the data. We will test this approach for \ST{} and \FLAT{} architectures in the upcoming task. It has to be stressed though that this does not lead to an exactly invariant neural network, and also that it is not guaranteed to approximate the symmetry well enough on the test set.

\section{Task I: predicting physical observables} \label{sec:reg}

This section deals with a regression task that has already been anticipated and revisits the study performed in~\cite{Zhou:2018ill}. From the ensemble averages in Eqs.~
\eqref{Scalar:n} and \eqref{Scalar:phi}, we can associate each configuration $\{k_{x, \mu}, l_{x, \mu}\}$ with two values \begin{align}
    n&=\frac{1}{N_x N_t}\sum_x k_{x, 2}\,, \label{Regression:n} \\
    |\phi|^2 &= \frac{1}{N_x N_t}\sum_x\frac{W(f_x+2)}{W(f_x)}. \label{Regression:phi2}
\end{align}
A configuration represents the input of the neural network whereas the output are two real numbers which approximate the true labels~$n$ and~$|\phi|^2$. While $n$ only depends linearly on one component of the fields and is therefore expected to be easy to learn, $|\phi|^2$ depends on a ratio of the highly non-linear function $W(f_x)$, given in Eq.~\eqref{Scalar:W}, which depends on the fields through $f_x$ in Eq.~\eqref{Scalar:f_x}.

\subsection{Physical parameters and lattice sizes} \label{sec:pars}

 The physics of the system is governed by three parameters, namely $\lambda$, $\eta$ and $\mu$. In the experiments run for this task, we will use specific values of these parameters: the coupling constant and the mass are kept fixed at values $\lambda=1$ and $\eta=4+m^2=4.01$, while the chemical potential $\mu$ ranges in the interval $[0.91, 1.05]$ with steps of \mbox{$\Delta \mu = 0.005$} and in the interval $[1.1, 1.5]$ with values separated by \mbox{$\Delta \mu = 0.1$}. The combination of physical parameters with $\mu \in \{ 0.91, \dots, 1.5 \}$ is analogous to the one employed in~\cite{Zhou:2018ill} to create the dataset. There, the authors trained neural networks on configurations generated with the endpoints of such a range of the chemical potential, i.e.~\mbox{$\mu = \{ 0.91, 1.05 \}$}, while the whole interval was utilized for the test set. This is a standard approach in phase classification tasks, as pioneered in~\cite{Carrasquilla:2017}, and shows the ability of the network to extract information from configurations pertaining to values of the physical parameters not used during training. We argue though that for this system it is sufficient to train on configurations produced at only one of the two chemical potentials, specifically $\mu = 1.05$, in order to endow the networks with good generalization properties. Moreover, we want to check if these properties hold not only for lower $\mu$, but also as the chemical potential is increased, which is the reason for creating the additional dataset with $\mu \in \{ 1.1,\dots,1.5 \}$.

There is another parameter that needs to be chosen, namely the lattice size. In~\cite{Zhou:2018ill}, it remained fixed to $200 \times 10$ throughout the study, which was convenient since an architecture of type \FLAT{} was employed, while here it is possible to create differently sized configurations, train on a given lattice size, and, in the case of \EQ{} and \ST{} architectures, test the generalization capabilities to different lattice sizes without retraining. The lattice sizes we use are the following: $50\times2$, $60\times4$, $100\times5$, $125\times8$ and $200\times10$, where the first number refers to the time extension $N_t=1/T$ and the second one to the spatial one, $N_x=L$. A different $N_t$ corresponds to a different temperature $T$, which influences the properties of the phase transition.

\subsection{Data generation and worm algorithm} \label{sec:worm}

\begin{figure}
	\centering
	\subfigure[~The head is pushed back to the previous site, effectively undoing the previous movement. This highlights that the direction~$\nu$ is always chosen randomly, hence these back-and-forth steps are possible.]{\includegraphics[width=0.45\textwidth]{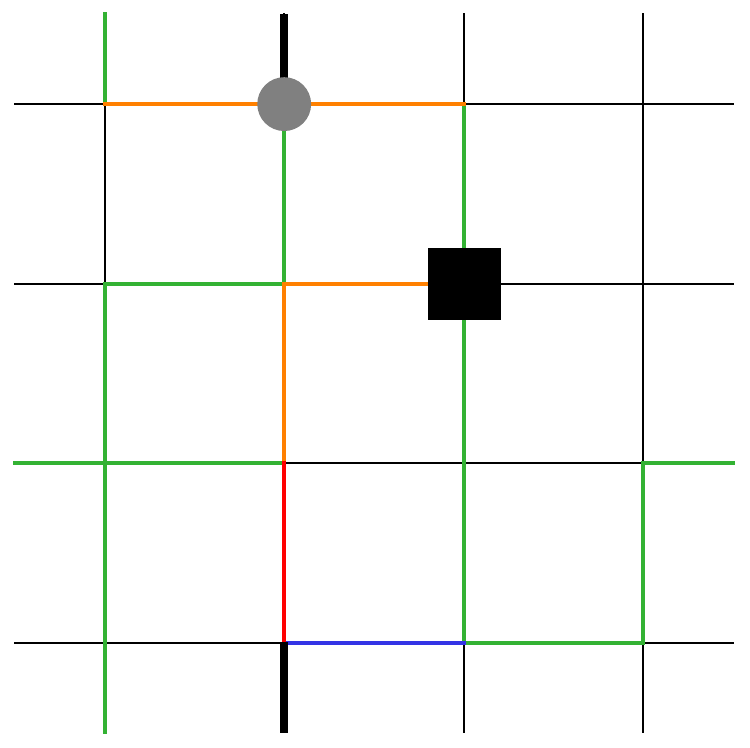}}
	\hfill
	\subfigure[~Another downward movement is depicted, with the link value decreased by~$1$.]{\includegraphics[width=0.45\textwidth]{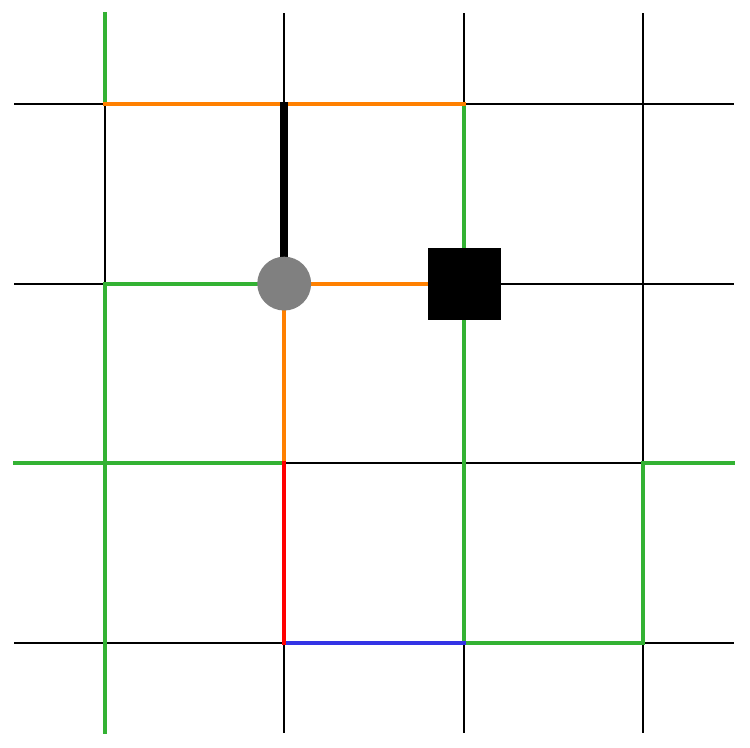}}
	\par\vspace{5mm}
	\subfigure[~The head shifts to the right rejoining the tail. The worm is closed, and, before starting a new one from a randomly chosen lattice site, a standard Monte Carlo update of the field~$l_{x,\nu}$ is executed.]
	{\includegraphics[width=0.45\textwidth]{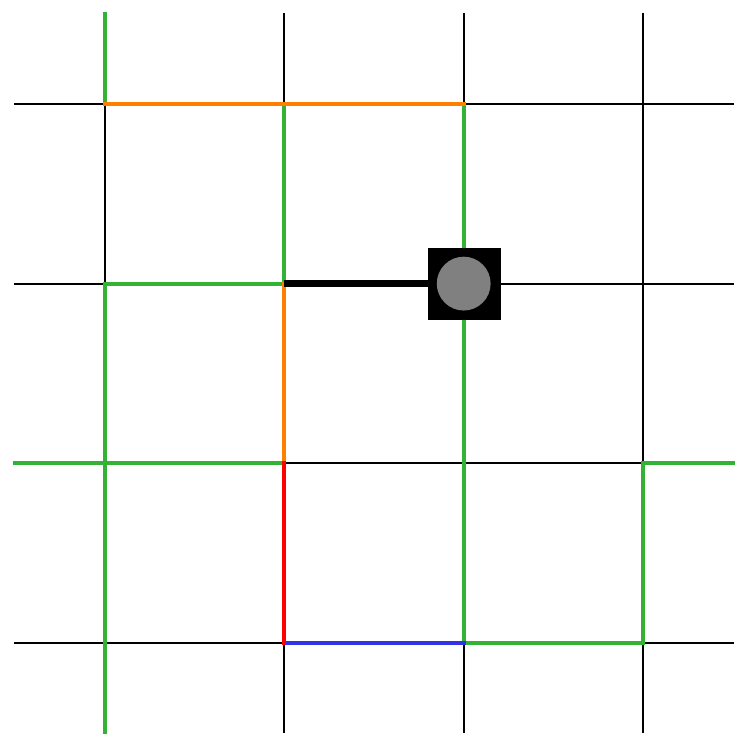}}
	\hfill
    \subfigure[~Legend of the previous pictures, clarifying the worm movements and the link values $k_{x,\nu}$.]
    {\includegraphics[width=0.45\textwidth]{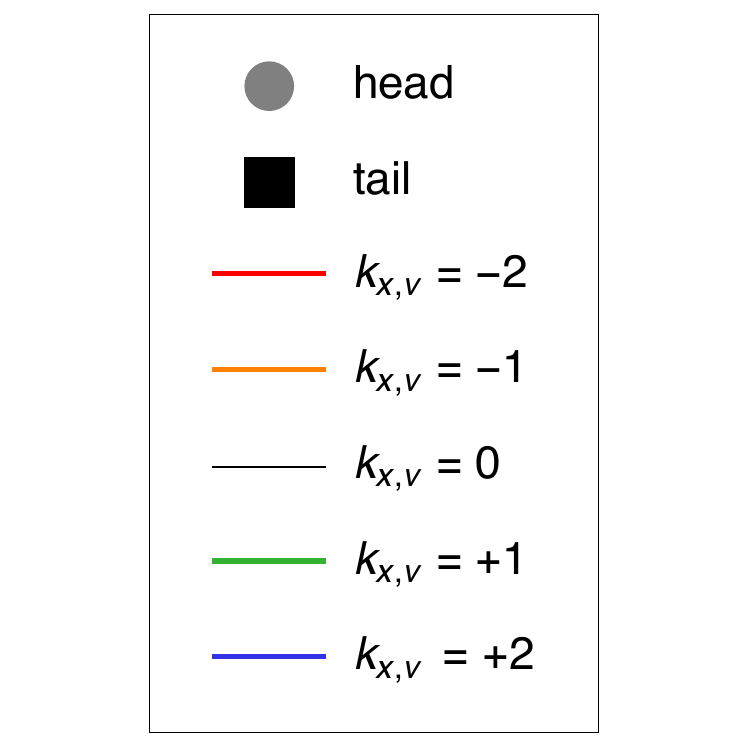}}
    
    \caption{The pictures above illustrate six possible consecutive movements of the worm in a two-dimensional lattice with size $N_t=N_x=4$.}
    \label{fig:worm_movement}
\end{figure}

The flux representation described in section~\ref{sec:scalar} is characterized by the integer field $k$ and the non-negative integer field $l$ constrained by Eq.~\eqref{Scalar:flux_conservation}. Notice how this conservation law only involves the link variable $k$. The method we intend to use to generate configurations is a standard Monte Carlo algorithm, which is suitable for $l$. In fact, we can propose a Metropolis step
\begin{align}
    l_{x, \nu} &\to l_{x, \nu} \pm 1
\end{align}
for each site and direction independently and compute its Metropolis acceptance probability as a ratio of Boltzmann weights of the action~\eqref{Scalar:lat_action}. On the other hand, applying the same approach to $k$ would disrupt the flux conservation~\eqref{Scalar:flux_conservation}, which would not be easy to restore afterwards. This is where an algorithm originally proposed by Prokof'ev and Svistunov in~\cite{Prokofev:2001zbb} comes into play. It performs a sequence of local Metropolis update steps each at an adjacent position with respect to the previous site. Concretely, a random site $x$ and a direction $\nu \in \{\pm1, \pm2\}$ are selected at the beginning, and the update
\begin{align}
    k_{x, \nu} &\to k_{x, \nu} + \text{sign}(\nu) \xi
    \label{Regression:worm_update}
\end{align}
is proposed, with $\xi = \pm 1$. Acceptance probabilities are calculated like in standard Monte Carlo as ratios of Boltzmann weights of the dual action, and if the step is accepted the proposal~\eqref{Regression:worm_update} is made at $x' = x + \nu$ keeping the same $\xi$, but randomly choosing a new $\nu$. This is repeated until $x'$ rejoins the initial site $x$. What we have described follows the prescriptions given in~\cite{Gattringer:2013df}. The resemblance to the movement of a worm led to this algorithm being known as worm algorithm and the paths formed by it as worms, where $x'$ represents the head and $x$ the tail. When the head meets the tail, the worm is said to be closed, otherwise, it is called open. For an open worm, the flux conservation is locally violated at its endpoints, while it is restored at all sites once the worm closes. In order to provide a clear understanding of the worm algorithm mechanism, a possible worm journey is portrayed in Fig.~\ref{fig:worm_movement}.

\begin{figure}
	\centering
    \subfigure[~We start from a physically allowed open path configuration, i.e. the flux is conserved at every site.
    All links are intended to be pointing in the positive directions of space (horizontal axis) and time (vertical axis). The head and tail of the worm are randomly selected.]{\includegraphics[width=0.45\textwidth]{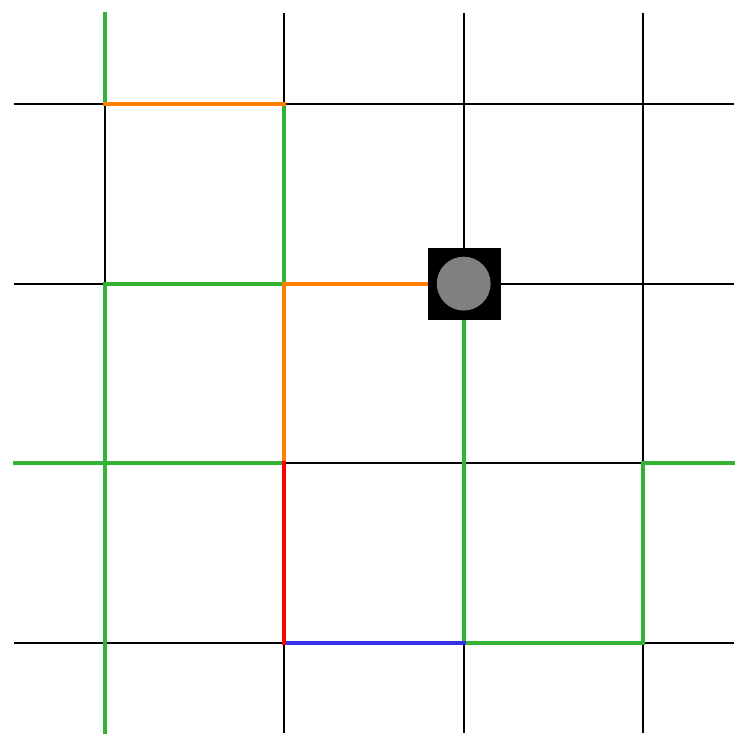}}
	\hfill
	\subfigure[~The movement of the head of the worm upwards with the sign~$\xi=+1$ is proposed and accepted, like all the next shifts. Links that have been just modified are highlighted with a greater thickness.]{\includegraphics[width=0.45\textwidth]{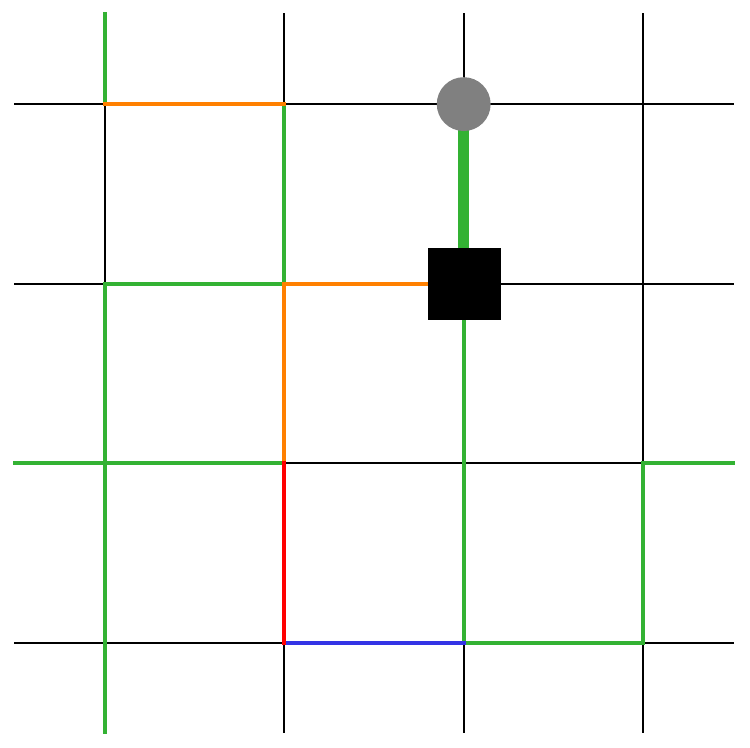}}
	\par\vspace{5mm}
	\subfigure[~The head is shifted to the left, therefore~$\text{sign}(\nu)=-1$ and the link is decreased according to Eq.~\eqref{Regression:worm_update}. It is important to stress that the value of~$\xi$ remains equal until the worm is closed.]{\includegraphics[width=0.48\textwidth]{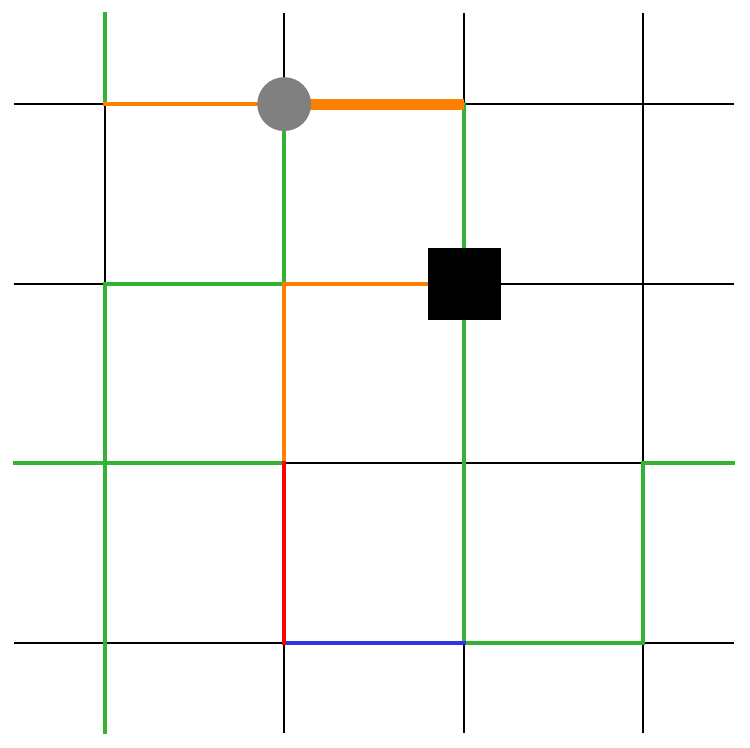}}
	\hfill
	\subfigure[~This upward movement illustrates the worm algorithm mechanism on a lattice equipped with periodic boundary conditions.]{\includegraphics[width=0.48\textwidth]{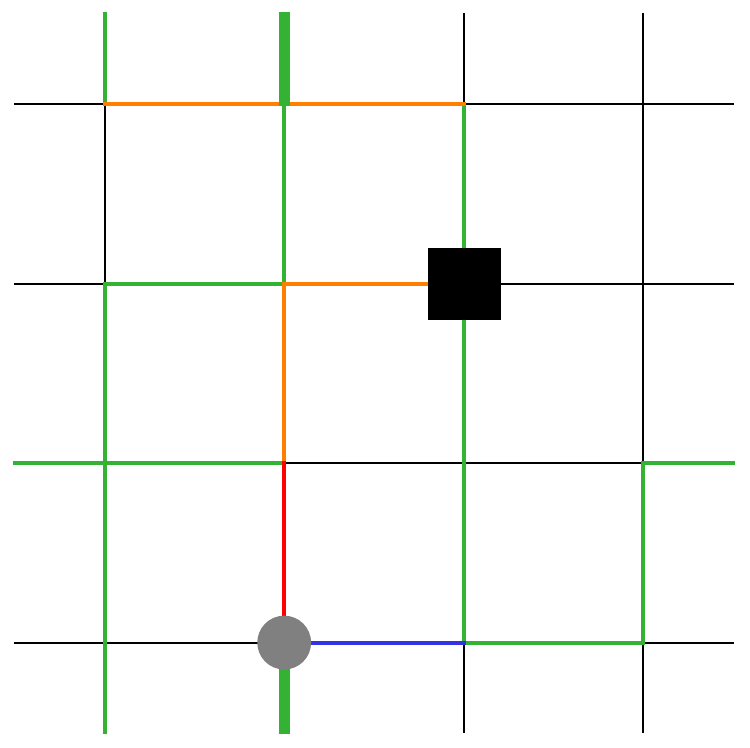}}
\end{figure}

The two algorithms for updating $l$ and $k$ are run alternately starting from a configuration where every field value is set to zero at each lattice point. Initially, the system experiences a thermalization phase that we are not interested in. Configurations and their labels at equilibrium form the dataset, with the caveat that autocorrelated data have to be avoided as much as possible, in particular because networks can learn autocorrelation instead of physical information. For this reason, we keep autocorrelation below a certain threshold by performing a sufficient number of sweeps between each data saving.

\subsection{Dataset} \label{sec:dataset}

The whole training set comprises a total of $N_{\mathrm{train}}=20\,000$ samples, and the entire validation set consists of $N_{\mathrm{val}}=2\,000$ data points, both generated at $\mu=1.05$ with a lattice size of $60\times4$. The test set contains $4\,000$ samples for every $\mu$ and lattice size that have been mentioned in~\ref{sec:pars}, amounting to a total of~\mbox{$6.8 \times 10^5$} samples. We will refer to the test set pertaining to chemical potentials $\mu \in \{ 0.91, \dots, 1.05 \}$ as test set A, while test set B encompasses the values $\mu \in \{ 1.1, \dots, 1.5 \}$. This distinction is introduced because test set A can be used to make a direct comparison with~\cite{Zhou:2018ill}, since the chemical potentials used are the same, while test set B informs about the generalization capabilities of the networks to $\mu$ higher than the one used during training. Thermalization effects are eliminated by discarding data generated in the first $1\,000$ sweeps, after which configurations and corresponding observables are saved every five sweeps. For a few combinations of chemical potential and lattice size, a high autocorrelation is observed. In these cases, the number of sweeps is increased to 50, such that autocorrelation becomes negligible.

We now take a closer look at the distributions of $k_{x,2}$ and $f_x$. These two quantities are crucial for the computation of the observables $n$ and $|\phi|^2$ respectively, hence checking their distributions can provide a better understanding of the dataset properties and can also shed more light on the possible reasons behind the success or failure of a model. Let us remind again that $k_{x,2}$ is an integer number, while $f_x$ can be either 0 or a positive even number, as can be checked by making use of its definition~\eqref{Scalar:f_x} and the flux conservation~\eqref{Scalar:flux_conservation}. In the rest of the section, we drop the lattice index $x$ and adjust the notation as follows: $k_{x,1} \rightarrow k_x$, $k_{x,2}\rightarrow k_t$, $f_x\rightarrow f$.

\begin{figure}
    \centering
    \includegraphics{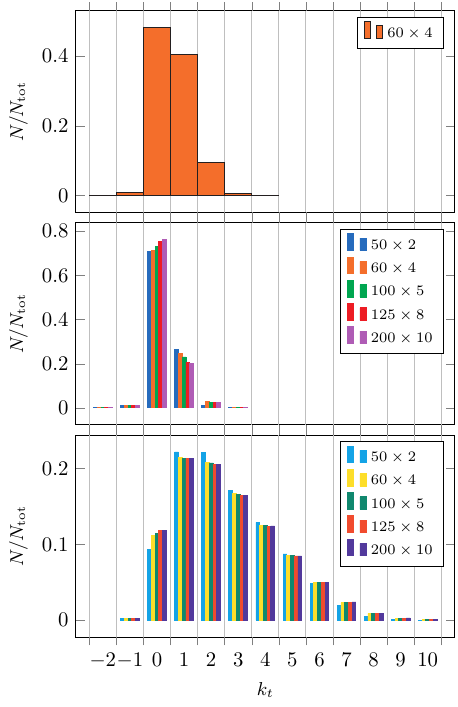}
    \caption{Distributions of the field component $k_t$. These histograms illustrate the distributions of $k_t$ in the training set (top), test set A (middle) and test set B (bottom). The test sets noticeably maintain a consistent distribution across the different lattice sizes. The training set and test set A feature mostly the same values of $k_t$, although with largely different distributions. Test set B is characterized by a less peaked distribution, where higher values of $k_t$ are reached. Densities below $10^{-4}$ are not shown. Image from \cite{Bulusu:2021rqz}.}
    \label{fig:data_distr_reg_kt}
\end{figure}

\begin{figure}
    \centering
    \includegraphics{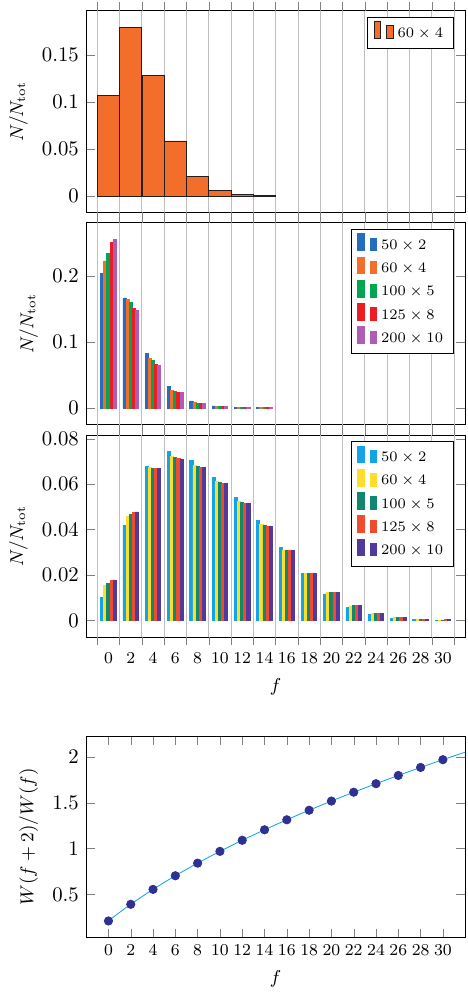}
    \caption{Distributions of $f$ and ratio of $W(f)$. The histograms depict the distributions of $f$ in the training set (top histogram), test set A (middle histogram), and test set B (bottom histogram). Similar remarks can be made for the distribution of $f$ in Fig.~\ref{fig:data_distr_reg_kt}. The last plot shows $W(f+2)/W(f)$ with $\eta=4.01$ and $\lambda=1$. The markers highlight even integer values of $f$, which directly enter the calculation of $|\phi|^2$. Weights lower than $10^{-4}$ are not shown. Image from \cite{Bulusu:2021rqz}.}
    \label{fig:data_distr_reg_f}
\end{figure}

In Fig.~\ref{fig:data_distr_reg_kt}, we report the distribution of $k_t$ for the training and testing phase. Specifically, the upper histogram displays the distribution of the training set, while the second and the third show the distributions of test sets~A and~B, respectively. The range of $k_t$ in the training set and test set A is almost identical, meaning that values used for testing have also been encountered by the networks during training, but the distributions are fairly dissimilar, which makes this task non-trivial. For what concerns test set B, the domain includes higher values, making this test in fact an analysis on the ability of the networks to extrapolate to entirely unseen data. Similar considerations can be made also for $f$, whose distributions are shown in Fig.\ref{fig:data_distr_reg_f}.

\subsection{Architecture search} \label{subsec:arch_sel}

In machine learning applications to physics, it is advisable to exploit any available knowledge about the system under examination to build a network, as learning can be facilitated and performance enhanced. Here, if we assume no prior knowledge of the exact expressions for $n$ and $|\phi|^2$, two key considerations can guide us in designing an appropriate architecture to deal with the present task.

The first one deals with the fact that the partition function in Eq.~\eqref{Scalar:flux_representation} contains products over all spacetime points, and observables are calculated as derivatives of $\ln Z$, which means that a general expression for observables involves sums over lattice sites. If the observable does not scale with the volume, therefore its definition contains a factor $1/(N_x N_t)$ on a 1+1D lattice, it is intensive and global average pooling is the appropriate layer, while if the observable does scale with the volume, it is extensive and global sum pooling is the suitable operation. Given Eqs.~\eqref{Scalar:n} and~\eqref{Scalar:phi}, the observables studied in this task are intensive, therefore we use a GAP. The dense part after this layer does not affect the intensive nature of the prediction.

The second point is about the translational symmetry of the action, which implies invariance under translations of the observables, as already discussed in section~\ref{sec:scalar}. This dictates some restrictions in the architecture that lead to a preference of the \EQ{} type. Whether such a preference is justified, is the subject of this chapter and will be empirically evaluated by testing the performance of the three architecture types presented in section~\ref{sec:archs}.

A comparison can be made, for example, by searching for a very good \EQ{} model, then add at least one spatial pooling layer to form its \ST{} counterpart, and finally replace the global pooling layer with a flattening step to obtain the corresponding \FLAT{} model. However, it is not guaranteed that the non-equivariant counterparts realized with this approach are representative of well-performing models of the same type, and the analysis could be biased towards equivariant models. In order to carry out a more meaningful and fair study, we define a set of possible hyperparameters and use an optimization strategy to identify the most promising models. The search space for each type can be found in table~\ref{tab:Regression:optuna_search_space_EQ} for \EQ{} architectures, table~\ref{tab:Regression:optuna_search_space_ST} for \ST{} and table~\ref{tab:Regression:optuna_search_space_FL} for \FLAT{}. In a preliminary phase, hyperparameters are tried out manually, which served to establish the search space in the first run. Its results are used to partially reduce the search space in the second run. In both these searches, 50 different combinations of hyperparameters are selected and the models are trained on each of the training sets that will be specified in section~\ref{sec:sets}. The extended search is intended to explore a larger hyperparameter space with 100 trials, also here with unique combinations of hyperparameters, to check if a better architecture was overlooked because of a too restricted search space. This search is exclusive to the largest training set. Not shown in the tables is the fact that an activation function is employed after every convolution and after every linear layer, except right before the prediction is made. In the first run, different activation functions are tested: \textit{ReLU, tanh, PReLU} and \textit{LeakyReLU}~\cite{Dubey:2022}. Due to the better performances obtained, the latter is the only one employed in the second run and in the extended search. An advantage of \textit{LeakyReLU} over \textit{ReLU} lies in the prevention of the well-known problem of dead neurons, which are not active after initialization or become inactive during training.

\begin{table}[tbp]
    \centering
    \scriptsize
    \caption{Search spaces for \EQ{} architectures. This table lists the possible number of convolutional (conv, \mbox{$s=1$}) and linear layers (lin), kernel sizes, the number of channels of the convolutional layers and the number of units in the linear layers. The insertion of spatial pooling layers with \mbox{$s=1$} has been avoided in these search spaces as it appears to degrade predictions. Table from \cite{Bulusu:2021rqz}}
    \vspace{2mm}
    \begin{tabular}{lllll}
        & conv & lin & kernel size & channels/nodes \\
        \hline
        run 1 & $[2,3]$ & $[0,1]$ & $\{ (1 \mspace{-5mu} \times \mspace{-5mu} 1), (2 \mspace{-5mu} \times \mspace{-5mu} 2) \}$ & $\{4,8,16,24,32,48,64,80\}$ \\
        run 2 & $[2,4]$ & $1$ & $\{ (1 \mspace{-5mu} \times \mspace{-5mu} 1), (2 \mspace{-5mu} \times \mspace{-5mu} 2) \}$ & $\{4,8,16,24,32,48,64,80\}$ \\
        \begin{tabular}{@{}c@{}} extended \\ search \end{tabular} & $[2,4]$ & $[0,3]$ & $\{ (1 \mspace{-5mu} \times \mspace{-5mu} 1), (2 \mspace{-5mu} \times \mspace{-5mu} 2) \}$ & $\{4,8,16,24,32,48,64,80\}$ \\
    \end{tabular}
    \label{tab:Regression:optuna_search_space_EQ}
\end{table}

\begin{table*}
    \centering
    \scriptsize
    \caption{Search spaces for \ST{} architectures. Here are presented the possible number of convolutional \mbox{$(s=1)$} and linear layers (abbreviated as in table~\ref{tab:Regression:optuna_search_space_EQ}), kernel sizes, the number of channels for the convolutional layers, the number of neurons in the dense part, the number of spatial pooling layers (SPL, \mbox{$s=2$}) and the spatial pooling mode (SPM). Table from \cite{Bulusu:2021rqz}}
    \vspace{2mm}
    \begin{tabular}{lllllll}
        & conv & lin & kernel size & channels/nodes & SPL & SPM \\
        \hline
        run 1 & $[2,4]$ & $[0,3]$ & $\{(1 \times 1), (2 \times 2)\}$ & $\{4,8,16,24,32,48,64,80\}$ & $\{1,2\}$ & $\{\mathrm{avg}, \mathrm{max}\}$ \\
        run 2 & $[2,4]$ & $[0,2]$ & $\{(1 \times 1), (2 \times 2)\}$ & $\{4,8,16,24,32,48,64,80\}$ & $\{1,2\}$ & avg \\
        \begin{tabular}{@{}c@{}} extended \\ search \end{tabular} & $[2,4]$ & $[0,3]$ & $\{(1 \times 1), (2 \times 2)\}$ & $\{4,8,16,24,32,48,64,80\}$ & $\{1,2\}$ & $\{\mathrm{avg}, \mathrm{max}\}$ \\
    \end{tabular}
    \label{tab:Regression:optuna_search_space_ST}
\end{table*}

\begin{table*}
    \centering
    \scriptsize
    \caption{Search spaces for \FLAT{} architectures. This table displays the possible number of convolutional \mbox{$(s=1)$} and linear layers, kernel sizes, the number of channels of the convolutional layers, the number of nodes in the linear layers, the number of spatial pooling layers \mbox{$(s=2)$} and the spatial pooling mode. The amount of convolutions is not selected directly, but is conditioned on how many $1\times 1$ convolutional layers are chosen by \textit{optuna}. Two $2 \times 2$ convolutions and their respective spatial pooling layer are always present. It is possible to additionally insert $1 \times 1$ convolutions before each $2 \times 2$ convolutional layer or between each of them and the subsequent spatial pooling. These restrictions in the position of each convolution are signaled with an asterisk next to ``kernel size''. Abbreviations are intended as in table~\ref{tab:Regression:optuna_search_space_ST}. Table from \cite{Bulusu:2021rqz}}
    \vspace{2mm}
    \begin{tabular}{lllllll}
        & conv & lin & kernel size$^*$ & channels/nodes & SPL & SPM \\
        \hline
        run 1 & $[2,6]$ & $[1,3]$ & $\{(1 \times 1), (2 \times 2)\}$ & $\{4,8,16,24,32,48,64,80\}$ & $2$ & $\{\mathrm{avg}, \mathrm{max}\}$ \\
        run 2 & $[2,6]$ & $[1,3]$ & $\{(1 \times 1), (2 \times 2)\}$ & $\{4,8,16,24,32,48,64,80\}$ & $2$ & avg \\
        \begin{tabular}{@{}c@{}} extended \\ search \end{tabular} & $[2,6]$ & $[1,3]$ & $\{(1 \times 1), (2 \times 2)\}$ & $\{4,8,16,24,32,48,64,80\}$ & $2$ & $\{\mathrm{avg}, \mathrm{max}\}$ \\
    \end{tabular}
    \label{tab:Regression:optuna_search_space_FL}
\end{table*}

The difference between \ST{} architectures and \EQ{} is the presence of spatial pooling layers with~\mbox{$s = 2$} in the convolutional part. The kernel of these layers is fixed to~$2 \times 2$, and each of them can be either a max pooling or an average pooling. The number and the position of these layers is part of the search space, but some limitations are applied. For instance, a spatial pooling has to be inserted between an activation function and a convolutional layer, meaning that it is not allowed to apply it directly to the input or immediately before global pooling. Also, if the number of convolutions is two, i.e. the minimum permitted in the search space, the only choice is to interpose one spatial pooling layer between them. 

We design \FLAT{} architectures drawing inspiration from traditional CNNs used for similar machine learning problems. The building block is formed by a~$2 \times 2$ convolution followed by a spatial pooling layer with a stride~$s = 2$. The architecture consists of two of these building blocks one after the other, with the optional insertion of a~$1 \times 1$ convolution before each~$2 \times 2$ convolution and between each of them and the following spatial pooling, with a maximum of six convolutions in total. Activation functions are inserted after each convolutional layer except before flattening.

For the hyperparameter optimization, we choose an automatized optimizer called \textit{optuna}~\cite{Akiba:2019}. The metric that has to be minimized is the validation loss averaged over three different parameter initializations. This averaging mitigates statistical fluctuations introduced by random parameter initializations, which is crucial because \textit{optuna} dynamically adjusts its search space, with early results influencing the probability distributions guiding later hyperparameter value selections. This whole procedure is repeated separately for every training set, since for small training set sizes simpler models may perform better, while more complicated ones are likely to be favored for larger sizes.

Once the \textit{optuna} runs are completed, the architectures scoring the lowest average validation loss are selected and ten new instances for each of them are trained from scratch, with the scope of further reducing the influence of random initializations. The best-performing architectures are reported for each type in table~\ref{tab:best_performing_architectures}. These architectures can be considered as representatives of their type, because many of the architectures that have been retrained exhibit similar performance, which is the case independently of the training set size. The notation \mbox{Conv($K \times K$, $N_\mathrm{in}$, $N_\mathrm{out}$)} refers to a two-dimensional convolution with equal kernel size $K$ for both directions, $N_\mathrm{in}$ input channels and $N_\mathrm{out}$ output channels. Circular padding is applied before every convolutional operation to ensure periodic boundary conditions, and a stride of one is chosen for each convolution. Average pooling layers with kernel size $K$ and stride $s$ are denoted as \mbox{AvgPool($K \times K$, $s$)}, and dense layers are written as \mbox{Linear($N_\mathrm{in}$, $N_\mathrm{out}$)} with $N_\mathrm{in}$ input nodes and  $N_\mathrm{out}$ output nodes.

\begin{table}[htbp]
\centering
\scriptsize
\caption{Best architectures for the prediction of $n$ and \mbox{$|\phi|^2$} for each architecture type. This table presents the most promising architectures obtained with the \textit{optuna} searches. The network inputs consist of field configurations ${k_{t, x}, l_{t, x}}$ provided in the form of tensors with dimensions $(N_t, N_x, 4)$. The output is formed by two nodes which approximate the observable values. The last row specifies the amount of trainable parameters for each architecture. Table from \cite{Bulusu:2021rqz}.}
\begin{tabular}{lll}
\textbf{\EQ{}} & \textbf{\ST{}}  & \textbf{\FLAT{}} \\
 \hline
Conv($1 \times 1$, 4, 64)  & Conv($1 \times 1$,  4, 80)         & Conv($1 \times 1$, 4, 64)           \\
LeakyReLU                  & LeakyReLU                          & LeakyReLU                          \\
Conv($1 \times 1$, 64, 48) & Conv($1 \times 1$, 80, 80)         & Conv($2 \times 2$, 64, 80)         \\
LeakyReLU                  & LeakyReLU                          & LeakyReLU                          \\
Conv($1 \times 1$, 48, 80) & Conv($1 \times 1$, 80, 48)         & AvgPool($2 \times 2$, $2$)         \\
LeakyReLU                  & LeakyReLU                          & Conv($1 \times 1$, 80, 48)         \\
Conv($2 \times 2$, 80, 80) & AvgPool($2 \times 2$, $2$)         & LeakyReLU                          \\
LeakyReLU                  & Conv($2 \times 2$, 48, 80)         & Conv($2 \times 2$, 48, 64)         \\
GlobalAvgPool              & LeakyReLU                          & LeakyReLU                          \\
Linear(80, 2)              & GlobalAvgPool                      & AvgPool($2 \times 2$, $2$)         \\
                           & Linear(80, 2)                      & Conv($1 \times 1$, 64, 24)         \\
                           &                                    & Flatten                            \\
                           &                                    & Linear(360, 24)                    \\
                           &                                    & LeakyReLU                          \\
                           &                                    & Linear(24, 2)                      \\
\hline
33202                      & 26370                              & 47394                              \\
\end{tabular}
\label{tab:best_performing_architectures}
\end{table}

\subsection{Training and testing} \label{sec:sets}

In subsection~\ref{sec:dataset}, the dataset has been introduced and its main characteristics have been discussed. In this section, we give the details of how the data are used during training and testing.

The training phase is repeated for each architecture in Table~\ref{tab:best_performing_architectures} with different training set sizes. The mean squared error (MSE) is used as loss function, averaging over the contribution of the two observables:
\begin{equation}
    \mathcal{L} = \displaystyle \frac{1}{2N_\mathrm{data}} \sum_{i=1}^{N_\mathrm{data}} \left[ (n_{i,\mathrm{true}} - n_{i,\mathrm{pred}})^2 + (|\phi|^2_{i,\mathrm{true}} - |\phi|^2_{i,\mathrm{pred}})^2 \right].
    \label{eq:reg_loss}
\end{equation}
Optimization is performed using the \textit{AMSGrad}~\cite{Reddi:2019} variant of the \textit{AdamW} optimizer~\cite{Loshchilov:2019} without weight decay. The reason for running the experiment with different training sets is to acquire information about sample efficiency. The utility of studying examples with fewer training samples is connected to real situations where generating configurations is highly time-consuming, e.g.~in large-scale simulations. The training set size ranges from~$100$ to~$20\,000$, with larger step sizes for a larger number of samples, as follows:
\begin{align}
    N_\mathrm{data} = \{&100, 150, 200, 250, 500, 750, 1\,000, 1\,500, 2\,000, 2\,500, 3\,000, 4\,000,  \nonumber \\
    &5\,000, 6\,000, 7\,000, 8\,000, 9\,000, 10\,000, 11\,000, 12\,000, 13\,000, 14\,000, \nonumber \\ &15\,000,16\,000, 17\,000, 18\,000, 19\,000, 20\,000\}.
    \label{eq:training_set_size}
\end{align}
The validation sets used in combination with each training set contain a number of samples amounting to~$1/10$ of the training set size Eq.~\eqref{eq:training_set_size}. In order to avoid being too close to a batch training situation, the batch size is set to $50$ for training set sizes smaller than $500$ and $100$ for bigger ones. We employ early stopping (see e.g.~Section 7.8 of \cite{Goodfellow:2016}) monitoring the validation loss with a patience of~$25$ to determine the number of epochs, which has a minimum and maximum threshold of $100$ and $1\,000$ respectively. Patience refers to the number of epochs the training process is allowed to continue without improvement in a chosen metric before stopping. The model parameters are checkpointed when the lowest validation loss is reached. An overview of the settings used in this procedure is given in table~\ref{tab:Regression:parameters_for_training}.

\begin{table}[tbp]
    \centering
    \small
    \caption{Loss, optimizer and early stopping settings for \mbox{\textit{PyTorch}}. Table from \cite{Bulusu:2021rqz}.}
    \begin{tabular}{llllll}
        loss & size\_avg & reduce & \multicolumn{3}{l}{reduction} \\
        MSELoss & None & None & \multicolumn{3}{l}{`mean'} \\
        \hline
        optimizer & lr & betas & eps & weight\_decay & amsgrad  \\
        AdamW & $0.001$ & $(0.9, 0.999)$ & $10^{-8}$ & 0 & True \\
        \hline
        \multicolumn{1}{l}{} & monitor & min\_delta & patience & \multicolumn{2}{l}{mode} \\
        EarlyStopping & `val\_loss' & $0$ & $25$ & \multicolumn{2}{l}{`min'}  \\
    \end{tabular}
    \label{tab:Regression:parameters_for_training}
\end{table}
Given the large number of maximum epochs and the early stopping criterion based on validation loss, the models are essentially trained until convergence to a very good minimum in the parameter space.
As previously mentioned, we train on $60 \times 4$ configurations generated with a chemical potential $\mu = 1.05$.
The impact of data augmentation in learning translational symmetry and improving the performance is checked for both \ST{} and \FLAT{} architectures. These are trained with and without data augmentation, which is accomplished by random shifts of the input data that depend on the architecture type. \ST{} architectures include a maximum of two spatial pooling layers with a \mbox{$2 \times 2$} kernel and a stride of~$2$, as reported in table~\ref{tab:Regression:optuna_search_space_ST}. The discussion in Subsec.~\ref{sec:layers} implies that \ST{} architectures feature translational equivariance under shifts that are multiples of~$4$, therefore data augmentation is achieved by means of shifts of~$[0,3]$ in both dimensions. At the same time, it brings to the conclusion that \FLAT{} architectures do not incorporate translational equivariance no matter what the applied shift to the input is, meaning that the possible shifts are~$[0,59]$ in the time direction and~$[0,3]$ in the space direction, where the limits come from the lattice extension.

Other than the distinction of test set A and test set B mentioned in Sec.~\ref{sec:dataset}, the models are evaluated both on the whole test set and on the subset that includes only~$60 \times 4$ configurations in the attempt to check the generalization ability to other lattice sizes and chemical potential separately. Because of the \FLAT{} architectures being restricted to one lattice size, the second test can only be performed for the \EQ{} and \ST{} types.

Let us point out that the winning \ST{} architecture can be applied to the~$50 \times 2$ lattice because it features only one spatial pooling layer with a \mbox{$2 \times 2$}~kernel and a stride~\mbox{$s = 2$} (see table~\ref{tab:best_performing_architectures}).

\subsection{Results}

In the following subsections, the details of various test results are discussed. The architectures are initially evaluated on the subset of test set A containing solely~\mbox{$60 \times 4$} configurations, later inspecting the performance on all lattice sizes. An analysis of the Silver Blaze phase transition is also carried out for various lattice sizes. At last, we will evaluate the models on test set B first on the~\mbox{$60 \times 4$} lattice then on all lattice sizes, exactly like with test set A, to study the extrapolation to chemical potential larger than the one used while training.

\subsubsection{Results on the same lattice size as training}

\begin{figure}[tbp]
	\centering
	\includegraphics{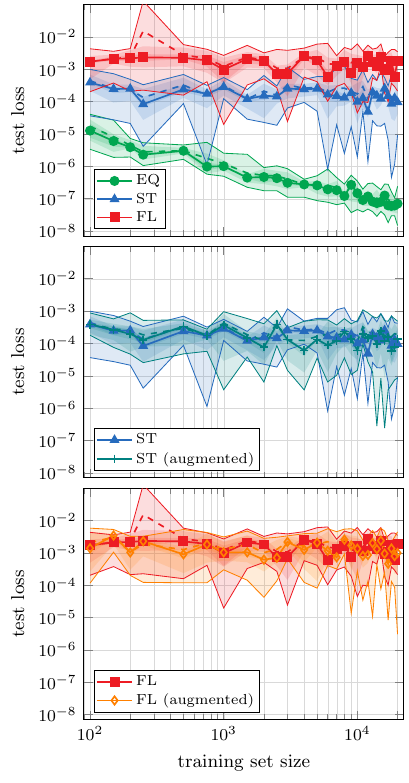}
	\caption{Test loss on the test set of all \mbox{$60 \times 4$} configurations as a function of the samples used to train. The top plot displays the results of the three architecture types trained without data augmentation. In the middle and bottom plot, the impact of training with data augmentation is shown respectively for \ST{} and \FLAT{} models. In each plot, the solid lines represent the best and worst loss, the dashed lines denote the average over the ten models that have been initialized independently, and the shaded regions indicate the $20\%$ quantiles. The symbols signal the median and are connected by a continuous line to guide the eye. The conventions just discussed will be used in similar plots later on. Image from \cite{Bulusu:2021rqz}.}
	\label{fig:test_loss_60_times_4}
\end{figure}

The same loss used for training in Eq.~\eqref{eq:reg_loss} is employed as a metric for the test set. Its values are plotted against the training set sizes listed in~\eqref{eq:training_set_size} in Fig.~\ref{fig:test_loss_60_times_4}. The different amount of training samples shows how the three architecture types behave under limited information for smaller training set sizes and the extent of the performance improvement as the information is increased. The top plot indicates that the \EQ{} models achieve a better test loss when the size of the training set becomes larger, as one would expect. On the other hand, the two non-equivariant architectures surprisingly do not benefit from having access to a greater number of samples in the training set. The center and bottom plots report on another noteworthy finding: data augmentation does not help the non-equivariant architectures to improve their results. If we only take into consideration the median of the test loss, we may come to the conclusion that the \ST{} and the \FLAT{} architectures are not able to reach a precision in the predictions as their equivariant counterpart. A non-equivariant model that has already converged to a very good local minimum in the loss function landscape is not going to benefit from the addition of more training samples, regardless of whether they are generated by Monte Carlo sampling or by data augmentation. The downward spikes in the bands of the \ST{} architecture, though, hint that it is possible for some models to achieve a good approximation of the observables and even compete with the \EQ{} models. Therefore, we conclude that although possible, it is less likely for the \ST{} and \FLAT{} models than for the \EQ{} models to learn a satisfactory approximation of~$n$ and~$\lvert \phi \rvert^2$.

\begin{figure}[htbp]
	\centering
	\includegraphics{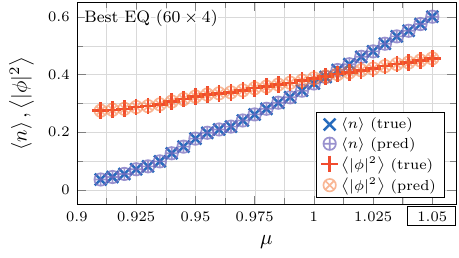}
	\caption{Predicted and true values for the ensemble averages $\left< n \right>$ and $\left< \lvert \phi \rvert^2 \right>$ as functions of the chemical potential $\mu$ on a $60 \times 4$ lattice. The predictions shown are generated by the \EQ{} model which achieved the smallest test loss. Notably, the model has been trained solely on configurations generated at $\mu = 1.05$ but exhibits remarkable generalization capabilities to other values of $\mu$. The training point is highlighted by a rectangle in this and subsequent plots. Image from \cite{Bulusu:2021rqz}.}
	\label{fig:reg_observables_over_mu}
\end{figure}

At this point, we select the best \EQ{} model according to the test loss, and evaluate it separately at all the chemical potentials present in test set A. In Fig.~\ref{fig:reg_observables_over_mu}, we plot the predictions of the observables averaged over the configuration ensemble that pertains to each $\mu$, together with the exact ensemble average. The figure illustrates that the network can generalize to all the chemical potentials in the given range, despite being trained uniquely on samples generated with $\mu = 1.05$. This remarkable generalization ability is explained by the fact that the model is not really generalizing from one value of the chemical potential to all the others, but rather from the training samples to other samples, each consisting of a configuration and the corresponding value of the observables. The samples present in the training set feature a range of values for $n$ and $|\phi|^2$ which is sufficiently large to cover most of the observable values in the test set. A similar observation has been made about the input distributions in subsection~\ref{sec:dataset}, where $k_t$ and $f_x$ span the same range in the training set and test set.

\begin{figure}[htbp]
	\centering
	\includegraphics{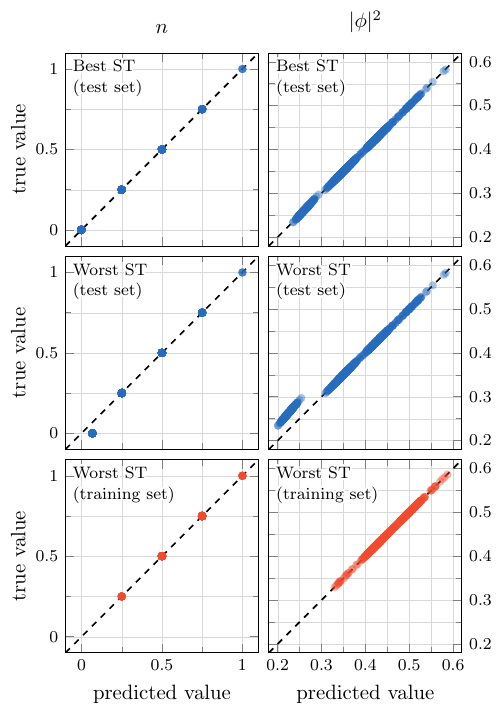}
	\caption{True versus predicted observables for the best and the worst \ST{} network, both trained on $18\,000$ samples. The two top plots illustrate that the best instance of the \ST{} architecture accurately estimates the observable values over the their whole range. Conversely, the middle plots demonstrate that the worst instance fails to predict correctly the smaller values of $n$ and $|\phi|^2$. The noticeable discrepancy between ground truth and prediction arises from the training set including only larger values of the observables, as shown in the bottom plot, and the worst model struggling to generalize beyond that range. The top and middle plots show $1\%$ of the test data, while the bottom plots display $4\%$ of the training data. Image from \cite{Bulusu:2021rqz}.}
	\label{fig:test_plots_mu_60_times_4}
\end{figure}

To illustrate this point, we employ \ST{} models that have been trained on $18\,000$ samples. In figure~\ref{fig:test_plots_mu_60_times_4}, the test results are reported in the form of a scatter plot of true values versus the predicted ones of both observables, which is done for the best and the worst performing (according to the test loss) \ST{} model respectively in the top and middle plot. The bottom plot displays the performance of the worst \ST{} model on the training data. In these scatter plots the model's predictions for each sample are drawn, not just the average over the ensemble. Both networks can predict the larger values of both observables, but the worst one largely mispredicts the smaller values, because they are not present in the training set. Therefore, we can conclude that the better \ST{} models are able to generalize to configurations and ranges of the two labels that were missing in the training phase, while the worse ones do not possess this ability. The relatively large values of the median in Fig.~\ref{fig:test_loss_60_times_4} suggest that the instances of the \ST{} architecture have a low probability of achieving good generalization. Concerning  the \FLAT{} models, their behavior is similar to the \ST{} models, but overall the predictions are less accurate.

\subsubsection{Results on different lattice sizes} \label{subsec:reg_diff_sizes}

Different lattice sizes are not suitable for the same \FLAT{} architecture that was used in the previous tests, therefore the comparison here is made between the \EQ{} and \ST{} type. Since data augmentation does not change the performance of the \ST{} architecture significantly, the test is run only for models trained without data augmentation. Additionally, we will set the training set size to $20\,000$~training samples for this comparison.

\begin{figure}[tbp]
	\centering
	\includegraphics{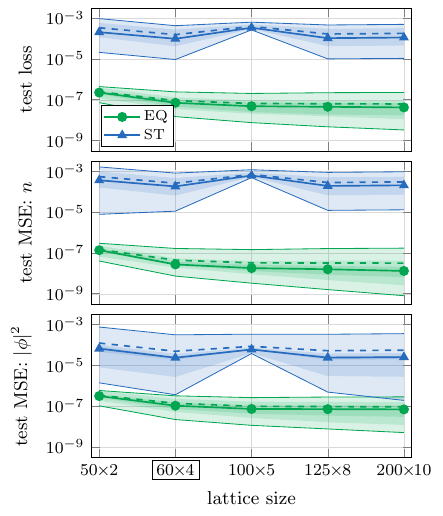}
	\caption{Overall test loss (top) and its contributions coming from each observable (middle and bottom) on various lattice sizes. Training was executed on \mbox{$60\times 4$} configurations. On average, the \EQ{} type consistently outperforms the non-equivariant counterpart. The performance does not deteriorate when testing on lattice sizes that were not used for training for neither of the two types, with the only exception of the \ST{} architecture (depicted in blue) when tested on the \mbox{$100 \times 5$}~lattice. What spoils the generalization capabilities of the models is the stride of the spatial pooling layer present in the \ST{} architecture, which allows the use of the first four rows but not of the fifth one, consequently discarding $20\%$ of the information which results in a higher loss. Image from \cite{Bulusu:2021rqz}.}
	\label{fig:test_loss_various_lattice_sizes}
\end{figure}

In Fig.~\ref{fig:test_loss_various_lattice_sizes} the total test loss in Eq.~\eqref{eq:reg_loss} is shown, while the losses of each observable are portrayed in the middle and bottom plot. Despite the \ST{} architecture maintaining a worse performance for all lattice sizes with respect to the \EQ{} architecture, both types display a remarkable generalization ability. The only exception is found for the \ST{} networks on the \mbox{$100 \times 5$}~lattice, where a kink in the blue curve is present in the prediction of both observables, an anomaly that does not manifest itself for \EQ{} architecture. This issue arises from the odd number in the lattice dimension, which in combination with the \ST{} architecture leads to an information loss. Let us review the effect of the \ST{} architecture in Table~\ref{tab:best_performing_architectures} on configurations with such a lattice size: the first three convolutions do not affect the input size because of the application of circular padding. Then, the \mbox{$100 \times 5$}~feature maps are passed to a spatial pooling layer with a $2 \times 2$~kernel and a stride of~$2$. This layer operates on the first four rows and discards the last one, effectively wasting~$20 \%$ of the information, and outputs a feature map with size~\mbox{$50 \times 2$}. For this reason, the \ST{} networks cannot use all of the data to make predictions, resulting in a less accurate test loss. An analogous observation can be made for the \mbox{$125 \times 8$}~lattice, but in that case the missing information only amounts to $1 / 125$ of the total available, whose effect is not visible in Fig.~\ref{fig:test_loss_various_lattice_sizes}.

\subsubsection{Silver Blaze phase transition}

The Silver Blaze~\cite{Cohen:2003kd} phenomenon is a second-order phase transition occurring at zero temperature~$T$, for which thermodynamical quantities are independent of the chemical potential~$\mu$ below a critical value~$\mu_c$~\cite{Gattringer:2013df}. Specifically, the observables $\langle n \rangle$ and $\langle \lvert \phi \rvert^2 \rangle$ remain constant when~\mbox{$\mu < \mu_c$}, and they start increasing once the chemical potential goes above the $\mu_c$ threshold. The particle density~$\langle n \rangle$ serves as an order parameter for the Silver Blaze phase transition. Due to the finite size of the lattices, the temperature cannot vanish completely, hence the transition is not guaranteed to be sharp. Given that the networks are trained to approximate two observables that are relevant to study this phenomenon, their predictions are expected to reproduce the Silver Blaze phase transition.

\begin{figure}[htbp]
	\centering
	\includegraphics{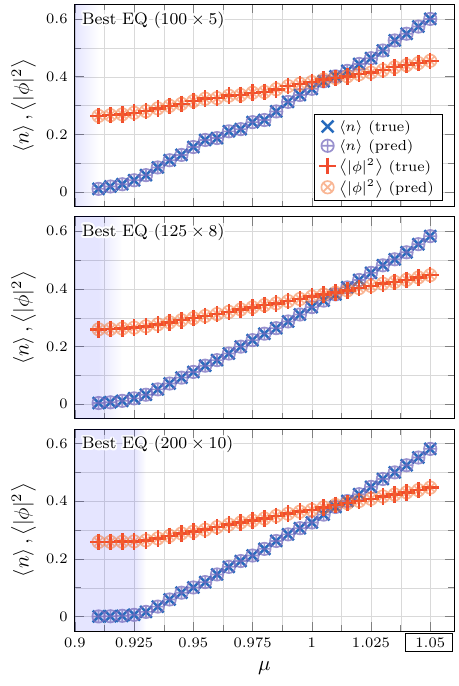}
	\caption{Predicted and true ensemble averages of the physical observables at each $\mu$ on the larger lattices. The \EQ{} model trained on $20\,000$ samples with the best performance on the test set has been selected to provide the predictions shown in the plot. Training was executed only at $\mu = 1.05$. Notably, the kinks in the curves offer an estimate of the Silver Blaze phase transition, highlighted by the color gradient from the shaded region to the white background. The kinks become more evident as the lattice size increases. Image from \cite{Bulusu:2021rqz}.} 
	\label{fig:silver_blaze_prediction_larger_lattice_sizes}
\end{figure}

In Figure~\ref{fig:silver_blaze_prediction_larger_lattice_sizes} are illustrated the predictions of the \EQ{} model that has been trained on $20\,000$~training samples and has achieved the lowest validation loss along with the true values. The symbols correspond to the ensemble average of each observable for every chemical potential in test set A on the~\mbox{$100 \times 5$},~\mbox{$125 \times 8$} and \mbox{$200 \times 10$} lattices respectively in the top, middle and bottom plot. On the larger lattices, the phase transition is visible, while on the smaller ones no phase transition is detected within the range of~$\mu$ under examination. The reason for this is that the critical value~$\mu_c$ decreases with increasing temperature.

The phase transition is accurately predicted also by the \ST{} models that are able to generalize to smaller values of the observables, such as the one that was used in Fig.~\ref{fig:test_plots_mu_60_times_4}, but this is not the case for all the \ST{} models.

\subsubsection{Extrapolation to larger chemical potentials}

The results obtained by the \EQ{} architecture and some \ST{} models on test set A have demonstrated their ability to generalize to different chemical potentials and lattice sizes. As we have argued, in such a test set the fields and the observables span values that are also found in the training set, which allows some networks to generalize well. We are now interested in investigating the behavior of the models when we increase the value of the chemical potential, reaching field and observable values that were not present while training. Test set B, as described in subsection~\ref{sec:dataset}, serves this scope, therefore we evaluate the models on it without retraining.

\begin{figure}[tbp]
	\centering
	\includegraphics{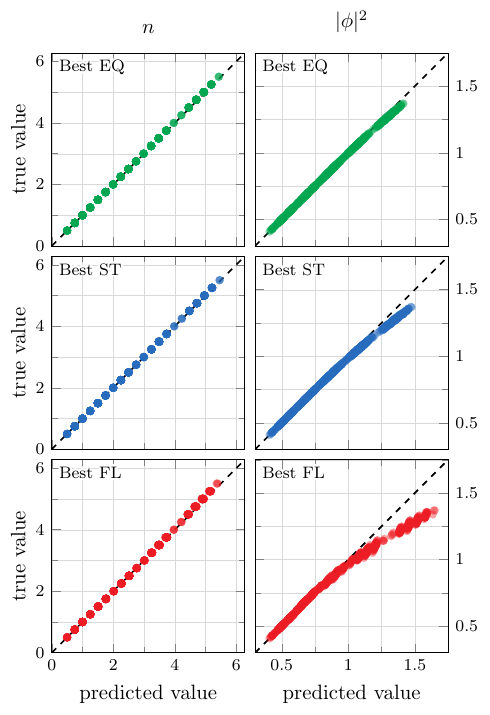}
	\caption{Scatter plot of predicted versus true observable values for the model achieving the lowest validation loss for each architecture type, evaluated on \mbox{$60 \times 4$} configurations generated with \mbox{$\mu \in [1.1, 1.5]$}. The particle number density is predicted with good accuracy also for values larger than the ones used during training by all three models, while for $|\phi|^2$ the generalization capability deteriorates as the values lie further away from those belonging to the training set. In the plots, $6.25\%$ of the test data has been employed. Image from \cite{Bulusu:2021rqz}.}
	\label{fig:reg_scatter_higher_mu}
\end{figure}

Figure~\ref{fig:reg_scatter_higher_mu} shows scatter plots of the true values against the predicted ones for both observables on the \mbox{$60 \times 4$} lattice. Each row displays the outcome respectively for the best \EQ{}, \ST{} and \FLAT{} type, according to the validation loss. All three architectures struggle with the prediction of higher values of $\lvert \phi \rvert^2$ more than with the prediction of higher $n$. This can be understood considering the simple linear dependence on $k_t$ in Eq.~\eqref{Regression:n} and the non-linear dependence on all fields in Eq.~\eqref{Regression:phi2}, as pointed out at the beginning of this section. However,
the predictions made by the best \EQ{} model exhibit a closer alignment with the optimal results represented by the identity line compared to the predictions generated by the non-equivariant counterparts, of which the \FLAT{} architecture displays a more pronounced deviation from the ground truth.
This leads to a noticeable discrepancy in the ensemble averages of the observables only for~\mbox{$\mu = 1.5$}, and this trend is also consistent across all considered lattice sizes for \EQ{} and \ST{} architectures. The selection criterion for the best architecture is based on the validation loss and should not involve test results, but we point out that there exist some models for each architecture that extrapolated better than the respective ones used in Fig.~\ref{fig:reg_scatter_higher_mu}.

\begin{figure}[tbp]
	\centering
	\includegraphics{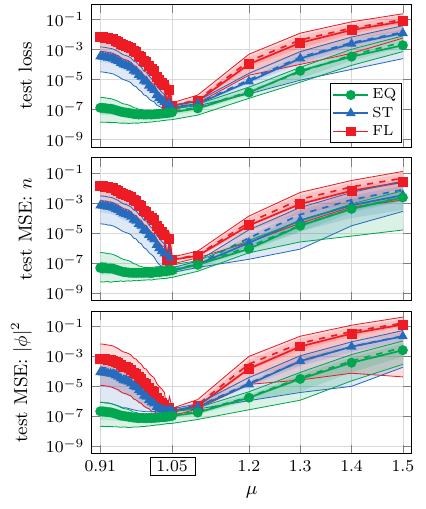}
	\caption{Total test loss and its contributions given by the physical observables $n$ and $|\phi|^2$ as functions of the chemical potential on the \mbox{$60 \times 4$} lattice. For each architecture type, the test results of an ensemble of the best architecture trained on $20\,000$ training samples are illustrated. Together with Fig.~\ref{fig:test_loss_60_times_4}, these plots confirm a substantial difference in the accuracy of the predictions for \mbox{$\mu \le 1.05$}. For larger chemical potentials, the loss seems to increase in a more similar way for each type, although in the range \mbox{$1.2 \le \mu \le 1.5$} the separation between the average of the total test loss of \EQ{} and \ST{} models is approximately one order of magnitude, just as is between the average loss of \ST{} and \EQ{} models. Image from \cite{Bulusu:2021rqz}.}
	\label{fig:reg_loss_over_mu}
\end{figure}

In Fig.~\ref{fig:reg_loss_over_mu} the total and individual test losses on \mbox{$60 \times 4$} configurations are plotted against the chemical potential. As expected, the predictions become worse for all the architectures as $\mu$ increases, but while for $\mu < \mu_c$ the difference between equivariant and non-equivariant architectures is considerable, for $\mu > \mu_c$ the loss grows more similarly for the three types, although maintaining the same ranking, with \EQ{} models achieving the best results, followed by \ST{} and then \FLAT{} networks. For instance, at \mbox{$\mu = 1.5$} the mean and median losses scored by the \EQ{} architecture are lower than those of the non-equivariant ones, but the lowest loss is reached by the \ST{} best model. A similar behavior is observed on the other lattice sizes when testing the \EQ{} and the \ST{} architectures, apart from the \mbox{$100 \times 5$} lattice, where the \ST{} models encounter the same problem shown in Fig.~\ref{fig:test_loss_various_lattice_sizes}. We note that Fig.~\ref{fig:reg_scatter_higher_mu} visualizes the results of the model with the lowest validation loss associated with each individual architecture. For the \EQ{} type, it is a model of the ensemble that has been trained with $20\,000$ samples, while for the \ST{} and the \FLAT{} types, it is a model that has been trained with $18\,000$ samples. On the other hand, in Fig.~\ref{fig:reg_loss_over_mu} we used the ensemble of models trained on $20\,000$ samples for every architecture.

\subsubsection{Results summary}

To summarize, the best translationally equivariant architecture outperforms the respective best model of the two non-equivariant types on the lattice size they have been trained on. Only a part of the \ST{} networks are found to generalize beyond values of observables they have been shown during training, while no \EQ{} network has exhibited limitations. The behavior of \FLAT{} models is analogous to the behavior of \ST{} networks, but their predictions are less accurate overall. The tests on different lattice sizes are not directly possible for \FLAT{} architectures, and are thus executed only for the \EQ{} and the \ST{} types. These are both capable of generalizing to different lattice sizes, although the \ST{} models score the higher average test loss on the \mbox{$60 \times 4$}~lattice, which can be explained by their lack of generalization to observable values absent in the training set. Another restriction is that \ST{} architectures are not suited to make predictions on any arbitrary lattice size: depending on the lattice dimension and the stride of the spatial pooling layers in the network, parts of the information can be lost. In contrast, \EQ{} architectures have no such restriction. Despite being trained solely on~\mbox{$\mu = 1.05$} on the \mbox{$60 \times 4$}~lattice, many models effectively predict the Silver Blaze phase transition on a different lattice size, with~\mbox{$\mu_c \ll 1.05$}. The \EQ{} models accomplish this result remarkably well. Data augmentation does not provide significant benefits in training \ST{} and \FLAT{} architectures, which is why this approach is not employed in the next two tasks.

As a last important consideration, the results found on test set A can be compared with the ones in~\cite{Zhou:2018ill}, where this regression task was performed on a test set generated with the same physical parameters. The architecture employed there belongs to the \FLAT{} category, and can therefore only be applied to the lattice size used for training, which in that case was chosen to be $200 \times 10$. Training was performed on configurations pertaining to $\mu = 0.91$ in addition to $\mu = 1.05$, thus providing information from both phases. Furthermore, the number of trainable parameters of the architecture in~\cite{Zhou:2018ill} has an order of magnitude of \mbox{$10^7$}, much more than our best model needs, which is about \mbox{$3 \times 10^4$}. Other well-performing models we found contain about an order of magnitude fewer parameters than our best one. The test loss achieved by the architecture in~\cite{Zhou:2018ill} is of the order of~$10^{-6}$, while the best \EQ{} model  achieves a loss of about an order of magnitude smaller.

\section{Task II: detecting flux violations} \label{sec:class}

\begin{figure*}
\subfigure[~Example field configuration]{\includegraphics[scale=.8]{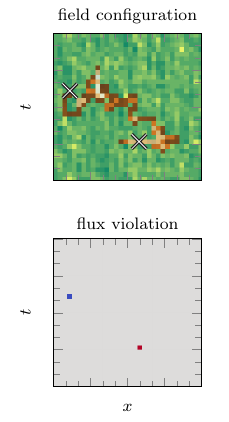}}
\hfill
\subfigure[~Feature maps of convolutional network in best \EQ{} and \ST{} models]{\includegraphics[scale=.8]{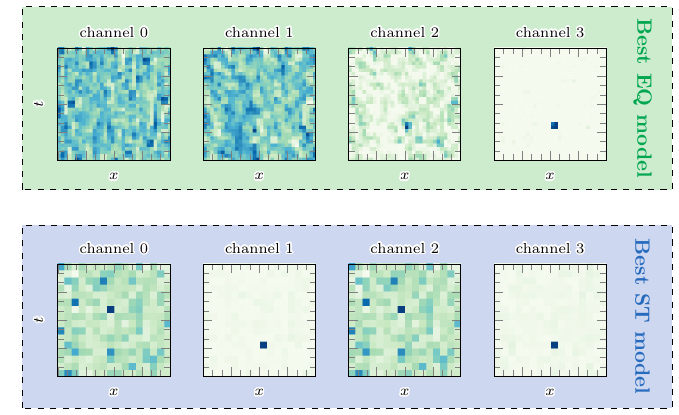}}
	\caption{Visualization of an open worm field configuration and of one of the best models' feature maps. (a) A field configuration illustrating an open worm (depicted in brown) created with movements like the ones in Fig.~\ref{fig:worm_movement}, and the corresponding flux violation computed with Eq.~\eqref{eq:flux}. The violations are located at the two open ends of the worm, marked with crosses in the upper plot. (b) First four channels out of 32 and 16 respectively for the best \EQ{} (top, green) and \ST{} (bottom, blue) models in one of the feature maps of the convolutional part. In some of the channels one of the two flux violations is detected (e.g., channels 2 and 3 for \EQ{} and 1 and 3 for \ST{}), while for others (e.g., 0 and 1 for \EQ{} and 0 and 2 for \ST{}) the output is not straightforwardly interpretable. Image from \cite{Bulusu:2021rqz}.}
	\label{fig:openworm_schematic}
\end{figure*}

The first task was successfully tackled by relatively simple CNN models, which were favored by the hyperparameter optimizer \textit{optuna}. This is not surprising if we consider that the observable $n$ can be reproduced exactly by an equivariant linear model with a single $1 \times 1$ convolution. Since information from next-neighboring sites is required for an exact representation of $|\phi|^2$, it is impossible for a simple $1 \times 1$ convolution to perfectly reproduce this observable, however we empirically observed that architectures featuring only $1 \times 1$ convolutions were able to satisfactorily approximate it.

In this section, we would like to extend our architecture comparison to a task where a trivial $1 \times 1$ convolutional layer cannot be trained to provide almost perfect predictions and information from multiple lattice sites is simultaneously required. The quantity we identified as a good candidate is the local flux
\begin{equation}
\mathcal{F}_x \equiv \sum_{\nu=1}^D \left( k_{x,\nu} - k_{x-\hat{\nu}, \nu} \right) \in \mathbb{Z}.
\label{eq:flux}
\end{equation}
This is also a local quantity, but unlike $|\phi|^2$ it requires field values from nearest neighbors, which means that a $1 \times 1$ convolution is not sufficient for a good approximation.

We tackle a classification task where the map between a given field configuration $X = \{ k_{x,\mu}, l_{x, \mu} \}$ and its label $y(X)$:
\begin{align}
y(X) = \left\{
\begin{array}{ll}
0 &\quad \mathcal{F}_x = 0, \quad \forall x, \\
1 &\quad \rm{else} \\
\end{array}
\right.
\end{align}
has to be found.
The worm algorithm delineated in subsection~\ref{sec:worm} generates physical field configurations, for which the flux constraint $\mathcal{F}_x = 0$, $\forall x$ is respected. Relaxing this condition and allowing the existence of open worms, hence flux violations, is achieved by a modification of the original worm algorithm. This is done after an initial thermalization performed with the alternate update of the fields $l$ and $k$. On top of the resulting configuration, a new worm is started and the configuration with the open worm is saved. The worm moves on the lattice and the new configuration replaces the previous one with probability $1/\ell$, where $\ell$ is the current worm length, until the worm closes. With this method, each open worm configuration has equal probability of being picked. \footnote{This can be proven by induction. Starting from a closed worm configuration, the head of the worm moves and the corresponding configuration is saved. After the next accepted movement (provided the worm stays open) the resulting configuration is stored with a probability of $1/2$, hence the previous configuration remains stored with a probability of $1/2$. After a third movement, the new configuration replaces the old one with a probability of $1/3$. The other two have a probability of $(1-1/3)=2/3$ of remaining saved. Since they had the same probability before the last step, they will equally share the probability of $2/3$, meaning that each of the three possible configurations has the same probability of being stored. This can be generalized to an arbitrary number of movements.}
As also visible in Fig.~\ref{fig:openworm_schematic}, the flux conservation is violated at the two ends of the open worm. In this work, open worms will serve as labels in the tasks the neural networks deal with, but they have physical relevance on their own, since they enter the calculation of $n$-point functions of $\phi$ \cite{Gattringer:2013ap, Rindlisbacher:2016zht}.

\subsection{Physical parameters, lattice sizes and dataset}

The dataset consists of two classes, closed worm and open worm configurations. The first is obtained with the original worm algorithm, while the second one results from a distinct run with the modified version described earlier. Both datasets are characterized by the same values of the physical parameters and lattice dimensions, specifically the coupling constant is fixed to $\lambda = 1$, the mass $m$ can take three values corresponding to $\eta = 4 + m^2 \in \{4.01, 4.04, 4.25\}$, the chemical potential can be set to $\mu \in \{1, 1.25, 1.5\}$, and the lattices are square, with $N_t = N_x \in \{8, 16, 32, 64\}$. Only a subset of these possible values is used for the training set, namely the two combinations $(\eta, \mu) \in \{(4.01, 1.5), (4.25, 1)\}$ and the smallest lattice size, i.e. $8 \times 8$, with a total number of $N_{\mathrm{train}}=4\,000$ instances equally distributed among each class and parameter combination. The validation set is structured in the same way, but contains $N_{\mathrm{val}}=400$ samples. All combinations of parameters and lattice sizes are employed to generate the test set, and for each of them 100 samples pertain to closed worm configurations and another 100 pertain to open worm configurations, with a total of $7\,200$ data. The thermalization phase is taken care of by discarding data in the first $2\,000$ sweeps, after which measurements are saved every $100$ sweeps. As in subsection~\ref{sec:dataset}, we checked the distribution of the input fields. The findings are very similar to what will be shown in the next task, where they are also discussed.

\subsection{Architecture search, training and testing}

We compare the performance of the three different architecture types reported in Fig.~\ref{fig:architectures} on this task. To ensure a fair comparison, we again rely on \textit{optuna} to perform a search for promising architectures based on validation loss, which is chosen to be the binary cross entropy loss
\begin{align}
    \mathcal{L} = -\frac{1}{N_\mathrm{data}} \sum_{i=1}^{N_\mathrm{data}} \left[y_i \log p(y_i) + (1 - y_i) \log (1 - p(y_i)\right],
\end{align}
where $y_i$ denotes the class (closed or open worm) and $p$ the probability of belonging to such a class. For every type, the maximum number of convolutions is $N_\mathrm{conv, max} = 3$, all with a maximum kernel size of $K = 3$, $N_\mathrm{ch} \in \{ 4, 8, 16, 32 \}$ possible channels and equipped with circular padding. After each convolutional layer a \textit{LeakyReLU} activation function is applied. Furthermore, every convolution except the last one can be followed by a pooling layer (average or max pooling) with a stride $s = 1$ in the case of \EQ{} architectures and $s = 2$ in the case of \ST{} and \FLAT{} architectures. Non-equivariant architectures need to feature at least one pooling layer with $s=2$. At the end of the convolutional part of the network, a global max pooling layer is used for \EQ{} and \ST{} architectures, while a flattening step characterizes the \FLAT{} ones. Other global pooling layers can be employed, like average pooling or sum pooling, but given that the task deals with the detection of local defects, global max pooling is a more natural choice. Another option that is investigated in the search is to explicitly set the bias terms to zero in every convolutional layer. The resulting output is further processed by a dense network with up to $N_\mathrm{dense, max}  = 2$ layers, each with $N_\mathrm{nodes} \in \{ 4, 8, 16, 32 \}$ possible nodes. Similarly to the convolutional part, also in the dense part of the architecture \textit{LeakyReLU} follows each linear layer. A final linear layer leads to a single node, followed by a sigmoid activation function. The bias in each linear layer can be set to zero in the search.

For each architecture type, we conduct two \textit{optuna} runs with 400 trials each. The combinations of hyperparameters that are picked are trained starting from five random initializations of the weights, in order to reduce the impact of random fluctuations in the hyperparameter search. The best-performing architecture of the two runs according to the validation loss is selected and trained 50 times with different parameter initializations to provide an ensemble of models.

The training process follows a similar prescription to the regression task in section~\ref{sec:reg}. We employ the \textit{AMSGrad} variant of the \textit{AdamW} optimizer without weight decay, the learning rate is chosen to be $\lambda_{lr} = 10^{-3}$, the batch size 100 and the epochs are set to 200. Early stopping based on validation loss is used with a patience of 50.

\begin{table}
\centering
\scriptsize
\caption{Best architectures for the detection of flux violations. This table reports the architectures for each type resulting from the \textit{optuna} searches. The input of every model are configurations reshaped as $(N_t, N_x, 4)$ tensors. The output is a binary classification probability. The row at the end indicates the number of learnable parameters for each architecture. The asterisk (*) signals layers where no bias is used. Table from \cite{Bulusu:2021rqz}.}
\vspace{2mm}
\begin{tabular}{l l  l}
\textbf{\EQ{}} & \textbf{\ST{}}  & \textbf{\FLAT{}} \\
 \hline
Conv($2 \times 2$, 4, 32)  & Conv${}^*$($2 \times 2$, 4, 16)    & Conv${}^*$($3 \times 3$, 4, 8)    \\
LeakyReLU                  & LeakyReLU                          & LeakyReLU                         \\
Conv($1 \times 1$, 32, 32) & MaxPool($2 \times 2$, $2$)         & MaxPool($2 \times 2$, $2$)        \\
LeakyReLU                  & Conv($1 \times 1$, 16, 16)         & Conv($2 \times 2$, 8, 32)         \\
GlobalMaxPool              & LeakyReLU                          & LeakyReLU                         \\
Linear(32, 32)             & Conv($1 \times 1$, 16, 8)          & AvgPool($2 \times 2$, $2$)        \\
LeakyReLU                  & LeakyReLU                          & Conv($2 \times 2$, 32, 32)        \\
Linear${}^*$(32, 1)        & GlobalMaxPool                      & LeakyReLU                         \\
Sigmoid                    & Linear${}^*$(8, 32)                & Flatten                           \\
                           & Linear(32, 1)                      & Linear${}^*$(128, 1)              \\
                           & Sigmoid                            & Sigmoid                           \\
\hline
2657                       & 953                                & 5600                              \\
\end{tabular}
\label{tab:class_archs}
\end{table}

In Table~\ref{tab:class_archs}, the best architectures resulting from the \textit{optuna} search for each type are displayed.

\subsection{Results}

\begin{figure}
    \centering
    \includegraphics{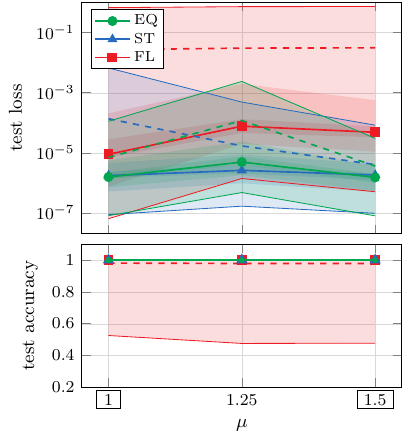}
    \caption{Test loss (top) and test accuracy (bottom) of the best \EQ{} (green), \ST{} (blue) and \FLAT{} (red) architecture types over the chemical potential $\mu$ on the $8 \times 8$ lattice for the detection of flux violations. The training phase was carried out using data generated at $\mu=1$ and $1.5$ exclusively. The shaded regions indicate where the ensemble of all 50 randomly initialized models ranges, while the conventions for quantiles, symbols and lines are identical to the ones in Fig.~\ref{fig:test_loss_various_lattice_sizes}. According to both metrics, the \EQ{} and \ST{} architectures achieve comparable performances and outclass the \FLAT{} architecture. Image from \cite{Bulusu:2021rqz}.}
    \label{fig:class_res_8x8}
\end{figure}

\begin{figure}
    \centering
    \includegraphics{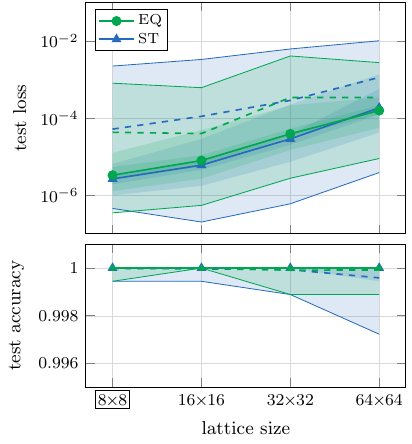}
    \caption{Test loss (top) and test accuracy (bottom) of the best \EQ{} (green) and \ST{} (blue) architecture types as functions of lattice size for the detection of flux violations. Training took place only on the $8 \times 8$ lattice. The colored bands and the solid and dashed lines have the same meaning as in Fig.~\ref{fig:class_res_8x8}. Both types of architecture demonstrate robust generalization across different lattice sizes, with a slightly reduced variance in the performance of the \EQ{} models. Image from \cite{Bulusu:2021rqz}.}
    \label{fig:class_res}
\end{figure}

The main test results are depicted in Figs.~\ref{fig:class_res_8x8} and \ref{fig:class_res}. 
In Fig.~\ref{fig:class_res_8x8}, we provide a comparison of the three architecture types evaluated on $8 \times 8$ lattices showing the dependence on the chemical potential. On average, the \EQ{} and \ST{} models exhibit an excellent accuracy and a remarkable test loss, while the ensemble of \FLAT{} models contains statistical outliers which contribute to worsen the average performance. Similar results are found when the test loss is plotted as a function of $\eta$. The comparison over all lattice sizes between \EQ{} and \ST{} types is plotted in Fig.~\ref{fig:class_res}, and shows that both architectures are able to effectively generalize to larger lattices, as well as to other physical parameters. \FLAT{} models are excluded from this test, because no retraining has been run and therefore they could be applied to $8 \times 8$ lattices, where training happened. While the \FLAT{} architecture scores worse results on average than the other two, the \EQ{} and \ST{} networks achieve a comparable performance, in contrast to the previous task. The breaking of translational symmetry caused by pooling layers with stride $s > 1$ does not ruin the performance of \ST{} models in distinguishing closed from open worm configurations.

The results in Figs.~\ref{fig:class_res_8x8} and \ref{fig:class_res} can be analyzed in more detail trying to understand and interpret how the \EQ{} and \ST{} models can reach such a high degree of accuracy. In order to do this, we inspect the feature maps of already trained \EQ{} and \ST{} models that appear in the convolutional part of the network, some examples of which are shown in Fig.~\ref{fig:openworm_schematic}~(b). We observe that in some of the channels of these feature maps flux violations in the proximity of one of the open ends of the worm are highlighted, as visible in Fig.~\ref{fig:openworm_schematic}~(a). Interestingly, it is sufficient for the network to detect a single defect to accurately predict the right class. This is reasonable if we consider that the networks did not have direct access to the local flux in Eq.~\eqref{eq:flux} during training, but were provided with global information about whether a configuration features a flux violation.

Finally, we highlight that the number of weights of the best architectures found with \textit{optuna} have an order of magnitude of $10^3$, as reported in Table~\ref{tab:class_archs}, which is rather small compared to the neural networks typically used.

\section{Task III: counting flux violations} \label{sec:multi_worm}

The task tackled in the previous section is generalized here to include more than one open worm, so that the networks need to solve the regression problem of finding a map between field configurations and the number of open worms present in it.

As discussed earlier, when an open worm is added, its endpoints show a flux violation, such that the flux in Eq.~\eqref{eq:flux} is $\mathcal{F}_x=\pm1$. If the algorithm for generating a new open worm is run on top of an open worm configuration, the second worm might cross the flux violations already present, where the Metropolis acceptance probability cannot be defined. Therefore, we realize a traffic-light system which forbids every new worm to move into a flux violation. With this caveat, the algorithm for generating an open worm is repeated until the desired number of open worms is reached, then the resulting configuration is saved. Afterwards, we go back to the last configuration without open worms and perform 100 waiting sweeps before adding open worms again.

Since lattice sites that are characterized by a flux violation are not accessible to new open worms, the flux can take three values: 0, $+1$ and $-1$. This implies that a configuration with $N_\mathrm{worms}$ open worms features $2N_\mathrm{worms}$ points where $\mathcal{F}_x=\pm1$. Therefore, the function that needs to be approximated can be expressed as
\begin{equation}
    y(X)=\displaystyle\frac{1}{2}\sum_x\left|\mathcal{F}_x\right|,
\end{equation}
where $X$ is a lattice configuration $\{k_{x, \mu}, l_{x, \mu\}}$. It is worth mentioning that this task can be viewed as a counting problem, such as crowd counting~\cite{Gao:2020}, although in a simplified version.

\subsection{Physical parameters, lattice sizes and dataset} \label{subsec:multi_worm_dataset}

We keep using the same set of physical parameters and lattice sizes of the previous section, with the addition of a number of open worms $N_\mathrm{worms} \in \{0, 1, \dots, 10\}$. In order to investigate the generalization capabilities of the neural networks as in the other tasks, the training set consists of data generated at a small subset of all the possible combinations, namely $(\eta, \mu) \in \{(4.01, 1.5), (4.25, 1)\}$, the lattice size $8 \times 8$ (the same that were used in the classification problem), and a number of worms $N_\mathrm{worms} \in\{0,5\}$. The training samples are $N_\mathrm{train}=20\,000$ and are distributed equally between two different numbers of open worms and physical parameters. The validation set is structured analogously but consists of $N_\mathrm{val} = N_{\mathrm{train}} / 10 = 2\,000$ samples. Every parameter combination is employed for the test set generation, with 100 samples each, leading to a total of $39\,600$ instances.

\begin{figure}[ht]
    \centering
    \includegraphics{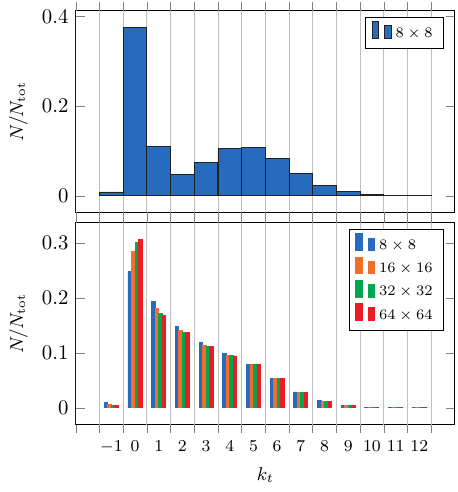}
    \caption{Distributions of the link field $k_t$. These two histograms feature the distributions of $k_t$ in the training set (top) and in the test set (bottom). Image from \cite{Bulusu:2021rqz}.}
    \label{fig:data_distr_reg2_kt}
\end{figure}

\begin{figure}[ht]
    \centering
    \includegraphics{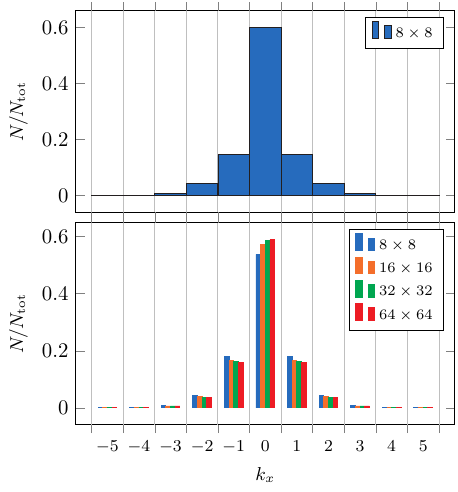}
    \caption{Distributions of the link field $k_x$. These two histograms feature the distributions of $k_x$ in the training set (top) and in the test set (bottom). Image from \cite{Bulusu:2021rqz}.}
    \label{fig:data_distr_reg2_kx}
\end{figure}

We now drop the lattice index $x$ to use the notation $k_t$ and $k_x$ for the two fields that are involved in the computation of the flux, as indicated by Eq.~\eqref{Scalar:flux_conservation}. The distributions of such fields are illustrated in Figs.~\ref{fig:data_distr_reg2_kt} and~\ref{fig:data_distr_reg2_kx}, respectively. We notice that for $k_t$ the distribution is visibly different for training and testing configurations, even though the range of possible values is the same. The selection of the two $(\eta,\mu)$ pairs used for training is specifically designed to encompass the lower and higher values of $k_t$. This choice ensures that the domain covered is the same for training and testing. Each of the two peaks in the $k_t$ training distribution in the top histogram in Fig.~\ref{fig:data_distr_reg2_kt} is associated with one of the two $(\eta,\mu)$ pairs. The behavior of $k_t$ is related to its coupling with the chemical potential, whereas $k_x$ lacks this feature and in fact does not exhibit large differences between training and testing distributions.

\subsection{Architecture search, training and testing}

An initial exploration phase is conducted to examine trends with different hyperparameter choices, alongside an empirical check that a global sum pooling is the most favorable layer after the convolutional part of the networks. This is because the quantity that needs to be predicted is extensive, as discussed in Section~\ref{sec:archs}. The insights gained during this preliminary stage are used to define the architecture search space of each of the three architecture types depicted in Fig.~\ref{fig:architectures}. To reduce bias in favor of a particular architecture type as much as possible, the search spaces are designed to be similar.

The search space for the \EQ{} architecture involves $N_\mathrm{conv} \in \{2, 3, 4\}$ convolutional layers with a kernel size $K \in \{1, 2, 3\}$, followed by a global sum pooling layer leading to a dense network consisting of $N_\mathrm{dense} \in \{0, 1, 2\}$ layers. For the \ST{} architecture, the search space is the same except for the insertion of $N_\mathrm{pool} \in \{1, 2\}$ spatial pooling layers with a stride of 2. Given that training takes place on $8 \times 8$ lattices, we limit the spatial pooling layers to a maximum of two, ensuring that the lattice is not reduced to only one site, as would happen with three spatial pooling layers, preserving the effectiveness of global sum pooling. The search space of the \FLAT{} architecture type includes two mandatory convolutions with a kernel size $K \in \{2, 3\}$, and features a spatial pooling layer after each of them. An optional $1 \times 1$ convolution is inserted before and after each mandatory convolution, resulting in $N_\mathrm{conv}^{'} \in \{2, 3, 4, 5, 6\}$ convolutions in total. The convolutional part is followed by a flattening step and a dense network with $N_\mathrm{dense}^{'}\in\{1, 2, 3\}$ layers. The maximum number of layers is increased compared to the other two architecture types to compensate for the potential absence of $1 \times 1$ convolutions. For all search spaces, the channels in the convolutions and the nodes in the dense layers take the following values: $N_\mathrm{ch/nodes}\in\{4, 8, 16, 32\}$.

Other hyperparameter choices are not part of the search space and are fixed based on the indications provided by the exploratory phase, with an outcome very similar to the other tasks. Circular padding is employed in all convolutions to take care of the periodicity of the lattice. We choose \textit{LeakyReLU} as activation function throughout the network, and apply it after every convolution and every linear layer except the last one. The convolutions and the linear layers do not have a bias term.

It is worth mentioning that another \textit{optuna} search was carried out for the \EQ{} type with the same structure we have described for \ST{} apart from having a stride $s=1$ instead of $s=2$. However, none of the models suggested in this run outperform the \EQ{} models found in the initial search.

\begin{table}
\caption{Best architectures for counting flux violations. This table reports the three most promising architectures obtained with an \textit{optuna} search, ordered according to the average validation loss over 20 instances trained from scratch. As in the previous tasks, the input tensor shape is $(N_t, N_x, 4)$, while the output is a real number that approximates the amount of open worms in the configuration. The final row informs about the number of learnable parameters for each architecture. Table from \cite{Bulusu:2021rqz}.}
\vspace{2mm}
\centering
\scriptsize
\begin{tabular}{l l l}
\textbf{1st \EQ{}} & \textbf{2nd \EQ{}} & \textbf{3rd \EQ{}} \\
 \hline
Conv($1 \times 1$, 4, 32) & Conv($2 \times 2$, 4, 8) & Conv($1 \times 1$, 4, 4) \\
LeakyReLU & LeakyReLU & LeakyReLU \\
Conv($2 \times 2$, 32, 8) & Conv($2 \times 2$, 8, 8) & Conv($2 \times 2$, 4, 8) \\
LeakyReLU & LeakyReLU & LeakyReLU \\
Conv($2 \times 2$, 8, 16) & Conv($1 \times 1$, 8, 4) & Conv($2 \times 2$, 8, 4) \\
LeakyReLU & LeakyReLU & LeakyReLU \\
Conv($1 \times 1$, 16, 8) & Conv($1 \times 1$, 4, 8) & Conv($3 \times 3$, 4, 1) \\
LeakyReLU & LeakyReLU & LeakyReLU \\
GlobalSumPool & GlobalSumPool & GlobalSumPool \\
Linear(8, 1) & Linear(8, 1) & \\
\hline
1800 & 456 & 308 \\
\end{tabular}
\medskip

\centering
\begin{tabular}{l l l}
\textbf{1st \ST{}} & \textbf{2nd \ST{}} & \textbf{3rd \ST{}} \\
 \hline
Conv($2 \times 2$, 4, 16) & Conv($2 \times 2$, 4, 4) & Conv($2 \times 2$, 4, 4) \\
LeakyReLU & LeakyReLU & LeakyReLU \\
Conv($1 \times 1$, 16, 32) & MaxPool($2 \times 2$, $2$) & AvgPool($2 \times 2$, $2$) \\
LeakyReLU & Conv($2 \times 2$, 4, 4) & Conv($3 \times 3$, 4, 16) \\
Conv($1 \times 1$, 32, 32) & LeakyReLU & LeakyReLU \\
LeakyReLU & GlobalSumPool & GlobalSumPool \\
AvgPool($2 \times 2$, $2$) & Linear(4, 1) & Linear(16, 32) \\
Conv($1 \times 1$, 32, 8) & & LeakyReLU \\
LeakyReLU & & Linear(32, 1) \\
GlobalSumPool & & \\
Linear(8, 32) & & \\
LeakyReLU & & \\
Linear(32, 1) & & \\
\hline
2336 & 132 & 1184 \\
\end{tabular}
\medskip

\centering
\begin{tabular}{l l l}
\textbf{1st \FLAT{}} & \textbf{2nd \FLAT{}} & \textbf{3rd \FLAT{}} \\
 \hline
Conv($2 \times 2$, 4, 4) & Conv($2 \times 2$, 4, 8) & Conv($2 \times 2$, 4, 32) \\
LeakyReLU & LeakyReLU & LeakyReLU \\
AvgPool($2 \times 2$, $2$) & AvgPool($2 \times 2$, $2$) & AvgPool($2 \times 2$, $2$) \\
Conv($3 \times 3$, 4, 8) & Conv($3 \times 3$, 8, 4) & Conv($3 \times 3$, 32, 4) \\
LeakyReLU & LeakyReLU & LeakyReLU \\
AvgPool($2 \times 2$, $2$) & AvgPool($2 \times 2$, $2$) & AvgPool($2 \times 2$, $2$)
\\
Flattening & Flattening & Flattening 
\\
Linear(8, 4) & Linear(4, 4) & Linear(4, 32)
\\
LeakyReLU & LeakyReLU & LeakyReLU \\
Linear(4, 32) & Linear(4, 32) & Linear(32, 16) \\
LeakyReLU & LeakyReLU & LeakyReLU \\
Linear(32, 1) & Linear(32, 1) & Linear(16, 1) \\
\hline
640 & 640 & 2704 \\
\end{tabular}
\label{tab:reg2_archs}
\end{table}

In order to evaluate the network performances, we make use of two metrics similarly to the previous task, the MSE loss and the accuracy. For the latter, the predictions are rounded to the closest integer. During the optimization process, the monitored metric is the validation loss. Two \textit{optuna} runs are executed with the aim of mitigating the risk of overlooking promising combinations of hyperparameters. For each of these, three models are trained for 200 epochs without early stopping to reduce the impact of random initializations. We used a batch size of 16, a learning rate $\lambda_{lr}=10^{-3}$ and the \textit{AMSGrad} variant of the \textit{AdamW} optimizer with zero weight decay.

From the set of 100 different architectures resulting from the two \textit{optuna} searches, we identify the top three for each architecture type based on the average validation loss across their three initializations. We train these architectures from 20 different random initializations using the same training and validation sets, and keeping the same hyperparameters, except for the amount of epochs which is increased to 500. We keep track of the validation loss and checkpoint the weight values of the best model. For each type, the architectures are sorted according to the average of the validation loss over the ensemble of models. The results are detailed in Table~\ref{tab:reg2_archs}.

\subsection{Results}

\begin{figure}
    \centering
    \includegraphics{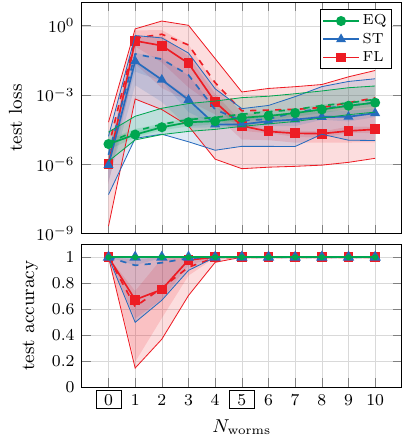}
    \caption{Test loss (top) and test accuracy (bottom) of the architectures which scored the lowest mean of the validation loss, evaluated on $8\times8$ configurations given as functions of the number of open worms. Training and validation are executed with $N_\mathrm{worms} = 0$ and $N_\mathrm{worms} = 5$, while the test set comprises a number of open worms in the interval $N_\mathrm{worms} \in [0, 10]$. Conventions for shaded regions, quantiles, symbols, solid and dashed lines are the same as in Fig.~\ref{fig:test_loss_various_lattice_sizes} except that the number of instances used to create a model ensemble is 20. While the accuracy of every \EQ{} model is perfect and its loss increases slightly with the number of open worms, the non-equivariant models fail to accurately predict a small number of flux violations. Image from \cite{Bulusu:2021rqz}.}
    \label{fig:reg2_val_mean_res_8x8}
\end{figure}

\begin{figure}
    \centering
    \includegraphics{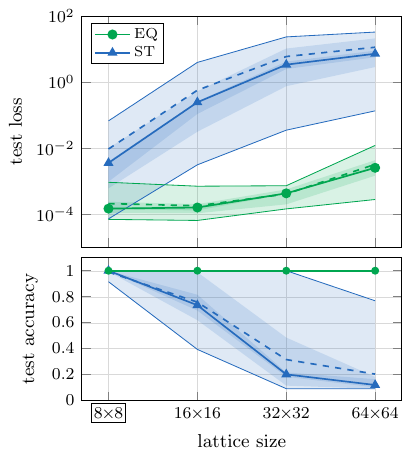}
    \caption{Test loss (top) and test accuracy (bottom) of the architectures which scored the lowest mean of the validation loss, evaluated on any number of worms in the range $N_\mathrm{worms} \in [0, 10]$ given as functions of the lattice size. Training and validation are executed on the smallest lattice ($8 \times 8$), while testing takes place on all lattice sizes. Except for a few \ST{} models accomplishing good results on $8 \times 8$ configurations, the \EQ{} architecture outperforms its non-equivariant counterpart across all lattice sizes. Image from \cite{Bulusu:2021rqz}.}
    \label{fig:reg2_val_mean_res}
\end{figure}

We divide the results in a test performed on the subset containing $8 \times 8$ configurations and another which includes the whole test set. The former test is executed for all the architecture types and its results are shown in Fig.~\ref{fig:reg2_val_mean_res_8x8}, while the \FLAT{} architecture is excluded from the latter, whose results are depicted in Fig.~\ref{fig:reg2_val_mean_res}.

The plots indicate that the \EQ{} architecture consistently outperforms its non-equivariant counterparts. An interesting observation comes from Fig.~\ref{fig:reg2_val_mean_res_8x8}, where \ST{} and even more so \FLAT{} models struggle particularly with the prediction of one, two and three of open worms. The reason for this can be associated to the fact that the training set includes samples with only $N_\mathrm{worms}\in\{0,5\}$, therefore the same architectures have also been trained on a dataset that features $N_\mathrm{worms}\in\{0, 1, 5\}$. Surprisingly, the outcome is similar to the one reported in Fig.~\ref{fig:reg2_val_mean_res_8x8}, suggesting that the problem for non-equivariant architectures is not simply a lack of sufficient information about the problem.

\begin{table*}
\centering
\tiny
\caption{\label{tab:reg2_metrics} Metrics of the best architectures for counting flux violations. For each type and metric, the result of the best architecture is highlighted in bold. Table from \cite{Bulusu:2021rqz}.
}
\begin{tabular}{l c c c c c c}
 & \multicolumn{2}{c}{\textbf{validation loss on $8\times8$}} & \multicolumn{2}{c}{\textbf{test loss on $8\times8$}} & \multicolumn{2}{c}{\textbf{test loss up to $64\times64$}} \\
& \textbf{mean} & \textbf{median} & \textbf{mean} & \textbf{median} & \textbf{mean} & \textbf{median} \\
\hline \\[-2ex]
\textbf{1st \EQ{}} & $\pmb{\num{4.676e-05}}$ & $\num{4.137e-05}$ & $\pmb{\num{2.108e-04}}$ & $\num{1.483e-04}$ & $\pmb{\num{1.008e-03}}$ & $\num{8.308e-04}$ \\
\textbf{2nd \EQ{}} & $\num{1.042e-04}$ & $\pmb{\num{2.440e-05}}$ & $\num{3.525e-04}$ & $\pmb{\num{8.783e-05}}$ & $\num{1.807e-03}$ & $\pmb{\num{7.936e-04}}$ \\
\textbf{3rd \EQ{}} & $\num{8.992e-03}$ & $\num{3.072e-04}$ & $\num{2.105e-02}$ & $\num{9.163e-04}$ & $\num{1.925e+00}$ & $\num{4.031e-02}$ \\
\hline \\[-2ex]
\textbf{1st \ST{}} & $\pmb{\num{2.331e-05}}$ & $\num{2.173e-05}$ & $\num{9.438e-03}$ & $\num{3.576e-03}$ & $\num{4.446e+00}$ & $\num{3.026e+00}$ \\
\textbf{2nd \ST{}} & $\num{8.479e-05}$ & $\num{4.372e-05}$ & $\pmb{\num{2.545e-04}}$ & $\pmb{\num{9.340e-05}}$ & $\pmb{\num{3.738e-03}}$ & $\pmb{\num{1.171e-03}}$ \\
\textbf{3rd \ST{}} & $\num{2.869e-04}$ & $\pmb{\num{2.171e-05}}$ & $\num{1.676e-02}$ & $\num{1.381e-03}$ & $\num{2.943e+00}$ & $\num{9.580e-01}$ \\
\hline \\[-2ex]
\textbf{1st \FLAT{}} & $\pmb{\num{2.602e-05}}$ & $\num{1.787e-05}$ & $\num{7.837e-02}$ & $\num{3.817e-02}$ & - & - \\
\textbf{2nd \FLAT{}} & $\num{4.004e-05}$ & $\num{1.117e-05}$ & $\pmb{\num{5.300e-02}}$ & $\pmb{\num{1.285e-03}}$ & - & - \\
\textbf{3rd \FLAT{}} & $\num{5.805e-05}$ & $\pmb{\num{1.031e-05}}$ & $\num{6.382e-02}$ & $\num{3.556e-02}$ & - & - \\
\end{tabular}
\end{table*}

\begin{figure}
    \centering
    \includegraphics{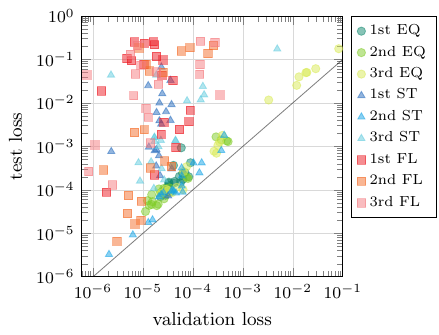}
    \caption{Test loss on the $8 \times 8$ lattice versus validation loss for every instance of each architecture. The test loss is shown for all 20 instances of the winning architectures displayed in table~\ref{tab:reg2_archs}. Training and validation were executed with $N_\mathrm{worms}\in\{0,5\}$ and $(\eta,\mu)\in\{(4.01,1.5),(4.25,1)\}$ on $8 \times 8$ configurations. The test set included other physical parameter values: $N_\mathrm{worms} \in [0, 10]$, $\mu \in  \{1.0, 1.25, 1.5 \}$, $\eta \in \{ 4.01, 4.04, 4.25\}$. The diagonal black line denotes where test loss and validation loss are equal. If a model has good generalization properties, the test loss has to be comparable with the validation loss, and therefore has to lie close to the black line, which happens for most of the \EQ{} models (green circles). Image from \cite{Bulusu:2021rqz}.}
    \label{fig:reg2_tloss_vloss_8x8}
\end{figure}

\begin{figure}
    \centering
    \includegraphics{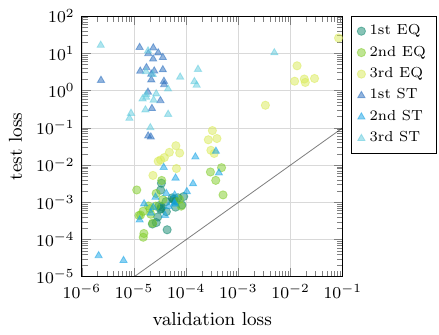}
    \caption{Test loss over the whole test set versus validation loss of every instance for \EQ{} and \ST{} architectures. We analyze our models in a similar fashion to Fig.~\ref{fig:reg2_tloss_vloss_8x8}, extending the study to lattice sizes up to $64 \times 64$. \EQ{} models (green circles) tend to be closer to the black line where test loss and validation loss are the same, showing their ability of generalizing to different lattice sizes and physical parameters. Image from \cite{Bulusu:2021rqz}.}
    \label{fig:reg2_tloss_vloss}
\end{figure}

In table~\ref{tab:reg2_metrics} the mean and median losses of the model ensemble on the validation and test sets are displayed for the three types. The best architecture for each metric and type is highlighted in bold. We notice that depending on whether we choose the mean or the median of the validation loss to select the best architecture we get a different outcome. An even more relevant observation is related to the generalization capabilities of the networks, specifically to the fact that performing well on the validation set does not guarantee that the networks achieve good results when testing. For this reason, we analyze the relationship between the validation and the test losses for every individual model of each type in Figs.~\ref{fig:reg2_tloss_vloss_8x8} and~\ref{fig:reg2_tloss_vloss}. 

Most \ST{} and \FLAT{} models are scattered vertically, meaning that for many of them the test loss is much worse than the validation loss, which is a sign of generalization issues. \EQ{} models, on the other hand, are distributed closer to the black line, where test and validation loss would match. These results suggest that if an \EQ{} architecture scores a low validation loss we can expect that the test loss is also low, which means that it generalized more reliably than the non-equivariant counterparts.

Remarkably, a non-equivariant architecture also exhibits this behavior, namely the 2nd \ST{} architecture. Its best two instances even achieve results better by almost an order of magnitude compared to the best \EQ{} models both when validating and testing. One factor that may have contributed to its success is the presence of a spatial max pooling layer, which can be useful for detecting local defects, as mentioned also in the previous task. Another possible reason is the very small number of parameters compared to all the architectures in table~\ref{tab:reg2_archs}, given the relative simplicity of this counting problem. This is confirmed also by the fact that \textit{optuna} favors in general small architectures for this task, as happened for the previous ones, with a number of weights between $\sim100$ and $\sim3\,000$.

\chapter{Gauge-symmetric neural networks in lattice gauge theory} \label{chap:lcnns}
\chaptermark{Gauge symmetry}

The results found in the previous Chapter have shown that incorporating translational symmetry in a CNN architecture gives a boost to the performance in a problem symmetric under translations. This serves as an additional motivation for the scope of this Chapter: designing neural networks that respect gauge symmetries by construction. After initially reviewing Yang-Mills theory and its discretization on the lattice, we define equivariance for lattice gauge theory. Layers such as convolutions, that are readily available in common frameworks, break gauge equivariance, therefore we introduce appropriate layers that possess such a property. The architectures resulting from stacking these layers respect gauge symmetry by construction, and they can in principle predict arbitrarily sized Wilson or Polyakov loops, making these networks capable of approximating any function on the lattice. We then show how these neural networks perform on regression tasks involving e.g.~Wilson loops of various sizes in 1+1D and in 3+1D. In the former case, we compare their results with standard CNNs that do not respect gauge symmetry.

\section{Yang-Mills theory and lattice gauge theory \label{sec:ltg}}

In this section, we introduce Yang-Mills theory and lattice gauge theory with particular focus on gauge symmetry. A broader introduction can be found in standard textbooks \cite{Gattringer:2010abc, Smit:2002aaa}.

Let us consider the $\SU(N_c)$ Yang-Mills theory at finite temperature $T$ in $D+1$ dimensions. The expectation value of an observables $\mathcal{O}$ is calculated by means of the path integral
\begin{align} 
    \left< \mathcal{O} \right> = \frac{1}{\mathcal{Z}} \int \mathcal{D} [A] \, \mathcal{O}(A_\mu) e^{- S_\mathrm{E}[A_\mu]}, \label{eq:tft_exp}
\end{align}
with the partition function
\begin{align}
    \mathcal{Z} = \int \mathcal{D} [A] \, e^{- S_\mathrm{E}[A_\mu]} \label{eq:tft_Z}
\end{align}
acting as a normalization constant, and the Euclidean Yang-Mills action
\begin{align}
    S_\mathrm{E}[A_\mu] = \frac{1}{2g^2} \intop_0^\beta d\tau \int d^{D} x \, \mathrm{Tr} \left[ F_{\mu\nu}(x) F_{\mu\nu}(x) \right].
    \label{eq:ym_action_euclid}
\end{align}
The inverse temperature of the system is $\beta = T^{-1}$ and $g \in \mathbb{R}$ is the Yang-Mills coupling constant.
The non-Abelian field strength tensor reads
\begin{align}
    F_{\mu\nu} = \p_\mu A_\nu - \p_\nu A_\mu + i \left[ A_\mu, A_\nu \right],
\end{align}
where $A_\mu(x)$ is a gauge field living in the algebra $\su(N_c)$. We write the $(D+1)$-dimensional spacetime point as $x = (\tau, \mathbf x)$. The $D$ spatial dimensions spanned by $\mathbf x$ are real, while the imaginary time dimension $\tau$ extends from 0 to the inverse temperature, $\tau \in [0, \beta)$. The metric on the spacetime is Euclidean: $\eta_{\mu\nu} = \mathrm{diag} (+, +, \dots, +)$. The gauge fields satisfy periodic boundary conditions along $\tau$, i.e.~\mbox{$A_\mu(\tau +\beta, \mathbf x) = A_\mu(\tau, \mathbf x)$}.

A characteristic property of a gauge theory is gauge invariance, which refers to the fact that observables are invariant under local symmetry transformations. For Yang-Mills theory, a gauge transformation $T_\Omega$ applied to the gauge field $A_\mu(x)$ can be written as
\begin{align}
    T_\Omega A_\mu(x) = \Omega(x) \left( A_\mu(x) - i \p_\mu \right) \Omega^\dg(x),
    \label{eq:gauge_tr}
\end{align}
where $\Omega(x)$ is an element of the $\SU(N_c)$ group, hence it is unitary and has unit determinant:
\begin{align}
    \Omega(x) \Omega^\dg(x) &= \one, \\
    \det \Omega(x) &= 1.
\end{align}
The application of this transformation to the field strength tensor yields
\begin{align}
    T_\Omega F_{\mu\nu}(x) = \Omega(x) F_{\mu\nu}(x) \, \Omega^\dg(x),
    \label{eq:F_tr}
\end{align}
where extra terms coming from the derivatives of the gauge field simplify exactly with the extra terms coming from the commutator. Given the cyclic property of the trace, the integrand in Eq.~\eqref{eq:ym_action_euclid} is invariant under gauge transformations, thus making the Yang-Mills action itself invariant under~\eqref{eq:gauge_tr}:
\begin{align}
    S_\mathrm{E}[T_\Omega A_\mu] = S_\mathrm{E}[A_\mu].
    \label{eq:ym_action_invariant}
\end{align}
As a consequence, the set of gauge fields $\{ T_\Omega A_\mu \}$ that are related via arbitrary gauge transformations forms a class of physically equivalent fields. Moreover, any physical observable $\mathcal{O}(A_\mu)$ must be invariant under gauge transformations, which guarantees that also its expectation value Eq.~\eqref{eq:tft_exp} has the same property.

Apart from perturbation theory, where the coupling constant $g$ is small,
an infinite dimensional functional integral of the form of Eq.~\eqref{eq:tft_exp} is generally intractable. As in the scalar field theory case, we can tackle the problem by approximating the continuous spacetime with a lattice
\begin{align}
    \Lambda = \bigg\{ x = \sum_{\mu=0}^D a^\mu n^\mu \hat{\mu} \, | n^\mu \in \mathbb{Z} \bigg\},
\end{align}
where $a^\mu$ are the spacings between the lattice sites in the direction $\hat{\mu}$. In the following, we set the lattice constant $a^\mu$ to unity for every direction. A naive discretization of the gauge fields fails to respect gauge symmetry. The regularization on the lattice circumvents this problem by replacing the gauge fields $A_\mu(x)$ with
\begin{align}
    U_{x,\mu}=\mathcal{P}\exp{\left(ig\int_x^{y}\mathrm{d}z^\nu A_\nu(z)\right)} \in \SU(N_c),
\end{align}
where $\mathcal P$ indicates the path ordering scheme. These quantities are defined along the edges of the lattice $\Lambda$ that connect the lattice site $x$ and its neighbor $y = x + \hat{\mu}$, hence the name gauge links. The definition above can be formally expressed by dividing the path into $N$ infinitesimal segments $(z_n, z_n + \mathrm{d}z_n), \, n\in \{0,1,\dots,N-1\}$ in the following way:
\begin{align}
    U_{x,\mu} &= \lim_{N\rightarrow\infty} \exp \left[-i\int_{z_{N-1}}^y \mathrm{d}z_{N-1}^\nu A_\nu(z)\right] \dots \exp \left[-i\int_{z_n}^{z_{n+1}} \mathrm{d}z_n^\nu A_\nu(z)\right] \nonumber \\
    &\phantom{=} \dots \exp \left[-i\int_x^{z_1} \mathrm{d}z_0^\nu A_\nu(z)\right] \nonumber \\
    &= \lim_{N\rightarrow\infty} [1-i\,\mathrm{d}z_0^\nu A_\nu(z_0)] \dots [1-i\,\mathrm{d}z_{N-1}^\nu A_\nu(z_{N-1})]+O(\mathrm{d}z^2),
\end{align}
where $z_0 = x$ and $z_N = y$. If we apply the transformation Eq.~\eqref{eq:gauge_tr} at the point $z_n$, we find
\begin{align}
    1-i\,\mathrm{d}z_n^\nu A'_\nu(z_n) &= \Omega(z_n) \Omega^\dagger(z_n) -i \mathrm{d}z_n^\nu \Omega(z_n)\left[A_\nu(z_n)\Omega^\dagger(z_n) + i\,\p_\nu\Omega^\dagger(z_n)\right] \nonumber \\
    &= \Omega(z_n) \left[1-i\,\mathrm{d}z_n^\nu A_\nu(z_n)\right]\Omega^\dagger(z_{n+1}) + O(\mathrm{d}z^2).
\end{align}
The derivative of $\Omega$ is defined as in Eq.~\eqref{Scalar:diff_quotient}. Using this transformation rule for each $z_n$, the terms involving $\Omega(z_n)$ simplify except at the endpoints, thus yielding the gauge transformation of the link
\begin{align}
    T_\Omega U_{x,\mu} = \Omega_x U_{x,\mu} \Omega^\dg_{x+\mu},
    \label{eq:link_tr}
\end{align}
where $\Omega_x$ indicates the transformation at lattice site $x$ and $x + \mu$ is a shorthand for $x + \hat{\mu}$. In terms of these new variables, the Yang-Mills action can be approximated
by the Wilson action
\begin{align}
    S_W[U] = \frac{2}{g^2} \sum_{x \in \Lambda} \sum_{\mu < \nu} \Tr \left[ \one - U_{x,\mu\nu} \right],
    \label{eq:wilson_action}
\end{align}
where the so-called plaquette variables
\begin{align}
    U_{x,\mu\nu} = U_{x,\mu} U_{x+\mu, \nu} U^\dg_{x+\nu, \mu} U^\dg_{x,\nu} = \mg{0.7}{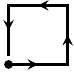} \,
    \label{eq:plaq}
\end{align}
are $1 \times 1$ Wilson loops on the lattice. We made use of the definition of a link pointing in the negative direction $-\rho$ as $U_{y,-\rho} = U^{-1}_{y-\rho,\rho} = U^\dagger_{y-\rho,\rho}$. Hereafter, Wilson loops are intended to be untraced matrices, unless specified otherwise. The plaquette variables transform locally at $x$ under the effect of Eq.~\eqref{eq:link_tr}, i.e.
\begin{align}
    T_\Omega U_{x,\mu\nu} = \Omega_x U_{x,\mu\nu} \Omega^\dg_x,
\end{align}
which implies that the Wilson action is invariant:
\begin{align}
    S_W[T_\Omega U] = S_W[U].
\end{align}
From the Wilson action~\eqref{eq:wilson_action}, it is possible to recover the Yang-Mills action~\eqref{eq:ym_action_euclid} by expanding $U_{x,\mu}$ in powers of $A_\mu(x)$ with lattice artifacts of order $O(a^2)$, where $a$ denotes the lattice spacing.
Lattice gauge theory therefore represents the appropriate formalism to describe gauge fields on a lattice while respecting gauge symmetry. Similarly to the continuum case, expectation values are given by
\begin{align}
    \left< \mathcal{O} \right> &= \frac{1}{Z} \intop \mathcal{D}[U] \,\mathcal{O}[U]\, e^{-S_W[U]}, \\
    Z &= \int \mathcal{D}[U] \, e^{-S_W[U]},
\end{align}
where $\mathcal{D}[U]$ is the Haar measure and $Z$ is the partition function of the new degrees of freedom. With the lattice formalism, the infinite-dimensional functional integrals are substituted by finite (although high) dimensional integrals. Furthermore, we restrict the lattice to a finite sublattice characterized by a number of
\begin{align}
    N_\mathrm{lat} = N_t \cdot N_s^D
\end{align}
lattice sites, with $N_t$ cells in the imaginary time direction and $N_s$ cells in the spatial directions. The lattice is also equipped with periodic boundary conditions in every dimension. This enables the computation of expectation values via Monte Carlo method, i.e.~we generate a series of configurations ${ U^{(n)} }$ according to the probability density $P[U] \propto \exp{\left( - S_W[U] \right)}$, and approximate the expectation value by
\begin{align}
    \left< \mathcal{O} \right> \approx \frac{1}{N_\mathrm{conf}} \sum^{N_\mathrm{conf}}_{n=1} \mathcal{O}[U^{(n)}].
\end{align}
Gauge field configurations can be generated for example with the Metropolis algorithm \cite{Metropolis:1953abc}, which is discussed in Appendix~\ref{app:montecarlo} for the $\SU(2)$ case, or using the heat bath method \cite{Creutz:1980bbb}. 

Physical observables are typically expressed in terms of (traced) Wilson loops. The simplest quantity is the Wilson action~\eqref{eq:wilson_action}, which is a function of $1 \times 1$ loops. It is possible to improve the approximation of the continuum action by including larger loops \cite{Niedermayer:1996eb, Iwasaki:1985we, Moore:1996wn, Lagae:1998pe, Ipp:2018hai}. Notably, the potential of a static quark pair can be determined from the expectation value of a Wilson loop with a large extent in the temporal direction \cite{Wilson:1974sk, Bali:2000gf}.

A local observable that is typically studied is the topological charge density \cite{Alexandrou:2017hqw}. In the continuum, its definition is
\begin{align}
    q(x) = \frac{1}{32 \pi^2} \epsilon_{\mu\nu\rho\sigma} \Tr \left[ F_{\mu\nu}(x) F_{\rho \sigma}(x) \right],
    \label{eq:top_charge_sec2}
\end{align}
which on the lattice can be approximated by products of Wilson loops. The plaquette approximation of Eq.~\eqref{eq:top_charge_sec2} reads
\begin{align}
    q^\mathrm{plaq}_x = \frac{\epsilon_{\mu\nu\rho\sigma}}{32 \pi^2}  \Tr \left[  \frac{U_{x,\mu\nu} \! - \! U^\dg_{x,\mu\nu}}{2i} \,  \frac{U_{x,\rho\sigma} \! - \! U^\dg_{x,\rho\sigma}}{2i}  \right].
    \label{eq:top_charge_plaq}
\end{align}
The discretization errors of the plaquette approximation can be reduced with the inclusion of appropriate larger loops in the discretization of the field strength tensor \cite{BilsonThompson:2002jk}. Other observables, such as the energy momentum tensor \cite{Caracciolo:1989pt}, can be formulated as (non-linear) combinations of arbitrarily-sized Wilson loops.

There exist also non-local observables, which can be written in terms of Polyakov loops \cite{Polyakov:1978vu}, that wrap around the periodic boundary of the lattice. The trace of the Polyakov loop can serve as an order parameter for quark confinement in a pure gauge theory at nonzero temperature, and correlators of the traced Polyakov loop are related to the static quark potential \cite{Kuti:1980gh, McLerran:1981pb}. Wilson loops and Polyakov loops are two topologically distinct objects, as can be seen for example by noticing that the former can be contracted to a single point, while the latter cannot.

In summary, all the aforementioned observables are in general non-linear functions either of Wilson or Polyakov loops. In the following section, we will introduce new layers of artificial neural networks capable of expressing such functions while preserving gauge symmetry on the lattice.

In the next section, we will introduce new layers of artificial neural networks that can express such lattice gauge equivariant functions and respect gauge symmetry on the lattice.

\section{Gauge equivariant layers}

Lattice gauge equivariant convolutional neural networks (\LGCNN{}s) can approximate a large class of gauge equivariant functions while preserving the symmetry properties of the underlying theory. Similar to how we introduced CNNs in the first chapter, we split \LGCNN{}s into more elementary layers and discuss each of them in this section.

\subsection{Input data}
\label{subsec:input}

The input data of each layer is a tuple $(\mathcal{U}, \mathcal{W})$, consisting of non-locally transforming gauge link variables $\mathcal{U}$ and locally transforming objects $\mathcal{W}$. A natural choice for the first part of the tuple is the set of gauge link variables $\mathcal{U} = \{ U_{x,\mu} \}$, whose transformation rule we remind to be Eq.~\eqref{eq:link_tr}. Concretely, we choose the fundamental representation of $\SU(N_c)$, in which links are treated as complex special unitary $N_c \times N_c$ matrices. The second part is a set of variables $\mathcal{W} = \{ W_{x,i} \}$ with $W_{x,i} \in \mathbb C^{N_c \times N_c}$ and index $i \in \{ 1, 2, \dots, N_\mathrm{ch} \}$, which can be interpreted as channels. The requirement for these input variables is to respect the following transformation rule:
\begin{align}
    T_\Omega W_{x,i} = \Omega_x W_{x,i} \Omega^\dg_x.
    \label{eq:gauge_transf}
\end{align}
We emphasize that these objects transform locally. A possible choice for $\mathcal{W}$ is the set of the Wilson loops $U_{x,\mu\nu}$ with positive orientation, i.e.~$\mu < \nu$.

We specify that in practice we represent both $\mathcal{U}$ and $\mathcal{W}$ as $N \times N$ complex matrices, even though the links can be described with less degrees of freedom since they are constrained by the unitarity and specialness conditions.

\subsection{Gauge equivariance}

Having established what the input data for \LGCNN{}s is, we can now define what we mean exactly by gauge equivariance and gauge invariance. A function $f$ taking as input the tuple $(\mathcal{U}, \mathcal{W})$ is gauge equivariant (or covariant) if
\begin{align}
    f(T_\Omega \, \mathcal{U}, T_\Omega \mathcal{W}) = T'_\Omega f(\mathcal{U}, \mathcal{W}),
    \label{eq:f_cov}
\end{align}
for a generic $\Omega \in \SU(N_c)$, with $T'_\Omega f$ denoting the effect of the gauge transformation on the function $f$. As for the other definition, a function $f$ of the tuple $(\mathcal{U}, \mathcal{W})$ is gauge invariant if
\begin{align}
    f(T_\Omega \, \mathcal{U}, T_\Omega \mathcal{W}) = f(\mathcal{U}, \mathcal{W}). \label{eq:f_inv}
\end{align}
This can be viewed as a special case of equivariance with $T'_\Omega = \one$.

An \LGCNN{} can be either an equivariant function as in Eq.~\eqref{eq:f_cov} or an invariant function as in Eq.~\eqref{eq:f_inv}. This means that such networks are suited to approximate these types of functions, which notably include physical observables, that are gauge invariant.

\subsection{Lattice gauge equivariant convolutions}

A fundamental component of neural networks is represented by linear layers. A possible realization of such a layer for lattice gauge theory applications is
\begin{align}
    L_x(\mathcal{W}) = \sum_{y \in \Lambda} \omega_{x,y} W_y,
    \label{eq:lin_1}
\end{align}
where $\omega_{x,y}$ are complex weights and $x$ and $y$ are lattice sites in $\Lambda$. No bias term has been included and the channel index in $\mathcal{W}$ has been dropped to simplify the notation.

This layer does not meet our requirements to be part of an \LGCNN{}, as it does not respect gauge equivariance. In fact, objects that transform locally at different lattice sites are added up, therefore $L_x(\mathcal{W})$ does not transform according to Eq.~\eqref{eq:gauge_transf} under arbitrary gauge transformations:
\begin{align}
    L_x(T_\Omega \mathcal{W}) = \sum_y \omega_{x,y} \,  \Omega_{y} W_{y} \Omega^\dg_{y} \nonumber \neq \Omega_x L_x(\mathcal{W}) \Omega^\dg_x.
\end{align}
In order to maintain a similar structure to a conventional linear layer and concurrently enabling the right transformation, we can modify Eq.~\eqref{eq:lin_1} introducing the parallel transporter $U_{x\rightarrow y}$ in the following way:
\begin{align}
    L'_x(\mathcal{U}, \mathcal{W}) = \sum_y \omega_{x,y} \,  U_{x\rightarrow y} W_{y} U^\dg_{x\rightarrow y}.
    \label{eq:lin_2}
\end{align}
The parallel transporter transforms according to
\begin{align}
    T_\Omega U_{x\rightarrow y} = \Omega_x U_{x\rightarrow y} \Omega^\dg_{y},
\end{align}
which guarantees that the definition of the linear layer in Eq.~\eqref{eq:lin_2} is equivariant:
\begin{align}
    L'_x(T_\Omega \, \mathcal{U}, T_\Omega \, \mathcal{W}) &= \Omega_x L'_x(\mathcal{U}, \mathcal{W}) \Omega^\dg_x= T_\Omega L'_x(\mathcal{U}, \mathcal{W}).
    \label{eq:linear_equi}
\end{align}

The parallel transporter can be built by taking the product of the sequence of links connecting $x$ and $y$. However, no particular path has been specified, which makes the gauge equivariant linear layer path dependent in general. A natural choice that can be made in the continuum is to set the path to be the geodesic (that is the shortest path) between $x$ and $y$, reminiscent of the formulation of gauge equivariant neural networks on manifolds \cite{Cohen:2019xsh}. On the lattice, a unique geodesic cannot be introduced unless two points lie on the same lattice axis, therefore we restrict the path to straight lines along the lattice axes. This practical solution corresponds to imposing $\omega_{x, y} = 0$ if $x$ and $y$ cannot be linked moving along a single axis.

Lattice gauge theory is also symmetric under spacetime translations, since they leave the Wilson action~\eqref{eq:wilson_action} unaltered. Given the results of the study in Chapter~\ref{chap:translations}, we equip the equivariant linear layers in Eq.~\eqref{eq:lin_2} with translational equivariance by requiring the weights to satisfy translational invariance, i.e.
\begin{align}
    \omega_{x+s, y+s} = \omega_{x, y},
    \label{eq:translational_inv}
\end{align}
where $s$ denotes a translation on the lattice. The gauge equivariant linear layer becomes a gauge equivariant convolution. As a consequence of Eq.~\eqref{eq:translational_inv}, convolutions are equivariant under translations. We define the shifted input for an arbitrary element and call $t$ the element of the translation group corresponding to the shift vector $s$. Its action on links and loops is
\begin{align}
    tW_x &\equiv W_{x+s}, \\
    tU_{x,\mu} &\equiv U_{x+s, \mu}.
\end{align}
Translational symmetry is preserved, as can be shown with the following:
\begin{align}
    L'_x(t \, \mathcal{U}, t \mathcal{W}) &= \sum_{y \in \Lambda} \omega_{x,y} U_{x+s \rightarrow y+s} W_{y+s} U^\dg_{x+s \rightarrow y+s} \nonumber \\
    &= \sum_{y' \in \Lambda} \omega_{x+s,y'} U_{x+s \rightarrow y'} W_{y'} U^\dg_{x+s \rightarrow y'} \nonumber \\
    &= L'_{x+s}(\mathcal{U}, \mathcal{W}) = t L'_{x}(\mathcal{U}, \mathcal{W}),
\end{align}
where we made use of Eq.~\eqref{eq:translational_inv} in the second equality. A linear layer with the property of translational equivariance is called convolution, therefore we define a gauge equivariant convolutional layer as
\begin{align}
    C_x(\mathcal{U}, \mathcal{W}) = \sum_s \, \omega_s \, U_{x \rightarrow x + s} W_{x+s} U^\dg_{x \rightarrow x + s},
    \label{eq:conv_2}
\end{align}
where we adopted a more compact notation with weight parameters $\omega_s \in \mathbb C$. In this form, it is more manifest that this operation is translational equivariant.

For general applications, it is necessary to take into account a generalization of such a convolutional layer to an arbitrary number of input and output channels. We name it lattice gauge equivariant convolution (\LConv) and write it as
\begin{align}
    \mathcal{C}_{x, i}(\mathcal{U}, \mathcal{W}) &= \sum_{j, \mu, k} \omega_{i, j, \mu, k} U_{x, k\cdot \mu} W_{x + k\cdot \mu, j} U^\dg_{x, k\cdot \mu},
    \label{eq:l-conv}
\end{align}
where $\omega_{i,j,\mu,k}$ are the complex learnable parameters of the convolution, and the indices run in the following intervals: $1 \le i \le N_\mathrm{ch,out}$, $1 \le j \le N_\mathrm{ch,in}$, $0 \le \mu \le D$ and $-K \le k \le K$, with $K$ being the kernel size. Note that with this choice the range of values for $k$ is such that the convolution is symmetric around the center of the kernel. In our experiments, we restrict ourselves to positive shifts along the lattice axes, therefore $0 \leq k \leq K$.

The unique parallel transporters along the lattice axes read
\begin{equation}
    U_{x, k\cdot \mu} = 
    \begin{cases}
    U_{x, \mu} U_{x+\mu, \mu} \cdots U_{x + (k - 1) \mu, \mu}, &\, k > 0, \\
    U_{x, -\mu} U_{x-\mu, -\mu} \cdots U_{x - (k - 1) \mu, -\mu}, &\, k < 0.
    \end{cases}
    \label{eq:extended_links}
\end{equation}

With the same arguments used for the equivariant linear layers,  we can show that the convolution~\eqref{eq:l-conv} is gauge equivariant
\begin{align}
    \mathcal{C}_{x, i}(T_{\Omega} \, \mathcal{U}, T_{\Omega} \, \mathcal{W}) =  \Omega_x \mathcal{C}_{x, i}(\mathcal{U}, \mathcal{W}) \Omega^\dg_x = T_{\Omega} \, \mathcal{C}_{x, i}(\mathcal{U}, \mathcal{W})
\end{align}
and transforms locally at $x$. Focusing on the effect on the data $( \mathcal{U}, \mathcal{W} )$, the \LConv{} operation modifies the $\mathcal{W}$ variables while the links remain the same:
\begin{align}
    \mathcal{C} : (\mathcal{U}, \mathcal{W}) \rightarrow (\mathcal{U}, \mathcal{W}')
\end{align}
with
\begin{align}
    {\mathcal{W}'} = \bigg\{ W'_{x, i} = \mathcal C_{x,i}(\mathcal{U}, \mathcal{W}), \, 1 \le i \le N_\mathrm{ch,out}, \, x \in \Lambda \bigg\}.
\end{align}

We initially disregarded the bias term for the linear layer~\eqref{eq:lin_1} for simplicity, but it is possible to include it to provide an even more general version of \LConv{}. The addition of a bias $\omega_0 \in \mathbb C$ as in
\begin{align}
    \mathcal{C}_{x, i}(\mathcal{U}, \mathcal{W}) = \sum_{j, \mu, k} \omega_{i, j, \mu, k} U_{x, k\cdot \mu} W_{x + k\cdot \mu, j} U^\dg_{x, k\cdot \mu} + \omega_{i,0} \one
\end{align}
is fully compatible with the equivariance properties of the \LConv{} layer.
It is also worth noticing that the \LConv{} operation is a linear map in the second argument $\mathcal{W}$, i.e.
\begin{align}
    \mathcal{C}_{x,i}(\mathcal{U}, \lambda_1 \mathcal{W}_1 + \lambda_2 \mathcal{W}_2) = \lambda_1 \mathcal{C}_{x,i}(\mathcal{U}, \mathcal{W}_1) + \lambda_2 \mathcal{C}_{x,i}(\mathcal{U}, \mathcal{W}_2), 
\end{align}
for arbitrary complex parameters $\lambda_1$ and $\lambda_2$, provided that no bias term is used, i.e.~$\omega_0 = 0$.

To conclude, we mention that dilated convolutions \cite{Yu:2016jsd} are covered by the definition~\eqref{eq:l-conv}, since these types of convolutions are a subset of standard convolutions.

\subsection{Lattice gauge equivariant bilinear layers} \label{subsec:lbl}

A particularly relevant operation for our applications, especially in combination with \LConv{}, is represented by bilinear functions. This infrequently-used layer can be adapted to satisfy gauge equivariance. A lattice gauge equivariant bilinear layer (\LBL) takes as input two tuples $(\mathcal{U}, \mathcal{W})$ and $(\mathcal{U}, \mathcal{W}')$ and combines them in the following way:
\begin{align}
    \mathcal{B}_{x,i}(\mathcal{U}, \mathcal{W}, \mathcal{W}') &= \sum_{j,k} \alpha_{i,j,k} W_{x,j} W'_{x,k}.
    \label{eq:lbilin}
\end{align}
The complex parameters $\alpha_{i,j,k}$ are learnable and the indices take values as follows: $1 \le i \le N_\mathrm{out}$, $1 \le j \le N_\mathrm{in, 1}$ and $1 \le k \le N_\mathrm{in, 2}$.

Eq.~\eqref{eq:lbilin} features a product of locally transforming variables, guaranteeing that gauge equivariance~\eqref{eq:f_cov} is respected:
\begin{align}
    \mathcal{B}_{x,i}(T_{\Omega} \, \mathcal{U}, T_{\Omega} \mathcal{W}, T_{\Omega} \mathcal{W}') &= \sum_{j,k} \alpha_{i,j,k} \left(\Omega_x W_{x,j} \Omega_x^\dg\right) \left(\Omega_x W'_{x,k} \Omega^\dg_x\right) \nonumber \\
    &= \Omega_x \sum_{j,k} \alpha_{i,j,k} W_{x,j} W'_{x,k} \Omega^\dg_x \nonumber \\
    &= T_{\Omega} \mathcal{B}_{x,i}(\mathcal{U}, \mathcal{W}, \mathcal{W}').
\end{align}
The bilinear operation can be extended by allowing $\mathcal{W}$ and $\mathcal{W}'$ to also include the unit element $\one$ and the Hermitian conjugates of $\mathcal{W}$ and $\mathcal{W}'$, meaning that $\mathcal{W}$ is replaced by $\tilde{\mathcal{W}} = \{ \tilde{W}_{x,l} \}$ with $\tilde N_\mathrm{ch} = 2 \, N_\mathrm{ch} + 1$ channels:
\begin{align}
    \tilde{W}_{x,l} = \begin{cases}
    W_{x,l}, &\quad 1 \leq l \leq N_\mathrm{ch}, \\
    W^\dg_{x,l - N_\mathrm{ch}} \mspace{-2mu}, &\quad N_\mathrm{ch} < l \leq 2 N_\mathrm{ch}  \\
    \one, &\quad l = 2 N_\mathrm{ch} + 1,
    \end{cases}
\end{align}
With this enlarged set, the \LBL{} layer can act as a residual module \cite{He:2016res} and includes a bias term associated with the parameter $\alpha_{i, 0, 0}$.

As in the case of the \LConv, the \LBL{} operation is seen as only modifying the $\mathcal{W}$ variables, while keeping the set of gauge links $\mathcal{U}$ unaltered:
\begin{align}
    \mathcal{B} : (\mathcal{U}, \mathcal{W}, \mathcal{W}') \rightarrow (\mathcal{U}, \mathcal{W}''),
\end{align}
with
\begin{align}
    {\mathcal{W}''} = \bigg\{ W''_{x, i} = \mathcal{B}_{x,i}(\mathcal{U}, \mathcal{W}, \mathcal{W}'), \, 1 \le i \le N_\mathrm{out}, \, x \in \Lambda \bigg\}.
\end{align}

In practice, the two layers introduced so far, \LConv{} and \LBL, are combined into a single module, called \LCB{}, in the implementation of \LGCNN{}s. In said layer, the terms contributing to the convolution are first appropriately parallel transported,
\begin{align}
    W'_{x+k \mu,j} =  U_{x,k \mu} W_{x+k \mu,j} U^\dg_{x,k \mu}, 
\end{align}
and afterwards the outcome is multiplied by the local terms $W_{x,j}$ with the bilinear operation just described, such that the resulting locally transforming variables are
\begin{align}
    W_{x,i} \rightarrow \sum_{j,j',k,\mu} \alpha_{i, j, j', k, \mu} W_{x,j} W'_{x+k \mu,j'}.\label{eq:lcb}
\end{align}
The \LCB{} operation is equivalent to stacking an \LConv{} and an \LBL{} layer, except that learnable weights are parameterized differently. 

\subsection{Gauge equivariant activation functions}

Another fundamental building block of neural networks which has to be generalized to satisfy gauge equivariance are non-linear activation functions. Even though the \LConv{} and the \LBL{} operations can lead to non-linear terms, more general functions can be expressed with the addition of a lattice gauge equivariant activation function (\LAct). We can define it as being applied to each lattice site via
\begin{align}
    a_{x,i}(\mathcal{U}, \mathcal{W}) = g_{x,i}(\mathcal{U}, \mathcal{W}) W_{x,i},
    \label{eq:activationfunction}
\end{align}
where the function $g$ respects
\begin{align}
    g_{x,i}(T_{\Omega} \, \mathcal{U}, T_{\Omega} \mathcal{W}) = 
    g_{x,i}(\mathcal{U}, \mathcal{W}),
    \label{eq:scalargaugeinvariant}
\end{align}
hence is gauge invariant.

The function $a$ is gauge equivariant:
\begin{align}
    a_{x,i}(T_{\Omega} \, \mathcal{U}, T_{\Omega} \mathcal{W}) &= \Omega_x  g_{x,i}(\mathcal{U}, \mathcal{W}) W_{x,i} \Omega^\dg_x \nonumber \\
    &= T_{\Omega} a_{x,i}(\mathcal{U}, \mathcal{W}).
\end{align}
The expression in Eq.~\eqref{eq:activationfunction} is very general. Usually, activation functions only depend on local quantities, which here corresponds to $g_{x,i}$ depending solely on the set of local objects $\{ W_{x,i} \}$.
For example, a widely-employed activation function in traditional neural networks is the rectified linear unit (ReLU)
\begin{align}
    \mathrm{ReLU}(x) = \theta(x) \, x
\end{align}
with $x\in \mathbb R$ and $\theta(x)$ denoting the Heaviside step function. An analogous function that respects gauge equivariance has the form
\begin{align}
    g_{x,i}(\mathcal{U}, \mathcal{W}) = \theta(\mathrm{Re} \Tr \left[ W_{x,i} \right]).
\end{align}
If we now take the real part of the trace of $a_{x,i}$ we obtain
\begin{align}
    \mathrm{Re} \Tr \left[ a_{x,i} (\mathcal{U} , \mathcal{W} )\right] &= \theta(\mathrm{Re} \Tr \left[ W_{x,i} \right]) \mathrm{Re} \Tr \left[ W_{x,i} \right] \nonumber \\
    &= \mathrm{ReLU}(\mathrm{Re} \Tr \left[ W_{x,i} \right]),
\end{align}
which is the behavior we wanted to recreate. Like \LConv{} and \LBL{}, \LAct{} only modifies $\mathcal{W}$ variables, i.e.
\begin{align}
    a : (\mathcal{U}, \mathcal{W}) \rightarrow (\mathcal{U}, \mathcal{W}'),
\end{align}
with
\begin{align}
    {\mathcal{W}'} = \bigg\{  W'_{x, i} = a_{x,i}(\mathcal{U}, \mathcal{W}), \, 1 \leq i \leq N_\mathrm{ch}, x \in \Lambda \bigg\}.
\end{align}
The scalar function in Eq.~\eqref{eq:scalargaugeinvariant} can also depend on variables evaluated at any lattice site and, moreover, on trainable parameters, such as in $\mathrm{PReLU}$~\cite{He:2015del}.

\subsection{Exponentiation layers}

The \LGCNN{} layers that have been discussed up to this point only alter the locally transforming objects $\mathcal{W}$ and leave the gauge links $\mathcal{U}$ unchanged. A gauge equivariant operation
\begin{align}
    \mathcal{E}: (\mathcal{U}, \mathcal{W}) \rightarrow (\mathcal{U}', \mathcal{W})
\end{align}
needs to modify the gauge links guaranteeing that they retain their gauge transformation behavior
\begin{align}
    T_\Omega U'_{x,\mu} &= \Omega_x U'_{x,\mu} \Omega^\dg_{x+\mu},
\end{align}
and concurrently preserve the unitarity and determinant constraints
\begin{align} 
    {U'}^{\dg}_{x,\mu} U'_{x,\mu} &= \one, \\
    \det U'_{x,\mu} &= 1,
\end{align}
for all the links in the set $\mathcal{U}' = \{ U'_{x,\mu} \}$.

The function $\mathcal{E}$ satisfies these properties if it acts on the links according to
\begin{align}
    U'_{x,\mu} = \mathcal{E}_{x,\mu} U_{x,\mu},
    \label{eq:exp_layer_U}
\end{align}
and $\mathcal{E}_{x,\mu}$ is an $\SU(N_c)$ group element which transforms locally:
\begin{align}
    T_\Omega \mathcal{E}_{x,\mu} = \Omega_x \mathcal{E}_{x,\mu} \Omega^\dg_x.
    \label{eq:exp_layerl_gauge}
\end{align}
A possible way of constructing $\mathcal{E}_{x,\mu}$ as a function of $\mathcal{W}$ objects is by means of the exponential map (\LExp)
\begin{align}
    \mathcal{E}_{x,\mu}(\mathcal{W}) = \exp{ \left( i  \sum_i \beta_{\mu, i} \left[W_{x,i} \right]_\mathrm{ah} \right) },
    \label{eq:exp_layer}
\end{align}
where $\beta_{\mu, i}$ are real trainable weights with indices $0 \le \mu \le D$ and $1 \le i \le N_\mathrm{ch}$. The notation $\left[W_{x,i} \right]_\mathrm{ah}$ indicates the anti-Hermitian traceless part of locally transforming objects, which for a generic matrix $X$ can be written as
\begin{align}
    \left[ X \right]_\mathrm{ah} = \frac{1}{2 i} \left(X \! - \! X^\dg \right) - \frac{1}{2 i N_c} \one \, \mathrm{Tr} \left( X \! - \! X^\dg\right).
    \label{eq:ah}
\end{align}
The \LExp{} layer~\eqref{eq:exp_layer} has local gauge transformation behaviour:
\begin{align}
    \mathcal{E}_{x,\mu}(T_\Omega \mathcal{W}) &= \exp{ \left( i  \sum_i \beta_{\mu, k} \Omega_x \left[W_{x,i} \right]_\mathrm{ah} \Omega^\dg_x \right) } \nonumber \\
    &= \sum_{n = 0}^\infty \frac{1}{n!} \left(i  \sum_i \beta_{\mu, k} \Omega_x \left[W_{x,i} \right]_\mathrm{ah} \Omega^\dg_x \right)^n \nonumber \\
    &= \Omega_x \sum_{n = 0}^\infty \frac{1}{n!} \left(i  \sum_i \beta_{\mu, k} \left[W_{x,i} \right]_\mathrm{ah} \right)^n \Omega^\dg_x \nonumber \\
    &= \Omega_x \exp{ \left( i  \sum_i \beta_{\mu, k} \left[W_{x,i} \right]_\mathrm{ah} \right) } \Omega^\dg_x \nonumber \\
    &= T_\Omega \mathcal{E}_{x,\mu}(\mathcal{W}),
    \label{eq:exp_layer_trans}
\end{align}
where we used the series expansion of the exponential map and the unitarity of $\Omega_x$. The \LExp{} operation also conserves the unitarity and determinant constraints:
\begin{align}
    \left( \sum_i \beta_{\mu, i} \left[ W_{x,i}\right]_\mathrm{ah} \right)^\dg &= \sum_i \beta_{\mu, i}\left[ W_{x,i}\right]_\mathrm{ah}, \\
    \Tr \left[ W_{x,i}\right]_\mathrm{ah} &= 0.
\end{align}
The determinant constraint is a consequence of the renowned relationship between the trace and the determinant: $\det \mathrm{e}^A = \mathrm{e}^{\Tr{A}}$.

This layer therefore updates the link configuration in a way that is consistent with gauge symmetry.

\subsection{Trace layers}

In the case an \LGCNN{} is required to predict physical observables, it is convenient to build a layer that automatically integrates out the group structure and outputs a gauge invariant quantity. This can be realized by computing the trace of the $\mathcal{W}$ objects and neglecting the set of gauge links. The trace layer (\Trace) is defined as
\begin{align}
    \mathcal{T}_{x,i}(\mathcal{U}, \mathcal{W}) &= \Tr \left[ W_{x,i} \right],
\end{align}
which is trivially gauge invariant
\begin{align}
    \mathcal{T}_{x,i}(T_\Omega \mathcal{U}, T_\Omega \mathcal{W})  &= \Tr \left[ \Omega_x W_{x,i} \Omega^\dg_x \right] \nonumber \\
    &= \mathcal{T}_{x,i}(\mathcal{U}, \mathcal{W}).
\end{align}
The output of \Trace{} are $N_\mathrm{lat} N_\mathrm{ch}$ complex numbers, which correspond to $2N_\mathrm{lat} N_\mathrm{ch}$ real numbers:
\begin{align}
    \mathcal{T}: (\mathcal{U}, \mathcal{W}) \rightarrow \mathbf y \in \mathbb R^{2N_\mathrm{lat} N_\mathrm{ch}},
\end{align}
which can be further processed by a CNN if the lattice structure is maintained, otherwise by a dense network, without ruining gauge symmetry.

\subsection{Plaquette layers}

In the tuple $(\mathcal{U}, \mathcal{W})$ introduced in Subsection \ref{subsec:input}, the link configuration $\mathcal{U}$ is the result of a Monte Carlo algorithm, as already mentioned, while the $\mathcal{W}$ variables need to be computed from the gauge links. The layers presented in this subsection and the next are used to calculate two common types of locally transforming objects. As opposed to the layers described so far, this type of operations is not characterized by trainable parameters and as such can be viewed as a preprocessing step.

Here, we introduce a layer
\begin{align}
    \mathcal{P} : (\mathcal{U}) \rightarrow (\mathcal{U}, \mathcal{W}),
\end{align}
called \Plaq{} that computes the plaquettes~\eqref{eq:plaq} given the set of links $\mathcal{U}$ as follows
\begin{align}
    \mathcal{P}_{x,\mu\nu}(\mathcal{U}) = U_{x,\mu\nu} = U_{x,\mu} U_{x+\mu,\nu} U^\dg_{x+\nu, \mu} U^\dg_{x,\nu},
\end{align}
thus yielding $\mathcal{W} = \left\{ \mathcal{P}_{x,\mu\nu} \right\}$. In order to avoid redundancy, it is possible to restrict the calculation to plaquettes with positive orientation, i.e.~$U_{x,\mu\nu}$ with $\mu < \nu$.

\subsection{Polyakov layers}

Another type of loop that has already been mentioned in Section~\ref{sec:ltg} is the Polyakov loop, a Wilson loop that wraps around the lattice periodic boundary. Analogously to \Plaq{}, we define the \Poly{} layer
\begin{align}
    \mathcal{L} : (\mathcal{U}) \rightarrow (\mathcal{U}, \mathcal{W}),
\end{align}
as the function that computes Polyakov loops at each lattice site from the set of gauge links $\mathcal{U}$ with the following operation
\begin{align}
    \mathcal{L}_{x,\mu}(\mathcal{U}) = \prod_k U_{x + k \cdot \mu,\mu} = U_{x,\mu} U_{x+\mu,\mu}  \dots  U_{x-\mu,\mu}.
\end{align}
The set of locally transforming objects is therefore $\mathcal{W} = \left\{ \mathcal{L}_{x,\mu\nu} \right\}$. We point out that Polyakov loops transform locally at the starting site $x$ because of the periodicity of the lattice. As shown in the next Section, plaquettes and Polyakov loops form a minimal basis to construct every possible Wilson loop on the lattice.

\section[Lattice gauge equivariant convolutional neural networks]{Lattice gauge equivariant convolutional neural networks \sectionmark{\LGCNN{}s}}
\sectionmark{\LGCNN{}s}
\label{sec:lcnns}

We list all the layers that have been introduced in the previous Section and can be combined using function composition to form a gauge equivariant neural network:
\begin{enumerate}
    \item Gauge equivariant convolutional layer (\LConv)
    \item Gauge equivariant bilinear layer (\LBL)
    \item Gauge equivariant activation function (\LAct)
    \item Exponentiation layer (\LExp)
    \item Trace layer (\Trace)
\end{enumerate}
We also defined the preprocessing layers
\begin{enumerate}
    \item Plaquette layer (\Plaq)
    \item Polyakov layer (\Poly)
\end{enumerate}
We now focus on how these layers can be assembled together into an architecture that is applicable to problems in lattice gauge theory.

A prototypical study is the prediction of physical observables, which are gauge invariant functions of Wilson loops, as discussed in section \ref{sec:ltg}. \LGCNN{}s should be able to tackle this task, which means that it should be possible for them to realize arbitrary Wilson loops by taking the set of gauge links $\mathcal{U}$ as input. This can be accomplished if we choose the initial $\mathcal{W}$ variables to be the set of plaquettes. Processing the initial tuple $(\mathcal{U}, \mathcal{W})$, \LGCNN{} architectures are indeed able to grow Wilson loops of any shape and size. We illustrate the mechanism in 1+1D by considering a simple \LGCNN{} architecture which computes a $2 \times 1$ Wilson loop from the set of gauge links $\mathcal{U}$:
\begin{align}
    \mathcal{U} \rightarrow \mathrm{\Plaq} \rightarrow \mathrm{L}\text{-}\mathrm{Conv} \rightarrow \mathrm{L}\text{-}\mathrm{Bilin}  \rightarrow (\mathcal{U}, \mathcal{W}').
\end{align}
With the first layer, the plaquettes $U_{x, 01}$, the only ones available in 1+1D, are computed at each lattice size. The tuple $(\mathcal{U}, \mathcal{W} = \{U_{x, 01}\})$ is passed to an \LConv{} layer with $N_\mathrm{ch, in} = 1$ input channels and $N_\mathrm{ch, out} = 2$ output channels. The learnable parameters can be adjusted to reproduce in one channel the original plaquettes $U_{x, 01}$ and in the other the plaquettes parallel transported from $x$ to $x + \hat{1}$. Then, the final bilinear layer combines the two channels into a single output channel multiplying the two plaquettes and giving as a result a $2 \times 1$ Wilson loop at each lattice site $x$.

The \LConv+\LBL{} combination as detailed above can be used as a building block in deeper \LGCNN{} architectures to compute Wilson loops of arbitrary size from elementary $1 \times 1$ loops. The number of times such a combination is repeated determines the maximum loop size that can be reached. For example, an \LConv+\LBL{} combination with an appropriate kernel size in the convolution can double the area of a loop, such that two \LConv+\LBL{} layers are required for $2 \times 2$ loops, three \LConv+\LBL{} layers can realize $2 \times 4$ or $4 \times 2$ loops and four \LConv+\LBL{} layers are necessary for $4 \times 4$ loops.

Although the method described above can be extended to the computation of arbitrarily large Wilson loops, not all loops can be realized this way. Polyakov loops, for example, are an exception, and the reason lies in their different topology: Wilson loops can be contracted to a single point, while Polyakov loops cannot. This issue can be conveniently circumvented by using as input $\mathcal{W}$ variables Polyakov loops, precomputed with the \Poly{} layer.

In addition to taking care of the loop shape and size, \LGCNN{}s are also able to provide non-linearity through the use of gauge equivariant activation functions in a similar fashion to how non-linear functions can be approximated in traditional neural networks. By analogy, the layer sequence \LConv{}+\LBL{}+\LAct{} allows to express a wide range of gauge equivariant non-linear functions in lattice gauge theory.

\begin{figure*}
    \centering
    \includegraphics[width=\textwidth]{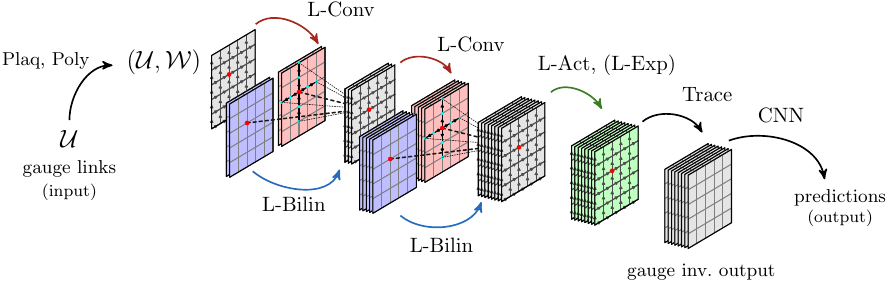}
    \caption{
        An example of \LGCNN{} architecture. Using gauge links $\mathcal{U}$, locally transforming objects $\mathcal{W}$ are computed via the \Plaq{} and \Poly{} layers. The tuple $(\mathcal{U}, \mathcal{W})$ is the input of the \LGCNN{}. A gauge-preserving convolutional operation is performed by \LConv{} using parallel transporters to connect nearby $\mathcal{W}$ objects (green dots) along the coordinate axes to a specific lattice site (red dot). The output of such a layer (blue) is combined with the original one (red) by \LBL{}, which multiplies locally transforming objects in a gauge-equivariant manner. The stacked configuration sheets signal that  \LBL{} can act on a preset number of channels. Non-linearity can be introduced via \LAct{}, while modifications of $\mathcal U$ are performed by \LExp{}, both in a gauge-equivariant way (green layer).
        The group structure is integrated out with a \Trace{} layer, and  its gauge invariant output can be further processed by a traditional CNN. The image displays 1+1D configurations but there is no restriction to the number of dimensions. Image from \cite{Favoni:2020reg}.}
    \label{fig:layers}
\end{figure*}

An \LGCNN{} architecture that can approximate a gauge invariant observable $\mathcal{O}[U]$ is given by
\begin{align}
\mathcal U &\rightarrow \mathrm{\Plaq} \rightarrow \left(\mathrm{L}\text{-}\mathrm{Conv} \rightarrow \mathrm{L}\text{-}\mathrm{Bilin} \rightarrow \mathrm{L}\text{-}\mathrm{Act} \right)^n \nonumber \\
&\rightarrow \textrm\Trace \rightarrow \mathrm{CNN} \rightarrow \mathcal{O}[U],
\label{eq:lcnn_obs}
\end{align}
where CNN indicates a traditional convolutional neural network and $\left( \dots \right)^n$ denotes the composition of $n$ \LConv{}+\LBL{}+\LAct{} blocks stacked one after the other. The \Trace{} layer makes the data gauge invariant before it enters the CNN. If translational symmetry is relevant for the prediction of the observable, we recommend using a translationally equivariant CNN as described in Chapter~\ref{chap:translations}, otherwise the CNN could even be replaced by a generic dense network without destroying gauge symmetry. An example of such an \LGCNN{} architecture is shown in Fig.~\ref{fig:layers}.

In contexts such as classical time evolution \cite{Ambjorn:1990pu} or gradient flow~\cite{Luscher:2010iy}, the set of gauge links~$\mathcal{U}$ are required to undergo modifications, which is possible via the application of \LExp{}. The $\mathcal W$ variable can be updated accordingly in the following \Plaq{} or \Poly{} layer.

We need to point out that depending on the problem, other gauge equivariant layers can be introduced. For example, in~\cite{Kanwar:2020xzo} and~\cite{Boyda:2020hsi} the coupling layer of a normalizing flow is made gauge equivariant by imposing it to be dependent on gauge invariant objects, specifically traced plaquettes made up by links that are not being updated by said coupling layer. An adaptation of the \LConv{} operation has also been used in~\cite{Lehner:2023bba} and~\cite{Lehner:2023prf} for the application of gauge equivariant neural networks to multigrid preconditioners for the Dirac operator~\cite{Brannick:2007ue}, together with gauge equivariant pooling and unpooling layers that are combined with subsampling such that gauge symmetry is respected. A generalization of \LGCNN{} layers (also including pooling operations) to preserve global symmetries as in the G-CNN construction~\cite{Cohen:2016} is given in~\cite{Aronsson:2023rli}.

\section[Proof for generation of arbitrary Wilson loops]{Proof for generation of arbitrary Wilson loops \sectionmark{Proof for generation of arbitrary loops}} \label{sec:proof}
\sectionmark{Proof for generation of arbitrary loops}

\begin{figure}
    \centering
    \includegraphics[width=\textwidth]{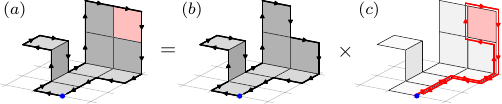}
    \caption{Sketch of the proof that \LGCNN{}s can generate arbitrarily sized Wilson loops. 
    An arbitrary Wilson loop (a) can be viewed as the boundary of a surface that can be tessellated into $n$ tiles, each of which has a \mbox{$1 \times 1$} unit lattice area. The initial point of the loop is indicated by a blue dot. This untraced Wilson loop can be decomposed into another one with $n-1$ tiles (b) and the loop surrounding the removed tile connected to the starting point via an appropriate path (c). This path has to run along the loop (b) and can be generated by consecutive \LConv{} layers. Image from \cite{Favoni:2020reg}.}
    \label{fig:proof}
\end{figure}

We provide a sketch of a proof
by induction that \LGCNN{}s can generate arbitrary Wilson loops, which is displayed in Fig.~\ref{fig:proof}.

By virtue of the 
non-Abelian version of Stokes' theorem \cite{Fishbane:1980eq}, it is possible to associate a contractible Wilson loop  with its enclosing surface, which can be tessellated into area units, i.e.~plaquettes. Such a surface is not unique in higher dimensions, hence a Wilson loop can be equivalently described by different tessellations. In order to be able to represent any Wilson loop, on topologies that are not simply connected it is necessary to include in the set of $\mathcal{W}$ loops that can not be contracted to a point, such as Polyakov loops.

An arbitrary Wilson loop can be decomposed into the same Wilson loop with a missing tile and another one, such that their product is the original Wilson loop. By means of the iterated application of \LConv{}s, it is always possible to generate the path that connects the origin (marked with a blue dot in Fig.~\ref{fig:proof}) with the removed plaquette, which can be computed via a \Plaq{} layer. The Wilson loop obtained this way can be multiplied by the Wilson loop with the missing tile in a \LBL{} layer. This argument is valid for any starting Wilson loop larger than a plaquette. This guarantees that \LGCNN{}s can in principle construct arbitrarily sized Wilson loops.
In \LGCNN{}s, generic Wilson loops are combined more efficiently by taking products of two larger sub-loops, instead of always multiplying with a loop of unit area.

The possibility of generating any closed loop enables the reconstruction of the complete gauge connection up to gauge transformations \cite{Giles:1981ej, Loll:1992fk}. With the usage of an \LAct{} layer that depends on the trace of loops, an \LGCNN{} is therefore able in principle to approximate arbitrary non-linear functions of Wilson loops. Recalling the universality of CNNs \cite{Yarotsky:2022, Zhou:2020ucn}, \LGCNN{}s can be seen as universal approximators of functions on the lattice.

\section{Computational experiments} \label{sec:comp_exp}

With an approach similar to the one in Chapter~\ref{chap:translations}, we want to demonstrate the advantage of designing neural networks that preserve gauge symmetry to tackle problems in lattice gauge theory. Specifically, we compare the performance of \LGCNN{}s with traditional CNN models on regression tasks consisting in the prediction of local gauge invariant observables $\mathcal{O}_x$. An observable that we consider is the real value of traced Wilson loops, given by
\begin{align}
W^{(m \times n)}_{x,\mu\nu} = \frac{1}{N_c} \mathrm{Re} \Tr \left[ U^{(m \times n)}_{x,\mu\nu} \right]
\label{eq:tr_w_loop}
\end{align}
where $U^{(m \times n)}_{x,\mu\nu}$ refers to an \mbox{$m\times n$} Wilson loop in the \mbox{$\mu\nu$} plane:
\begin{align}
U^{(m \times n)}_{x, \mu \nu} = U_{x,m \cdot \mu} U_{x+m\cdot \mu, n \cdot \nu} U^\dg_{x + n \cdot \nu, m\cdot \mu } U^\dg_{x, n\cdot \nu}.
\end{align}
The concatenation of gauge links $U_{x,n \cdot \mu}$ is expressed in Eq.~\eqref{eq:extended_links}. These observables are defined at every lattice site and are gauge invariant. The rationale for these experiments is that relevant observables in lattice gauge theory can be represented by non-linear functions of Wilson loops of arbitrary size.

Another observable that we focus on is the topological charge density $q_x$. The definition used here is the plaquette discretization given by Eq.~\eqref{eq:top_charge_plaq}.

\subsection{Datasets \label{sec:gauge_datasets}}

This section provides detailed information about the datasets to which the neural networks have access. The experiments are conducted on data generated in 1+1D and 3+1D with various lattice sizes and coupling constants $\beta$ using an $\SU(2)$ Markov chain Monte Carlo (MCMC) code, detailed in Appendix~\ref{app:montecarlo}.  The coupling $\beta$ is related to the Yang-Mills coupling $g$ in the continuum by
\begin{align}
    \frac{\beta}{2N_c}=\frac{1}{g^2}.
\end{align}

In order to achieve a higher computational efficiency, the set of $1 \times 1$ loops was precomputed and chosen as the locally transforming variables $\mathcal{W} = \{U_{x, \mu\nu}\}$ that is part of the input tuple $(\mathcal{U}, \mathcal{W})$ for the networks. In the following regression problems, such tuples have to be mapped to the following gauge invariant observables: $W^{(1\times 1)}_{x, \mu\nu}$, $W^{(1\times 2)}_{x, \mu\nu}$, $W^{(2\times 2)}_{x, \mu\nu}$, $W^{(4\times 4)}_{x, \mu\nu}$, and topological charge $q^\mathrm{plaq}_x$. The indices $\mu\nu$ take specific values: in 1+1D all loops lie in the $tx$ plane, therefore we compute $W^{(m\times n)}_{x, 01}$. In 3+1D we choose the $xy$ plane for all loops, thus $W^{(m\times n)}_{x, 12}$. Furthermore, the definition of the topological charge in the form of Eq.~\eqref{eq:top_charge_plaq} is only valid in 3+1D and is therefore only included in the 3+1D datasets.

\renewcommand*\arraystretch{1.4}
\begin{table}
    \caption{\label{tab:gauge_datasets} Datasets for regression tasks in 1+1D and 3+1D. The coupling constant $\beta$ ranges from $\beta_\mathrm{min} = 0.1$ to $\beta_\mathrm{max} = 6.0$, with intermediate values separated by constant steps of $\Delta \beta = (\beta_\mathrm{max} - \beta_\mathrm{min}) / N_\beta$, with $N_\beta = 10$. Table from \cite{Favoni:2020reg}.}
        \begin{tabular}{p{20mm} | l | l | p{25mm}}
            \textbf{1+1D} & & & \\
            \hline
            &Training & Validation &  Test\\
            \hline
            Lattice $N_t \times N_s$& $8 \! \times \! 8$ & $8 \! \times \! 8$ & ${8 \! \times \! 8}$, ${16 \! \times \! 16}$, ${32 \! \times \! 32}$, ${64 \! \times \! 64}$\\ 
            \hline
            Examples & $10^4$ & $10^3$ & $10^3$ per lattice \\
            \hline
            Labels & \multicolumn{3}{l}{$W^{(1 \times 1)}_{x,01}$, $W^{(1 \times 2)}_{x,01}$,  $W^{(2 \times 2)}_{x,01}$, $W^{(4 \times 4)}_{x,01}$ } \tabularnewline
            \hline
            Coupling  & \multicolumn{3}{l}{$\beta \in \{ 0.1, \dots, 6.0\}$} \tabularnewline
            \hline
            \hline
            \textbf{3+1D} & & & \\
            \hline
            &  Training &  Validation & Test\\
            \hline
            Lattice $N_t \times N_s^3$& $4 \! \times \! 8^3$ & $4 \! \times \!  8^3$ & ${4 \! \times \! 8^3}$, ${6 \! \times \! 8^3}$, ${6 \! \times \! 12^3}$, ${8 \! \times \! 16^3}$\\ 
            \hline
            Examples & $10^4$ & $10^3$ & $10^3$ per lattice  \\
            \hline
            Labels & \multicolumn{3}{l}{ $W^{(2 \times 2)}_{x,12}$, $W^{(4 \times 4)}_{x,12}$, $q^\mathrm{plaq}_x$ } \tabularnewline
            \hline Coupling  & \multicolumn{3}{l}{$\beta \in \{ 0.1, \dots, 6.0\}$} \tabularnewline
        \end{tabular}
\end{table}

\renewcommand*\arraystretch{1.0}

In order to create the datasets, lattice configurations are randomly initialized and then undergo $2 \times 10^3$ sweeps such that the system reaches equilibrium. The tuple $(\mathcal{U}, \mathcal{W})$ is then saved together with the desired observables $\{ \mathcal{O}_i \}$ every $10^2$ sweeps. In both the 1+1D and the 3+1D case, the training set consists of $10^4$ samples and the validation set of $10^3$ samples, which are generated only for the smallest lattice size. The larger lattice sizes are used only for the test sets, while the 10 different values of the coupling constant $\beta$ are employed both during training and testing. All details are reported in Table~\ref{tab:gauge_datasets}.

In the next two sections we discuss the hyperparameter choices made for the comparison between \LGCNN{} architectures and traditional CNNs, which are used as baseline models that do not respect gauge symmetry by design.

\subsection{\LGCNN{} networks}

The fully detailed composition of \LGCNN{} architectures is contained in Tables~\ref{tab:arch_lcnn_2d} and~\ref{tab:arch_lcnn_4d} in Appendix~\ref{app:lcnns}. In this section, we will go over their most relevant characteristics.

As mentioned in subsection~\ref{subsec:lbl}, \LConv{} and \LBL{} layers are combined into the \LCB{} operation. The architectures feature up to four \LCB{} layers stacked one after another, followed by a \Trace{} layer and a single linear layer applied individually at each lattice site (corresponding to a $1 \times 1$ convolution~\cite{Szegedy:2015abs}), with a total number of weights of varying from $\approx 100$ to $\approx 40\,000$. More specifically, we employed only one \LCB{} layer for the smallest Wilson loop, $W^{(1 \times 1)}$. The prediction of larger Wilson loops requires an increasing number of \LCB{} operations. Given that an \LCB{} layer doubles the maximum possible loop area, the repetition of $n$ \LCB{} layers that take plaquettes as initial input allows to predict loops of area $2^n$. In order to generate loops with a size different from powers of 2, it is possible to store loops of different sizes in different channels. As an illustration, constructing $3 \times 3$ loops can be done by  combining loops in the following sequence: $1 \times 1 \rightarrow (1 \times 1, 1 \times 2) \rightarrow (1 \times 1, 1 \times 2, 2 \times 2) \rightarrow (1 \times 3, 2 \times 3) \rightarrow 3 \times 3$, where $(\dots,\dots)$ indicates multiple channels. Compared to the general structure proposed in~\eqref{eq:lcnn_obs}, \LAct{} layers are superfluous for the observables featured here. The \LGCNN{} output is defined at every lattice site, allowing a pointwise comparison with the ground truth. We have observed that leaving out the lattice average, i.e.~a GAP layer, leads to much easier convergence of the networks during training. The inclusion of a GAP makes training much harder, often leading to models that do not converge.

\subsection{Baseline networks} \label{subsec:cnns}

The baseline models are CNNs that break gauge symmetry, whose details are displayed in Tables~\ref{tab:arch_base_w1x2},  \ref{tab:arch_base_w2x2} and~\ref{tab:arch_base_w4x4} in Appendix~\ref{app:cnns}. These networks consist of stacks of two-dimensional convolutional layers followed by an activation function, a GAP at the end of the convolutional part and a dense part leading to the predictions. With the aim of comparing the architectures fairly, the number of trainable parameters of the CNNs is chosen similar to \LGCNN{} architectures, such that their size and their ability to make predictions is in principle comparable. The amount of convolutions ranges from one or two up to six, and the number of trainable parameters varies from \mbox{$\approx 300$} to \mbox{$\approx 100\,000$}. Like the \LGCNN{} networks, the CNN models are translationally equivariant, which is achieved by choosing a stride of one in all convolutions, as discussed extensively in Section~\ref{sec:layers}. Also, periodic boundary conditions are satisfied imposing circular padding. This choice is justified by recalling that lattice gauge theory is symmetric under translations. We also note that both architecture types can be applied to different lattice sizes.

Another aspect we considered in order to ensure a fair comparison concerns the input of the CNNs. While in principle providing the networks only with gauge links $\mathcal{U}$ is sufficient, one might argue that  \LGCNN{}s are favored because they have also access to plaquettes in the input layer. For this reason, the input of the baseline networks also includes plaquettes. Empirically, the input tuple for these architectures that leads to the best performances is $(\mathcal{U}, \mathcal{W}, \mathcal{W}^\dg)$, which is the one assumed to be used for all the experiments with baseline models.

We observed that CNNs without a GAP layer performed much worse, in contrast to the results found for \LGCNN{}s. In order to enable a comparison, we take the lattice average over the final output layer of the \LGCNN{} models.

\subsection{Training details}

For both architecture types, we reduce the impact of random initializations by training ten models with different initial random weights. In all tasks, we use the \textit{AdamW} optimizer with no weight decay and early stopping. In 1+1D, the choices for the learning rate, the maximum number of epochs and the patience are chosen to be respectively $3\times 10^{-2}$, 100 and 25, while for \LGCNN{}s these number are different across the various tasks: for the prediction of $1 \times 1$ and $1 \times 2$ Wilson loops, the learning rate, the maximum number of epochs and the patience are respectively equal to $3\times 10^{-3}$, 20 and 5; for $2 \times 2$ and $4 \times 4$, these hyperparameters are set to $1 \times 10^{-3}$, 100 and 25 respectively.

Given the higher computational cost in 3+1D, training is run only for five models of the architecture reported in table~\ref{tab:arch_lcnn_4d}. For the prediction of $W^{(2 \times 2)}$ and $W^{(4 \times 4)}$, the learning rate is set to $3 \times 10^{-3}$, while it is equal to $3 \times 10^{-4}$ for the topological charge $Q_P$.

For all architectures in 1+1D, the batch size is always set to 50, and it is reduced to 10 in 3+1D because of memory limits.

\subsection{Traced Wilson loops in 1+1D}

The first task we investigate is almost trivial: predicting the real value of the traced $1 \times 1$ Wilson loop as in Eq.~\eqref{eq:tr_w_loop}. Since part of the input data consists of all possible $1 \times 1$ Wilson loops, the only required computations are first to perform the trace operation and then to take its real part.

\begin{figure}
    \centering
    \includegraphics[width=.8\textwidth]{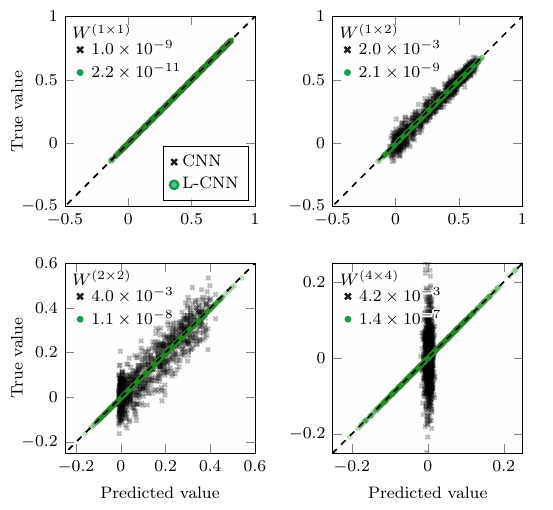}
    \caption{Scatter plot showing predictions versus ground truth of best \LGCNN{} and CNN models for traced Wilson loops of various sizes tested on $8 \times 8$ lattices. The dashed line represents points where perfect predictions lie. The performance of the traditional CNN model strongly deteriorates as the loop size increases, to the point where predictions of $4 \times 4$ loops mostly lie close to the average of the training set, signaling that the CNNs did not learn any meaningful physical information.
    On the contrary, the \LGCNN{} model scores a very low MSE (shown at the top left of each plot) for every loop size. Image from \cite{Favoni:2020reg}.}
    \label{fig:d2_scatter}
\end{figure}

\begin{figure}
    \centering
    \includegraphics[width=.8\textwidth]{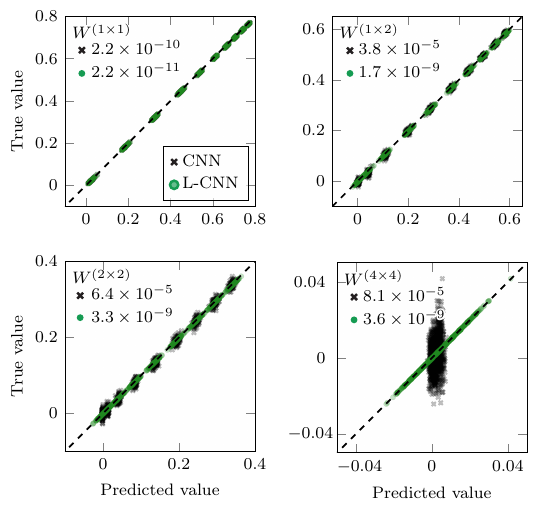}
    \caption{Scatter plot showing predictions versus ground truth of best \LGCNN{} and CNN models for traced Wilson loops of various sizes tested on $64 \times 64$ lattices. This is the same comparison made in Fig.~\ref{fig:d2_scatter}, but on the larger lattice size available in the test set. Similar conclusions to the previous scatter plot can be drawn. Each cluster visible in the plots is related to a specific value that the coupling constant \mbox{$\beta$} takes. Image from \cite{Favoni:2020reg}.}
    \label{fig:d2_scatter_large}
\end{figure}

Scatter plots comparing predictions to true values on $8 \times 8$ and $64 \times 64$ are shown in the top left plot of Figs.~\ref{fig:d2_scatter} and~\ref{fig:d2_scatter_large}  respectively. The models chosen for these plots are the best models based on validation loss on $8 \times 8$ lattices. It is clear that both the \LGCNN{} and the CNN models can perform the task of computing $W^{(1 \times 1)}$, although the \LGCNN{} performs better by a few orders of magnitudes in terms of MSE. This should not come as a surprise, considering that the \LGCNN{} architecture includes a \Trace{} layer, whereas the baseline model has to learn this operation during training. It should also be stressed that both architectures generalize well across different lattice sizes, even though training has only been performed on the smallest lattice size. As in Chapter~\ref{chap:translations}, our models do not need to be retrained when tested on larger lattices, and this applies to all tasks analyzed in this section. More details about the results obtained by the different architectures in the prediction of $W^{(1 \times 1)}$ are provided in Appendix~\ref{app:results} in Table~\ref{tab:results_w1x1}.

Traced Wilson loops of sizes $1\times 2$, $2 \times 2$ and $4 \times 4$ offer more sophisticated regression tasks. In this case, we try different \LGCNN{} architectures that are small, medium and large in terms of number of parameters for each task (see Table~\ref{tab:arch_lcnn_2d}). Analogously to $W^{(1 \times 1)}$, we display a comparison for these larger loops in the top right and bottom plots of Figs.~\ref{fig:d2_scatter} and~\ref{fig:d2_scatter_large}, where CNNs clearly do not solve adequately the more complicated task of computing larger Wilson loops. Conversely, all three \LGCNN{} architectures perform well on the $1 \times 2$ and $2 \times 2$ tasks, and larger \LGCNN{} models perform generally better in the $4 \times 4$ regression task. In some cases, the models do not converge during training, which shows why it is essential to repeat the training process multiple times with different random initializations of the weights. Another noteworthy observation is that baseline models struggle to predict negative values of traced Wilson loops, as can be seen in the top right and bottom left plots of Fig.~\ref{fig:d2_scatter} for $W^{(1 \times 2)}$ and $W^{(2 \times 2)}$. In the case of $W^{(4 \times 4)}$, the CNN seems to collapse and its predictions lie very close to the training set average. This shows that the baseline architecture is unable to learn any meaningful connection between the input and output data.

We report the performance of the different architectures used for the regression tasks with Wilson loops in Appendix~\ref{app:results} in Tables~\ref{tab:results_w1x1}, \ref{tab:results_w1x2}, \ref{tab:results_w2x2} and~\ref{tab:results_w4x4}.

\subsection{Traced Wilson loops in 3+1D}

The Wilson loop regression task is further complicated by moving from 1+1D to 3+1D lattices. In the previous section we have already demonstrated that traditional CNNs are not able to satisfactorily learn even moderately sized Wilson loops in 1+1D. Furthermore, four-dimensional convolutional layers are currently not built-in modules in \textit{PyTorch}, leading to a suboptimal formulation of traditional CNN architectures in 3+1D. We therefore focus only on testing \LGCNN{} models in this section.

The regression task is very similar to the one in 1+1D: we train models to predict the value of traced $2 \times 2$ and $4 \times 4$ Wilson loops. One immediate complication that arises when going from 1+1D to 3+1D lattices is that memory requirements grow quickly. If we consider a $D+1$-dimensional lattice with $N_s$ lattice sites for each spatial coordinate and $N_t$ lattice sites in the temporal one, the amount of real numbers in the input layer is (the first summand coming from the links, the second from the loops)
\begin{align}
N_\mathrm{input} = 2 \, N^2_c \, N_t \, N_s^D \, \left((D+1) + (D+1) \frac{D}{2} \right),
\label{eq:dataset_input_size}
\end{align}
and we see that a single sample of the $4 \cdot 8^3$ training set is larger by two orders of magnitude compared to a sample of the $8 \cdot 8$ training set. As a result, training a large \LGCNN{} architecture takes longer, which is why the batch size, the total number of epochs and the number of differently initialized instances are reduced compared to the 1+1D case. Test results are summarized in Appendix~\ref{app:results} in Table \ref{tab:res_4d}. Interestingly, smaller architectures seem to perform better in 3+1D compared to larger architectures. Overall, it is clear that \LGCNN{}s are generally able to represent traced Wilson loops in higher dimensions. We also observe that successfully trained models generalize across different lattice sizes in 3+1D.

We conclude by mentioning that it may seem trivial for \LGCNN{}s to solve regression tasks involving Wilson loops, since they are intrinsically constructed to form arbitrarily sized loops. As extensively discussed in the supplementary material of~\cite{Favoni:2020reg}, the number of possible loops grows exponentially with their perimeter, therefore problems where the contribution from different Wilson loop shapes is unknown are challenging. An example is given by the so-called improved actions, which have been conceived to alleviate the influence of discretization effects through the inclusion of variously sized Wilson loops~\cite{Symanzik:1983dc,Symanzik:1983gh, Luscher:1984xn}. In particular, for fixed-point actions the renormalization group is used to remove lattice artifacts to obtain, for example, on-shell quantities~\cite{Hasenfratz:1993sp}. The application of \LGCNN{}s to a fixed-point action in SU(3) has improved the previously existing state-of-the-art results~\cite{Holland:2024muu}. Another situation in which the contribution of various Wilson loops has to be determined is represented by trivializing maps, which map a theory to its strong-coupling limit, reducing critical slowing down. The approach suggested by L\"uscher in~\cite{Luscher:2009eq} can be extended order by order identifying the relevant loops and computing their coefficient, an approach that can be systematized as proposed in~\cite{Bacchio:2022vje}. In principle, \LGCNN{}s can be used in this approach, as we will describe in Chapter~\ref{chap:ngf}.

\subsection{Topological charge in 3+1D}

The final observable we consider is the topological charge density $q_x^\mathrm{plaq}$ in the plaquette approximation (see Eq.~\eqref{eq:top_charge_plaq}). Since the numerical values of this observable are rather small, it is reasonable to rescale the dataset by means of a constant factor, i.e. 
\begin{align}
    \tilde q^\mathrm{plaq}_x = C q^\mathrm{plaq}_x, \qquad C > 0,
\end{align}
in order to facilitate training. For this experiment, we employ $C=100$. In the following, $Q_P$ will denote the topological charge, which is equal to the sum over the lattice of the charge density
\begin{align}
    Q_P = \sum_{\mathbf x} q_\mathbf{x}^{\mathrm{plaq}}.
    \label{eq:top_charge}
\end{align}

The topological charge density is rather simple to describe in comparison to larger Wilson loops: it can be written as a linear combination of multiplications of two $1 \times 1$ Wilson loops. This regression task can be solved by relatively shallow \LGCNN{} models. The architecture employed is reported in Table~\ref{tab:arch_lcnn_4d}, and its results are shown in Table~\ref{tab:res_4d}. Only one of the five randomly initialized models did not converge during training. In all other cases, we observed convergence to solutions which can predict the topological charge to a high degree of accuracy. These results are in line with the MSEs exhibited by \LGCNN{} architectures in the two-dimensional tasks.

\begin{figure}
    \centering
    \includegraphics[width=.7\textwidth]{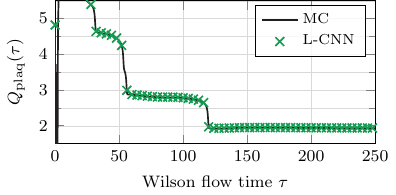}
    \caption{Predictions of the best \LGCNN{} model for the topological charge on an \mbox{{$8 \times 24^3$}} configuration at $2 / g^2 = 0.2$ evolved with Wilson flow. MC indicates the Wilson flow obtained with Monte Carlo updates. Image from \cite{Favoni:2020reg}.}
    \label{fig:qp_flow}
\end{figure}

Having at our disposal the topological charge enables us to perform another more elaborated test than just a performance check with the MSE. A typical lattice study is the evolution of $Q_P$ under Wilson flow or more generally gradient flow~\cite{Luscher:2010iy, Alexandrou:2017hqw}. In the continuum limit, the topological charge takes integer values due to restrictions that arise from the topology. We design a Wilson flow starting from configurations $U_{x, \mu}$ obtained from the MCMC simulation in Appendix~\ref{app:montecarlo} and then moderately smeared, such that the initial condition for the Wilson flow is
\begin{align}
    U_{\mathrm x,\mu}(\tau = 0) = U_{\mathrm x, \mu},
\end{align}
with $\tau$ denoting the auxiliary Wilson flow time. The update procedure employed to realize Wilson flow is
\begin{align}
    U_{\mathbf x,\mu}(\tau + \Delta \tau) &= \exp{\big( i \Delta \tau \, \omega_{\mathbf x,\mu}(\tau) \big)} U_{\mathbf x,\mu}(\tau),
\end{align}
where $\omega$ is an algebra element given by
\begin{align}
    \omega_{\mathbf x, \mu}(\tau) = - \sum_{|\nu|} \left[ U_{\mathbf x,\mu\nu}(\tau) \right]_{\mathrm{ah}},
\end{align}
with the sum $\sum_{|\nu|}$ running over the positive and the negative direction of the axis $\nu$ and the anti-Hermitian is computed as in Eq.~\eqref{eq:ah}. The steps $\Delta \tau$ are chosen to be small, leading to a gradual decrease of the Wilson action:
\begin{align}
    S_W[ \mathcal{U}(\tau + \Delta \tau) ] < 
    S_W[ \mathcal{U}(\tau)], \qquad \Delta \tau \ll 1.
\end{align}
Using $\Delta \tau = 0.005$ and training uniquely on the un-flowed configurations $U_{x, \mu}$ with a lattice size of $4 \times 8^3$, the \LGCNN{} models reach a high degree of accuracy on $8 \times 24^3$ lattices, as can be seen in Fig.~\ref{fig:qp_flow}.

We notice that as the configurations are flowed the values of the topological charge get closer to integers, which is the expected behavior, whose details are related to the lattice size and the initial condition at $\tau = 0$.

\chapter{Neural gradient flow} \label{chap:ngf}

In this chapter, we design a method for generating configurations distributed according to a given action starting from a set of random configurations. We describe how the mapping between the initial and the final configurations can be approximated by solving a differential equation with neural networks, specifically with L-CNNs in the context of gauge theories. Since the initial configurations are random, the mapping amounts to a trivializing flow, as introduced by L\"{u}scher in \cite{Luscher:2009eq}. The advantage of this framework is that sampling new configurations is very cheap once the neural network has been successfully trained. The differential equation that is solved is equivalent to a gradient flow \cite{Alexandrou:2017hqw, Luscher:2010iy} type of equation, which is why we call this technique neural gradient flow. We check the ability of a neural network to solve such an equation on a toy model consisting of single-link configurations in SU(2). We employ a simple network that mimics the behavior of L-CNNs and test it initially on the same interval of gradient flow time used during training, then we extend the test to much later times. Finally, we discuss the adjoint sensitivity method, which avoids memory saturation when performing backpropagation, and give our derivation in SU(N).

\section{Adaptation of NODEs to lattice gauge theory} \label{sec:ngf}

The starting point of this chapter is a technique that was proposed in~\cite{Chen:2018}. It aims to solve the first order ordinary differential equation for the time-dependent vector $\mathbf{x}(t) \in \mathbb{R}^D$
\begin{equation}
\label{node}
    \frac{\mathrm{d}\mathbf{x}}{\mathrm{d}t}=\mathbf{f}(\mathbf{x}(t),\theta,t),
\end{equation}
where the function $\mathbf{f}(\mathbf{x}(t),\theta,t)$ is not known a priori, hence its parameters $\theta$ have to be optimized to describe the dynamics of the vector. A neural network can be used to represent $\mathbf{f}$, which is why this setup takes the name neural ordinary differential equation (NODE). It is possible to interpret NODEs as a special case of residual networks~\cite{He:2016res} in the limit of the discrete step between each layer going to zero, becoming the continuous time $t$ in Eq.~\eqref{node}. Given an input state $\mathbf{x}_0=\mathbf{x}(t_0)$ as initial condition at $t=t_0$, we can formally solve the NODE integrating Eq.~\eqref{node}:
\begin{align}
    \mathbf{x}(T)=\mathbf{x}_0+\int_{t_0}^{T}\mathrm{d}t\,\mathbf{f}(\mathbf{x}(t),\theta,t).
\end{align}
Common ODE solvers such as Euler and Runge-Kutta (see e.g.~ Appendix A.7.1 of \cite{Thijssen:2007})
can be employed to get the prediction $\mathbf{x}(T)$ at the final time $T$. In order to be able to train the network, we can provide a dataset containing $N_\text{samples}$ initial states $\mathbf{x}_0^i=\mathbf{x}^i(t_0)$ and target final states $\mathbf{\tilde{x}}_T^i$, then use the MSE (cf.~with Eq.~\eqref{eq:reg_loss})
\begin{align}
    \mathcal{L}(\theta)=\displaystyle \frac{1}{N_{\text{samples}}} \sum_{i=1}^{N_\text{samples}} (\mathbf{\tilde{x}}_T^i-\mathbf{x}^i(T))^2
    \label{eq:ngf_loss}
\end{align}
as loss function to optimize the weights $\theta$. If other intermediate states of the actual dynamics $\mathbf{\tilde{x}}_j^i$ at time $t_j$ are known, it is possible to include in the MSE also the discrepancies between them and the predictions the network yields, thus leading to
\begin{align}
    \mathcal{L}(\theta)=\displaystyle \frac{1}{N_{\text{samples}}} \sum_{i=1}^{N_\text{samples}} \sum_{j=1}^{n}(\mathbf{\tilde{x}}_j^i-\mathbf{x}^i(t_j))^2,
\end{align}
with $t_n = T$. Resorting exclusively to the final state using the loss in Eq.~\eqref{eq:ngf_loss} requires that the dataset provides sufficient information for a successful reconstruction of the underlying dynamics.

Another option is to treat Eq.~\eqref{node} as a continuous normalizing flow~\cite{Chen:2018}, where the goal is the approximation of a probability distribution that can be achieved for example using the Kullback-Leibler (KL) divergence
\begin{align}
    \text{KL}(q(\mathbf{x}) \, || \, p(\mathbf{x})) = \int \mathrm{d}\mathbf{x} \, p(\mathbf{x}) \log{\frac{p(\mathbf{x})}{q(\mathbf{x})}}
    \label{eq:kl_div}
\end{align}
between the target distribution $p(\mathbf{x})$ and the probability distribution of the model $q(\mathbf{x})$.

As already mentioned in Chapter~\ref{chap:lcnns}, in lattice gauge theory expectation values of observables are calculated via the path integral
\begin{align}
    \langle\mathcal{O}\rangle = \frac{1}{Z} \int \mathrm{D}[U] \, \mathcal{O}(U) \mathrm{e}^{-S[U]},
\end{align}
with $Z$ being the partition function
\begin{align}
    Z = \int \mathrm{D}[U] \mathrm{e}^{-S[U]}.
\end{align}
If the links are redefined as $U = \mathcal{F}(V)$, the formula for the expectation value becomes
\begin{align}
    \langle\mathcal{O}\rangle = \frac{1}{Z} \int \mathrm{D}[V] \, \mathcal{O}(\mathcal{F}(V)) \mathrm{e}^{-S_{\mathcal{F}}[V]},
    \label{eq:trivial_path_int}
\end{align}
where the action $S_{\mathcal{F}}[V]$ incorporates the change of measure:
\begin{align}
    S_{\mathcal{F}}[V] = S(F(V)) - \log \det J(\mathcal{F}),
\end{align}
with $J(\mathcal{F})$ indicating the Jacobian.
The transformation $\mathcal{F}$ is said to be a trivializing map if $S_{\mathcal{F}}[V]$ is equal to a real-valued constant. In fact, the probability distribution for the redefined action is given by
\begin{align}
    p(V) = \frac{1}{Z} \mathrm{e}^{-S_{\mathcal{F}}[V]},
\end{align}
which is also constant, therefore it is possible to compute expectation values via~\eqref{eq:trivial_path_int} sampling from a uniform distribution. A model that approximates the transformation $\mathcal{F}(V)$ is characterized by a probability distribution $q(V)$ and can learn to reproduce a uniform distribution using the KL divergence~\eqref{eq:kl_div}, which is the approach used in \cite{Bacchio:2022vje} to find trivializing maps.

In order to find trivializing maps, L\"{u}scher proposed to solve an equivariant gradient flow equation that can be written as
\begin{align}
    \frac{\mathrm{d}U_{x,\mu}(\tau)}{\mathrm{d}\tau} = iH_\mu[U_{x,\mu}(\tau),\tau]\,U_{x,\mu}(\tau),
    \label{eq:luscher_flow}
\end{align}
where $U_{x,\mu}$ is a configuration of $\SU(N_c)$ gauge links and $\tau$ is the gradient flow time. It is natural to parameterize $H_\mu$ as the gradient of the flow action $\tilde{S}$ (hence the name gradient flow), i.e.
\begin{align}
    iH_\mu[U_{x,\mu}(\tau),\tau] = -\p_{x, \mu} \tilde{S}[U(\tau)],
    \label{eq:force}
\end{align}
where we introduced the group derivative
\begin{align}
    \p_{x, \mu} f(U) = \left. t_a \frac{\mathrm{d} f(\mathrm{e}^{\epsilon t^a}U_{x, \mu})}{\mathrm{d} \epsilon} \right|_{\epsilon = 0},
    \label{eq:group_der}
\end{align}
with $t_a$ being the group generators.
The action can be written as a linear combination of Wilson loops
\begin{align}
    \tilde{S}[U, \tau] = \sum_i c_i(\tau) W_i(U),
\end{align}
where the coefficients $c_i$ have to be determined. In \cite{Luscher:2009eq}, this action was expanded as
\begin{align}
    \tilde{S}[U, \tau] = \sum_{n=0}^\infty \tau^n \tilde{S}^{(n)}
\end{align}
in the flow time $\tau\in[0,1]$, and the leading order $\tilde{S}^{(0)}$ and its first correction $\tilde{S}^{(1)}$ were calculated analytically. In \cite{Bacchio:2022vje}, using a predetermined set of loops, the coefficients $c_i$ were machine-learned as a function of the flow time. The ability of \LGCNN{}s to generate in principle any Wilson loop allows even more flexibility in tackling this problem. We therefore intend to solve the following flow equation
\begin{equation}
    \frac{\mathrm{d}U_{x,\mu}(\tau)}{\mathrm{d}\tau} = iH_\mu[U_{x,\mu}(\tau),\theta,\tau]\,U_{x,\mu}(\tau),
    \label{eq:ngf}
\end{equation}
where $H_\mu[U(\tau),\theta,\tau]$ is an \LGCNN{} parametrized by the learnable parameters $\theta$. We can show that $H$ needs to be traceless and Hermitian. This derives from the requirement that the links remain in the group during the flow. The unitarity constraint
\begin{align}
    U(\tau)U^\dg(\tau) = \one
    \label{eq:unitarity}
\end{align}
needs to be preserved at every flow time $\tau$. We dropped the indices $x$ and $\mu$ for simplicity. Taking the derivative of both sides of the equation, we obtain
\begin{align}
    \displaystyle\frac{\mathrm{d}}{\mathrm{d}\tau}(UU^\dg) = \frac{\mathrm{d}U}{\mathrm{d}\tau} U^\dg + U \frac{\mathrm{d}U^\dg}{\mathrm{d}\tau}= 0,
    \label{eq:unitarity_der}
\end{align}
where the dependence on time is implied. The Hermitian of Eq.~\eqref{eq:ngf} is
\begin{align}
    \frac{\mathrm{d}U^\dg}{\mathrm{d}\tau} = -iU^\dg H^\dg.
\end{align}
When inserted in Eq.~\eqref{eq:unitarity_der}, we obtain
\begin{align}
    iHUU^\dg-iUU^\dg H^\dg = 0,
\end{align}
which can be simplified using Eq.~\eqref{eq:unitarity}, bringing us to the Hermiticity of $H$, i.e.
\begin{align}
    H=H^\dg.
\end{align}
The other property that has to be maintained under neural gradient flow is that the links must be represented by special matrices:
\begin{align}
    \det U(\tau) = 1.
    \label{eq:special}
\end{align}
We can recall Jacobi's formula, which relates the determinant of a square matrix $A(t)$ dependent on a real variable $t$ with its trace as follows:
\begin{align}
    \frac{\mathrm{d}}{\mathrm{d}t}\left(\det{A(t)}\right) = \det{A(t)} \, \Tr\left(A^{-1}(t) \, \frac{\mathrm{d}A(t)}{\mathrm{d}t} \right).
\end{align}
Jacobi's formula for the link variable $U(\tau)$ gives
\begin{align}
    \frac{\mathrm{d}}{\mathrm{d}\tau}\left(\det{U(\tau)}\right) = \det{U(\tau)} \, \Tr\left(U^{-1}(\tau) \, \frac{\mathrm{d}U(\tau)}{\mathrm{d}\tau} \right).
\end{align}
Using the specialness of $U(\tau)$~\eqref{eq:special} and Eq.~\eqref{eq:ngf}, we find that
\begin{align}
    0 = 1 \cdot \Tr \left(U^{-1}(iHU) \right),
\end{align}
where we imply the dependence on time. Since the trace is cyclic, $U$ and $U^{-1}$ simplify and we are left with the tracelessness of $H$:
\begin{align}
    \Tr H=0.
\end{align}
We can also derive the transformation property of $H$ during gradient flow. Let us consider the formal solution of Eq.~\eqref{eq:ngf}:
\begin{align}
    U_{x,\mu}(\tau) = U_{x,\mu}(\tau_0) + \int_{\tau_0}^\tau \mathrm{d}\tau' \, \left(i\,H_\mu(\tau')\, U_{x,\mu}(\tau')\right).
    \label{eq:ngf_int}
\end{align}
We can apply a time-independent left transformation $\Omega_x \in \SU(N_c)$ and a time-independent right transformation $\Omega_x^\dg$ to both sides obtaining
\begin{align}
    \Omega_x U_{x,\mu}(\tau)\,\Omega_{x+\mu}^\dg = \, &\Omega_x U_{x,\mu}(\tau_0)\,\Omega_{x+\mu}^\dg \nonumber \\
    &+ i \int_{\tau_0}^\tau \mathrm{d}\tau' \left(\Omega_x \, H_\mu(\tau') \, \Omega_x^\dg \Omega_x \, U_{x,\mu}(\tau') \, \Omega_{x+\mu}^\dg \right),
\end{align}
where we inserted the product $\Omega_x^\dg \Omega_x=\one$ between $H$ and $U$.
This expression informs us that if we want the transformation property of the links in Eq.~\eqref{eq:link_tr} to hold for every $\tau$, $H$ has to be a locally transforming object, meaning that it transforms according to
\begin{align}
    T_\Omega H = \Omega_x H \, \Omega^\dg_x.
\end{align}

Given the properties just discussed, $H$ can be modeled with an \LGCNN{}, in particular by taking from the output tuple $(\mathcal{U'}, \mathcal{W'})$ the objects of type $\mathcal{W'}$, since they follow the same gauge transformation of $H$, namely Eq.~\eqref{eq:gauge_transf}. The approach we propose to train such an \LGCNN{} is to generate a dataset with initial gauge link configurations $U_{x,\mu,0}^i = U_{x, \mu}^i(\tau_0)$ and target output configurations $\tilde{U}_{x, \mu, T}^i$, and then make use of the loss function
\begin{align}
    \mathcal{L}(\theta) = \frac{1}{N_{\text{samples}}} \sum_{x, \mu, i} \|\tilde{U}_{x, \mu, T} - U_{x, \mu}^i(T)\|^2,
\end{align}
to minimize the distance between the target configurations and the predicted configurations $U_{x, \mu}^i(T)$. We use the Frobenius norm to compute matrix distances. For a matrix $A \in \mathbb{C}^{n \times n}$ with elements $a_{lm}$ it is defined as
\begin{align}
    \|A\|^2 = \sum_{l,m=1}^n |a_{lm}|^2.
    \label{eq:frobenius}
\end{align}

In order to perform the evolution of the links, we need to solve the integral in Eq.~\eqref{eq:ngf_int}. Standard ODE integrators are not suited to compute trajectories on the $\SU(N_c)$ manifold. Instead, we resort to the exponential map, which we apply iteratively according to
\begin{align}
   U_{x,\mu}^i(\tau_{j+1})=\exp\left(iH_\mu[U^i(\tau_j), \theta, \tau_j]\Delta \tau\right)U_{x,\mu}^i(\tau_j).
   \label{eq:discrete_flow}
\end{align}
Since $H$ is traceless and Hermitian, we can view it as an $\su(N_c)$ algebra element, which the exponential map projects onto $\SU(N_c)$, thus guaranteeing that the links are flowed on the $\SU(N_c)$ manifold.

\section[Single SU(2) link toy model]{Single SU(2) link toy model \sectionmark{SU(2) toy model}}
\sectionmark{SU(2) toy model}

\begin{figure}[t]
    \centering
    \includegraphics[width=\textwidth]{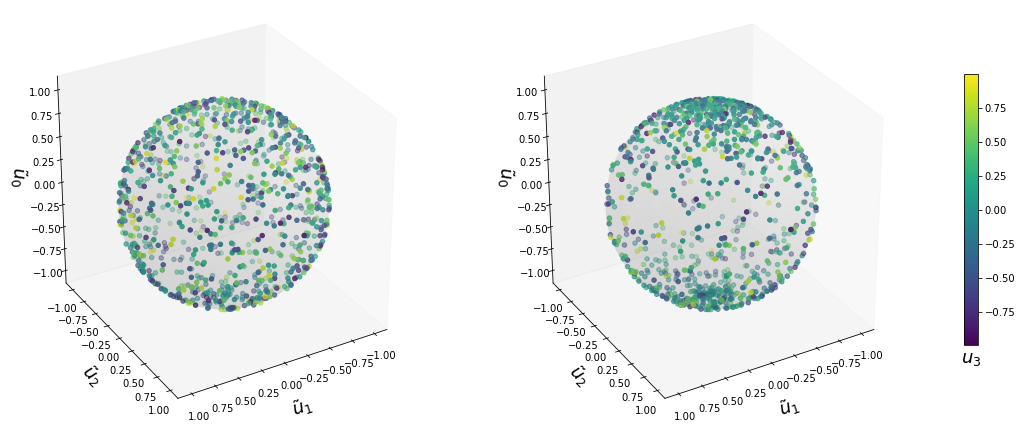}
    \caption{Projection on the 3D unit sphere of $1\,000$ matrices taken from the dataset. Left: visualization of the initial conditions $U_0^i$. Right: visualization of the target matrices $\tilde{U}_T^i$, resulting from the application of gradient flow to the initial conditions according to the action $S[U]=\mathrm{Re}\,\mathrm{Tr}\, (U^2)$ up to $T = 1$. Image from~\cite{Favoni:2022mcg}.}
    \label{fig:dataset}
\end{figure}

We test the framework delineated in the previous section on a toy model consisting of one $\SU(2)$ link governed by the action
\begin{align}
    S[U] = -\mathrm{Re}\,\Tr (U^2).
    \label{eq:action}
\end{align}
The input configurations are randomly distributed in such a way that the $\SU(2)$ manifold is sufficiently well covered, as shown in the left plot of Fig.~\ref{fig:dataset}. In order to get the target configurations, we flow the initial conditions according to the action. This can be done introducing an infinitesimal variation of the link
\begin{align}
    \delta U = i \delta A \, U,
\end{align}
and calculating the corresponding variation of the action
\begin{align}
    S[U+\delta U] &= -\Re \, \Tr ((U + \delta U)^2) \nonumber \\
    &= -\Re \, \Tr (U^2 + \delta U \, U + U \delta U) + O(\delta U^2) \nonumber \\
    &= -\Re \, \Tr (U^2 + 2\delta U \, U) + O(\delta U^2) \nonumber \\
    &= S[U] + \delta S[U, \delta U] + O(\delta U^2),
\end{align}
which can be written explicitly as
\begin{align}
    \delta S[U, \delta U] &= -2 \Re \, \Tr(\delta U \, U) \nonumber \\
    &= -2 \Re \, \Tr(i \delta A \, U^2).
\end{align}
Given the three Lie algebra generators $t_a = \sigma_a/2$ and Pauli matrices $\sigma_a$
\begin{align}
    \sigma_1 = \begin{pmatrix}
        0 & 1 \\
        1 & 0
    \end{pmatrix}, \quad
    \sigma_2 = \begin{pmatrix}
        0 & -i \\
        i & 0
    \end{pmatrix}, \quad
    \sigma_3 = \begin{pmatrix}
        1 & 0 \\
        0 & -1
    \end{pmatrix},
    \label{eq:pauli_mat}
\end{align}
we can write $\delta A = \delta A_a \, t_a$ and rework the variation of the action as
\begin{align}
    \delta S[U, \delta A] &= -2 \Re \left[\delta A_a \, \Tr(i t_a \, U^2)\right] \nonumber \\
    &= 2 \delta A_a \, \Im \left[\Tr(t_a \, U^2)\right],
\end{align}
where repeated indices are intended to be summed over and we also used $\Re(iw) = -\Im w$ for a complex number $w$.
We have thus obtained the variation of the action in terms of the infinitesimal coefficients $\delta A_a$
\begin{align}
    \frac{\delta S[U]}{\delta A_a} = 2 \Im \left[\Tr(t_a \, U^2)\right].
\end{align}
This term can be interpreted as the force, and, analogously to L\"{u}scher's setup provided by Eqs.~\eqref{eq:luscher_flow} and \eqref{eq:force}, we can write the flow as
\begin{equation}
    \frac{\mathrm{d}U}{\mathrm{d}\tau}=i t_a \frac{\delta S[U]}{\delta A_a} U(\tau).
    \label{eq:toy_model_flow}
\end{equation}
This differential equation can be solved analogously to Eq.~\eqref{eq:discrete_flow} as
\begin{align}
    U(\tau_{j+1})=\exp\left(i t_a \frac{\delta S[U]}{\delta A_a}\Delta \tau\right)U(\tau_j)
    \label{eq:true_flow}
\end{align}
in the limit of $\Delta \tau \rightarrow 0$.

Summarizing, the input of the network is a set of approximately uniformly distributed $\SU(2)$ matrices $U^i_0$ and the target is the set of flowed matrices $\tilde U^i_T$. The action~\eqref{eq:action} is characterized by two minima, i.e.~$\pm \one$, which are the elements towards which Eq.~\eqref{eq:true_flow} flows the initial links. More precisely, the link flows towards $+\one$, which we call north pole, if $\Tr \, U > 0$, while if $\Tr \, U < 0$ the dynamics pushes the link closer to the south pole $-\one$. Links with $\mathrm{Tr} U = 0$ are unstable stationary states. \LGCNN{}s can reconstruct the dynamics of the links via Eq.~\eqref{eq:discrete_flow}, and given its similarity to Eq.~\eqref{eq:true_flow}, it is clear that the objective of the network is to learn the force.

A visualization of a dataset $\SU(2)$ is not immediate, but we adopt a projection onto the three-dimensional unit sphere with the following steps. First, we rewrite the matrix $U$ as a linear combination of the identity and the Pauli matrices:
\begin{equation}
    U=u_0\one+i\sigma_au_a.
\end{equation}
Since the Pauli matrices are traceless and satisfy the anticommutation relation
\begin{align}
    \{\sigma_i, \sigma_j\} = 2 \delta_{ij} \one,
\end{align}
the coefficients $u_k$, with $k \in \{0,1,2,3\}$ can be found by
\begin{align}
    u_0=\frac{1}{2}\mathrm{Tr}\,(U), \qquad u_i=\frac{1}{2i}\mathrm{Tr}\,(U\sigma_i), \quad i \in \{1, 2, 3\}.
\end{align}
The group $\SU(2)$ is characterized by three degrees of freedom, which suggests that these coefficients are subject to a constraint. Indeed, the matrix $U$ can be rewritten via the exponential map and its power series as
\begin{align}
    U &= \exp \left(i A_a t_a \right) \nonumber \\
    &= \displaystyle \sum_{n=0}^\infty \frac{(i A_a t_a)^n}{n!} \nonumber \\
    &= \sum_{n=0}^\infty \frac{(-1)^n (A_a t_a)^{2n}}{2n!} + i \sum_{n=0}^\infty \frac{(-1)^n (A_b t_b)^{2n}}{(2n+1)!} A_c t_c \nonumber \\
    &= \sum_{n=0}^\infty \frac{(-1)^n (\frac{A}{2})^{2n}}{2n!} \one + i \sum_{n=0}^\infty \frac{(-1)^n (\frac{A}{2})^{2n+1}}{(2n+1)!} e_a \sigma_a \nonumber \\
    &= \cos \left(\frac{A}{2} \right) \one +i \sin \left(\frac{A}{2} \right) e_a \sigma_a,
\end{align}
where we have introduced $A = \sqrt{A_a A_a}$ and $e_a = A_a/A$ (see derivation e.g. in Appendix C of~\cite{Muller:2019bwd}). Most of the generators were simplified by the property $4t_a^2 = \one$. We can now equate this result with the coefficients above, obtaining
\begin{align}
    u_0 = \cos \left(\frac{A}{2} \right), \qquad u_a = \sin \left(\frac{A}{2} \right) e_a, \quad a \in \{1, 2, 3\}.
\end{align}
Summing over the square of the coefficients, it follows the group constraint
\begin{align}
    u_0^2+u_a u_a=1.
\end{align}
Moreover, we can see that each parameter is bounded:
\begin{align}
    -1 \leq u_k \leq 1 \, \quad k \in \{0, 1, 2, 3\}.
\end{align}
We now slightly modify the parametrization by normalizing $u_0$, $u_1$ and $u_2$:
\begin{align}
    \tilde{u}_j=u_j/\sqrt{u_0^2+u_1^2+u_2^2}, \quad j\in\{0,1,2\}.
\end{align}
With this choice, the triple $\{\tilde{u}_0, \tilde{u}_1, \tilde{u}_2\}$ lies on the three-dimensional unit sphere, and the remaining parameter $u_3$ is associated with a color map, as shown in Fig.~\ref{fig:dataset}. In the left plot, a subset of the initial matrices is visualized, and on the right plot we see the outcome under the effect of Eq.~\eqref{eq:discrete_flow}. Choosing $\tilde{u}_0$ as the vertical axis, allows to identify $U = +\one$ with the north pole and $U = -\one$ with the south pole, while matrices for which $\Tr \, U = 0$ lie on the equator.

Since for this problem configurations contain only one link, it is not clear how to define locally transforming objects that represent the input of an \LGCNN{}. For this reason, we design a network which guarantees that the flowed matrices stay on the $\SU(2)$ manifold. The first step to realize it is the separation of $U$ into its real and imaginary parts, which represent the input of a dense network with real weights. Its output consists of eight real numbers that are assembled to form a complex matrix. This matrix is in general neither special nor unitary, so some kind of projection is needed. We make use of Eq.~\eqref{eq:ah}, which projects it onto the $\su(2)$ Lie algebra. In order to get a group element, we finally apply the exponential map. This whole construction makes sure that the evolution stays on the $\SU(2)$ group. Let us point out that the action~\eqref{eq:action} is invariant under the transformation $U \rightarrow \Omega U \Omega^\dg$ with $\Omega\in \SU(2)$, however this symmetry property is not implemented in the network just described and is ignored in the following for simplicity.

\begin{figure}[t]
    \minipage{0.5\textwidth}
    \centering
    \subfigure[~True versus predicted trajectories]{\includegraphics[width=\textwidth]{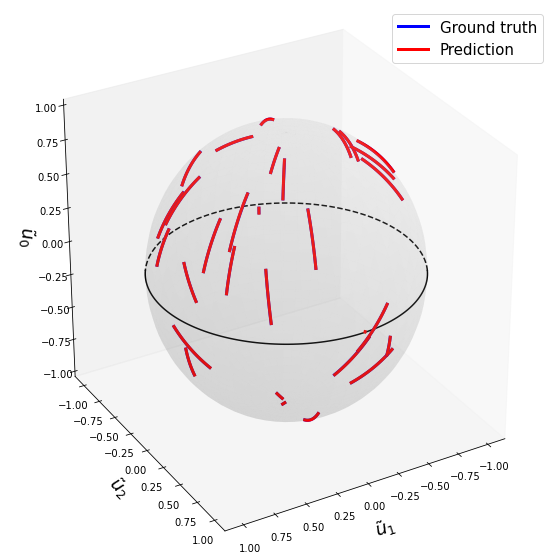}}
    \endminipage
    \hfill
    \minipage{0.5\textwidth}
    \centering
    \subfigure[~Loss as a function of flow time]{\includegraphics[width=.9\textwidth]{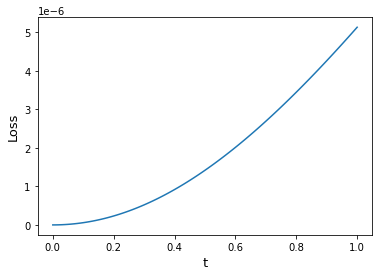}}
    \endminipage
    \caption{Test results. (a) Evolution of 30 samples projected on the 3D unit sphere. The true trajectories and the predicted ones are so close that the predictions cover the ground truth values. (b) Frobenius norm over the dataset as a function of flow time. Image from~\cite{Favoni:2022mcg}.}
    \label{fig:test_res}
\end{figure}

\begin{figure}[t]
    \minipage{0.5\textwidth}
    \centering
    \subfigure[~True versus predicted trajectories]{\includegraphics[width=\textwidth]{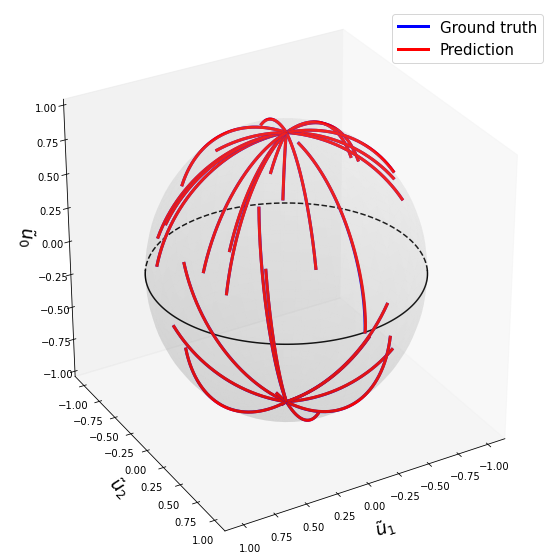}}
    \endminipage
    \hfill
    \minipage{0.5\textwidth}
    \centering
    \subfigure[~Loss as a function of flow time]{
    \includegraphics[width=.9\textwidth]{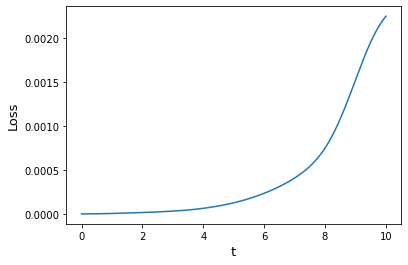}}
    \endminipage
    \caption{Test results for final flow time $\tau=10$ based on training up to $\tau=1$. (a) Extrapolated evolution of 30 samples projected on the 3D unit sphere. The true and predicted trajectories lie too close to each other to be able to tell them apart. (b) Frobenius norm as a function of flow time. The loss increases significantly at larger times, by up to three orders of magnitude compared to the higher values reached in the original interval $\tau \in[0, 1]$. Image from~\cite{Favoni:2022mcg}.}
    \label{fig:extrap_res}
\end{figure}

We provide now the details of our experiment. We choose the steps in gradient flow time to be sufficiently small for the discretized path~\eqref{eq:true_flow} with the continuous flow~\eqref{eq:toy_model_flow}, namely $\Delta \tau=0.01$, and the final gradient flow time $T=1$. The training set is made up by $50\,000$ samples and the batch size is 100. We train for 100 epochs using a learning rate of $10^{-3}$. The best dense network found consists of four hidden layers with 16, 64, 32 and 16 nodes respectively, each with a learnable bias term and followed by a $\tanh(x)$ activation function, with the exception of the last linear layer. Once the learning process ends, a test on $4\,000$ samples is run within the same window of gradient flow time, e.g.~$\tau \in [0,1]$. Fig.~\ref{fig:test_res} displays the outcome of this test. In the left plot, we show the projections on the three-dimensional unit sphere of the true trajectories obtained with Eq.~\eqref{eq:true_flow}, marked with blue, and of the predicted trajectories given by the NODE, in red. The two types of trajectories are visually indistinguishable, and indeed the MSE values reported as a function of flow time on the right plot confirm how accurate the predictions are. On the other hand, the increase of the Frobenius norm raises the question of how the network performs if the final flow time is chosen outside the training interval. To answer this, an additional test is run on the same $4\,000$ samples flowed up to $\tau=10$ without retraining the model. The results are shown in Fig.~\ref{fig:extrap_res}.

\begin{figure}
    \minipage{0.5\textwidth}
    \centering
    \subfigure[~True versus predicted trajectory]{
    \includegraphics[width=\textwidth]{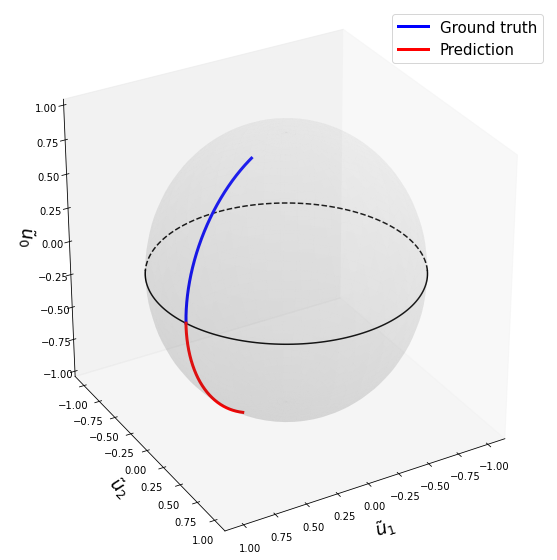}}
    \endminipage
    \hfill
    \minipage{0.5\textwidth}
    \centering
    \subfigure[Loss as a function of flow time]{
    \includegraphics[width=.9\textwidth]{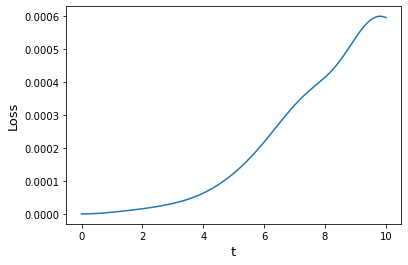}}
    \endminipage
    \caption{Mispredictions and their effect on the results. (a) We show that for one sample, the predicted trajectory goes in the opposite direction of the true trajectory, thus leading to an increase in the loss. (b) The Frobenius norm calculated on all other samples in the test set increases similarly to Fig.~\ref{fig:extrap_res}(b), but its value at the final flow time is smaller by about an order of magnitude.
    }
    \label{fig:mispred}
\end{figure}

We observe a clear deterioration in the model's performance with a drop of roughly three orders of magnitude for the loss compared to the test run in the range $\tau \in [0,1]$. A closer inspection of the test results leads us to the identification of two sources of the loss increase. One can be attributed to a specific sample, for which the predicted direction is wrong, as reported in Fig.~\ref{fig:mispred}. Provided that for $\tau \rightarrow \infty$ all matrices have flowed to one of the minima, the contribution of such mispredictions to the Frobenius norm $\mathcal{L}$ is given by
\begin{align}
    \lim_{\tau \rightarrow \infty} \Delta \mathcal{L} = \frac{N_{\text{mispred}}}{N_{\text{samples}}} \, \|\pm \one - (\mp \one)\|^2 = 8 \, \frac{N_{\text{mispred}}}{N_{\text{samples}}}.
\end{align}
In our case, $\Delta \mathcal{L}$ would be $2 \cdot 10^{-3}$, which is approximately the gain at $\tau=10$ if we discard the problematic sample. The other significant contribution to the loss comes from samples for which the predictions overshoot the ground truth. The interpretation we give is that the learnt force is too high, which makes these samples flow faster than required towards a minimum. Both effects are associated with samples whose initial condition lies within a thin neighborhood of the equator, i.e.~$\mathrm{Tr} \, U \approx 0$. For these samples, the force at the initial flow times can be very small, which is why the network struggles in determining the correct dynamics. Despite these flaws, the results are encouraging, considering that we employed a simple dense network which does not incorporate, for example, the symmetry of the dynamics with respect to the plane $\Tr \, U = 0$. More generally, the results can be improved with the use of a network that preserves the aforementioned symmetry of the system $U \rightarrow \Omega U \Omega^\dg$ with $\Omega\in \SU(2)$.

\section{Adjoint sensitivity method}

A technical aspect of NODEs that still needs to be addressed is backpropagation. In machine learning contexts, it refers to the computation of the gradient of the loss function with respect to the model parameters. Backpropagation is a highly efficient implementation of the chain rule, but for NODEs it would require to store in memory the whole evolution of the system. To see this explicitly, we can consider the NODE for $x \in \mathbb{R}$
\begin{align}
    \frac{\mathrm{d}x}{\mathrm{d}t} = f(x(t), \theta)
    \label{eq:asm_node}
\end{align}
with initial condition $x_0 = x(t_0)$. Formally, it is possible to solve it by
\begin{align}
    x(T) = x_0 + \int_{t_0}^{T} \, f(x(t), \theta) \, \mathrm{d}t.
\end{align}
We can choose the loss function to be the MSE between the target final state $\tilde{x}_T$ and the predicted final state $x(T)$
\begin{align}
    \mathcal{L}(x(T)) = (\tilde{x}_T - x(T))^2.
\end{align}
The gradients are given by
\begin{align}
    \frac{\mathrm{d}\mathcal{L}}{\mathrm{d}\theta} = \frac{\p \mathcal{L}}{\p x(T)} \frac{\mathrm{d} x(T)}{\mathrm{d}\theta},
\end{align}
but the term
\begin{align}
    \frac{\mathrm{d} x(t)}{\mathrm{d}\theta} = \int_{t_0}^T \mathrm{d}t' \left(\frac{\p f(x(t'), \theta)}{\p \theta} + \frac{\p f(x(t'), \theta)}{\p x(t')} \frac{\mathrm{d} x(t')}{\mathrm{d}\theta} \right).
\end{align}
requires to store the whole evolution of the system in memory, which can quickly be saturated if the network's depth is increased. We have used the fact that $x(t) = x(t, \theta)$ for $t_0 < t < T$, therefore its derivative is
\begin{align}
    \frac{\mathrm{d} f}{\mathrm{d} \theta}=\frac{\p f}{\p \theta}+\frac{\p f}{\p x}\frac{\mathrm{d} x}{\mathrm{d} \theta}.
    \label{eq:f_grad}
\end{align}
In programs which incur memory saturation issues, it is common practice to trade off memory with computation. Here, the adjoint sensitivity method~\cite{Chen:2018} serves the scope perfectly, as it computes the evolution backwards rather than saving it in memory. Historically, this method has encountered various applications ranging from meteorology \cite{Errico:1997} to geophysics \cite{Plessix:2006}.

We can formulate the adjoint sensitivity method as a constrained optimization problem using Lagrange multipliers \cite{Cao:2002}. We can rewrite Eq.~\eqref{eq:asm_node} in the form of a constraint:
\begin{equation}
    \dot{x} - f(x(t), \theta) = F(\dot{x}(t), x(t), \theta) = 0.
    \label{constraint}
\end{equation}
We introduced the notation $\dot{x}$ as an equivalent form of $\mathrm{d}x / \mathrm{d}t$. Our goal is to find $\argmin_{\theta}\mathcal{L}(x(T))$ such that the constraint~\eqref{constraint} remains satisfied. This can be fulfilled introducing the Lagrange multiplier $\lambda(t)$ and the modified loss function
\begin{equation}
    \psi = \mathcal{L}(x(T)) - \int_{t_0}^{T}\lambda(t)F(\dot{x}(t), x(t), \theta) \mathrm{d}t.
    \label{eq:psi}
\end{equation}
Since $F = 0$, the gradients of the original loss function coincide with the gradients of the modified loss function:
\begin{equation}
     \frac{\mathrm{d} \psi}{\mathrm{d} \theta} = \frac{\mathrm{d} \mathcal{L}(x(T))}{\mathrm{d}\theta}.
\end{equation}
Therefore, we can compute the gradients $\mathrm{d} \mathcal{L}/ \mathrm{d}\theta$ by taking the derivative of Eq.~\eqref{eq:psi}, obtaining
\begin{equation}
    \frac{\mathrm{d} \psi}{\mathrm{d} \theta} = \frac{\p \mathcal{L}}{\mathrm{d} x(T)} \frac{\mathrm{d} x(T)}{\mathrm{d} \theta} - \frac{\mathrm{d}}{\mathrm{d}\theta} \int_{t_0}^{T} \lambda F \mathrm{d}t.
\end{equation}
We focus on the second term and rework it as follows:
\begin{align}
    \frac{\mathrm{d}}{\mathrm{d}\theta}\int_{t_0}^T \lambda F \, \mathrm{d}t &= \frac{\mathrm{d}}{\mathrm{d}\theta} \int_{t_0}^T \lambda (\dot{x}-f) \, \mathrm{d}t \nonumber \\
    &= \frac{\mathrm{d}}{\mathrm{d}\theta} \left[\lambda(T) x(T)-\lambda(t_0) x(t_0) - \int_{t_0}^T (\dot{\lambda} x + \lambda f) \, \mathrm{d}t \right] \nonumber \\
    &= \lambda(T) \frac{\mathrm{d}x(T)}{\mathrm{d}\theta} - \int_{t_0}^T \left(\dot{\lambda} + \lambda \frac{\p f}{\p x}\right) \frac{\mathrm{d}x}{\mathrm{d}\theta} \, \mathrm{d}t - \int_{t_0}^T \lambda \frac{\p f}{\p \theta} \mathrm{d}t,
    \label{grad}
\end{align}
where we integrated by parts in the second equality and used Eq.~\eqref{eq:f_grad} to get to the third equality. We stress that $\lambda$ does not depend on the network's parameters $\theta$ and that
\begin{align}
    \frac{\mathrm{d} x(t_0)}{\mathrm{d}\theta} = \frac{\mathrm{d} x_0}{\mathrm{d}\theta} = 0.
\end{align}
The result for the gradients reads
\begin{align}
    \frac{\mathrm{d} \psi}{\mathrm{d} \theta} = &\left(\frac{\p \mathcal{L}}{\mathrm{d} x(T)} - \lambda(T)\right) \frac{\mathrm{d}x(T)}{\mathrm{d}\theta} + \int_{t_0}^T \left(\dot{\lambda} + \lambda \frac{\p f}{\p x}\right) \frac{\mathrm{d}x}{\mathrm{d}\theta} \, \mathrm{d}t + \int_{t_0}^T \lambda \frac{\p f}{\p \theta} \mathrm{d}t.
    \label{eq:asm_grad_psi_final}
\end{align}
In order to get rid of the dependence on $\mathrm{d}x / \mathrm{d}\theta$, we impose the following equations
\begin{empheq}[left=\empheqlbrace]{align}
    &\lambda(T) = 
    \displaystyle \frac{\p\mathcal{L}}{\p x(T)}, 
    \label{eq:adj_bv} \\[5pt]
    &\dot{\lambda}(t) = -\lambda(t) 
    \displaystyle \frac{\p f(x(t), \theta)}{\p x(t)}, 
    \label{eq:adj_ode}
\end{empheq}
which define respectively the initial condition and the equation of motion for the adjoint $\lambda$. The remaining term in Eq.~\eqref{eq:asm_grad_psi_final} yields the gradients we were looking for, i.e.
\begin{equation}
    \frac{\mathrm{d}\mathcal{L}(x(T))}{\mathrm{d}\theta} = -\int_{T}^{t_0} \lambda(t) \frac{\p f(x(t), \theta)}{\p \theta}\mathrm{d}t,
    \label{eq:asm_final_grad}
\end{equation}
without any dependence on $\mathrm{d}x / \mathrm{d}\theta$. We point out that in case the function $f$ also depends explicitly on the flow time $t$, the derivation and the results would be identical.

In light of the above results, the implementation of the adjoint sensitivity method consists of the following steps. First, the NODE~\eqref{eq:asm_node} is solved with the forward pass by using an ODE integrator such as the one given in Eq.~\eqref{eq:discrete_flow}. At $t=T$, we evaluate the initial condition of the adjoint in Eq.~\eqref{eq:adj_bv}, $\p \mathcal{L} / \p x(T)$. With a standard ODE solver such as Runge-Kutta, we integrate Eqs.~\eqref{eq:asm_node} and~\eqref{eq:adj_ode} backwards from $T$ to $t_0$. In this way, we obtain the terms $\lambda(t)$ and $\p f(x(t), \theta) / \p \theta$ that contribute to the gradients in Eq.~\eqref{eq:asm_final_grad} without the need of using backpropagation and storing in memory the whole evolution of the system. 

\subsection{Calculation for U(1)}

In principle, the adjoint sensitivity method just presented can be directly applied also in the context of groups, but the backward evolution of the system would not fulfill the group constraints in general, therefore we extend the derivation of the adjoint sensitivity method such that the constraints are respected by construction. We do it first for the U(1) group in order to develop some intuition for the calculation for $\SU(N_c)$. We can view a U(1) element as a complex variable $z$ living on the unit circle in the complex plane, which in polar coordinates can be written as
\begin{align}
    z = \mathrm{e}^{i\varphi}, \qquad  \varphi \in [0, 2\pi).
\end{align}
We define the complex conjugate of $z$:
\begin{align}
    \bar{z} = \mathrm{e}^{-i\varphi},
\end{align}
which acts as the inverse group element of $z$:
\begin{align}
    z \bar{z} = 1.
    \label{eq:U1_circle}
\end{align}
We aim to solve the following neural gradient flow equation:
\begin{equation}
    \displaystyle\frac{\mathrm{d}z}{\mathrm{d}t} = ig(z(t),  \bar{z}(t), \theta)\,z(t),
    \label{eq:ngf_U1}
\end{equation}
where $g \in \mathbb{R} \implies g = \bar{g}$, with initial condition $z_0=z(t_0)$ and target final state $\tilde{z}_T$. This NODE is very similar to the one used in the real case, Eq.~\eqref{eq:asm_node}, except we have provided a more specific expression to the right hand side. This choice is justified because it guarantees that the constraint~\eqref{eq:U1_circle} is satisfied during the flow:
\begin{align}
    \frac{\mathrm{d}}{\mathrm{d} t} (z \bar{z}) = \dot{z} \bar{z} + z \dot{\bar{z}} = (igz) \bar{z} - z(i\bar{g}\bar{z}) = i (g - \bar{g}) = 0,
\end{align}
where the complex conjugate of Eq.~\eqref{eq:ngf_U1} is
\begin{equation}
    \displaystyle\frac{\mathrm{d}\bar{z}}{\mathrm{d}t} = -i\bar{g}\bar{z}.
\end{equation}
The function $g$ can be modeled by a neural network, whose input is the variable $z$. In complex analysis, though, $g$ has to be considered as a function of $\bar{z}$ as well, which will be crucial when taking derivatives. The loss function is chosen to be the real-valued MSE
\begin{align}
    \mathcal{L}(z(T), \bar{z}(T)) = |z(T) - \tilde{z}_T|^2.
\end{align}
The gradients contain two terms due to the dependence on $z$ and $\bar{z}$:
\begin{align}
    \frac{\mathrm{d}\mathcal{L}}{\mathrm{d}\theta} = \frac{\p \mathcal{L}}{\p z(T)} \frac{\mathrm{d} z(T)}{\mathrm{d}\theta} + \frac{\p \mathcal{L}}{\p \bar{z}(T)} \frac{\mathrm{d} \bar{z}(T)}{\mathrm{d}\theta}.
\end{align}
The partial derivatives with respect to $z$ and $\bar{z}$ are called Wirtinger derivatives and are related to partial derivatives with respect to $u = \Re z$ and $v = \Im z$ as follows:
\begin{empheq}[left=\empheqlbrace]{align}
    &\frac{\p}{\p z} = \frac{1}{2} \left(\frac{\p}{\p u} -i \frac{\p}{\p v}\right), \label{eq:wirt_z} \\[10pt]
    &\frac{\p}{\p \bar{z}} = \frac{1}{2} \left(\frac{\p}{\p u} + i\frac{\p}{\p v}\right). \label{eq:wirt_zbar}
\end{empheq}
The NODE~\eqref{eq:ngf_U1} is reworked to obtain the constraint
\begin{equation}
    \dot{z} - igz = G = 0.
    \label{eq:U1_constr}
\end{equation}
As discussed in the real case, what we are trying to solve is a constrained minimization problem. The Lagrange multiplier technique is adapted to the complex case by including the complex conjugate in the following way:
\begin{equation}
    \psi = \mathcal{L}(z(T),\bar{z}(T)) - \int_{t_0}^{T} \left(\lambda G + \bar{\lambda} \bar{G}\right) \mathrm{d}t.
\end{equation}
Because of the constraint~\eqref{eq:U1_constr}, the gradients of $\psi$ coincide with the gradients of $\mathcal{L}$ and are given by
\begin{align}
    \frac{\mathrm{d}\mathcal{L}(z(T), \bar{z}(T))}{\mathrm{d}\theta} = \frac{\mathrm{d}\psi}{\mathrm{d}\theta}
    = &\frac{\p \mathcal{L}}{\p z(T)}\frac{\mathrm{d}z(T)}{\mathrm{d}\theta} + \frac{\p \mathcal{L}}{\p \bar{z}(T)}\frac{\mathrm{d}\bar{z}(T)}{\mathrm{d}\theta} \nonumber \\
    & -\frac{\mathrm{d}}{\mathrm{d}\theta}\int_{t_0}^{T} \left(\lambda G + \bar{\lambda}\bar{G}\right)\mathrm{d}t.
    \label{eq:U1_grads_lagr}
\end{align}
We notice the following useful property, which is a consequence of Eq.~\eqref{eq:U1_circle}:
\begin{equation}
    \frac{\mathrm{d}}{\mathrm{d}\theta}(z\bar{z}) = 0 \implies z\frac{\mathrm{d}\bar{z}}{\mathrm{d}\theta} = -\bar{z}\frac{\mathrm{d}z}{\mathrm{d}\theta}.
    \label{eq:der_relation}
\end{equation}
A key consideration is that $z(t) = z(t, \theta)$, such that the chain rule leads to a result similar to Eq.~\eqref{eq:f_grad}:
\begin{equation}
    \frac{\mathrm{d} g}{\mathrm{d} \theta}=\frac{\p g}{\p z}\frac{\mathrm{d} z}{\mathrm{d} \theta}+\frac{\p g}{\p \bar{z}}\frac{\mathrm{d} \bar{z}}{\mathrm{d} \theta}+\frac{\p g}{\p \theta}.
    \label{eq:g_grad}
\end{equation}
Elaborating the derivative of the integral in \eqref{eq:U1_grads_lagr},
we obtain

\begin{align}
    \displaystyle \frac{\mathrm{d}}{\mathrm{d}\theta} \int_{t_0}^{T} \lambda G \, \mathrm{d}t + \text{c.c.} &= \frac{\mathrm{d}}{\mathrm{d}\theta} \int_{t_0}^{T} \lambda(\dot{z} - igz) \, \mathrm{d}t + \text{c.c.} \nonumber \\
    &= \frac{\mathrm{d}}{\mathrm{d}\theta} \left[\lambda(T)z(T) - \lambda(t_0)z(t_0) - \int_{t_0}^{T} (\dot{\lambda}z + i\lambda gz)\mathrm{d}t\right] + \text{c.c.} \nonumber \\
    &= \lambda(T)\frac{\mathrm{d}z(T)}{\mathrm{d}\theta} - \frac{\mathrm{d}}{\mathrm{d}\theta} \int_{t_0}^{T} (\dot{\lambda}z + i\lambda gz)\,\mathrm{d}t + \text{c.c.},
    \label{eq:grad_calc}
\end{align}
where we performed an integration by parts and indicated with ``c.c." the complex conjugate of the entire expressions. The derivative of the integrand in the last line yields
\begin{align}
    \frac{\mathrm{d}}{\mathrm{d}\theta} (\dot{\lambda}z + i\lambda gz) &= \left[\dot{\lambda} + i\lambda \left(\frac{\p g}{\p z}z + g\right)\right] \frac{\mathrm{d}z}{\mathrm{d}\theta} + i \lambda \left(\frac{\p g}{\p \bar{z}} z \frac{\mathrm{d}\bar{z}}{\mathrm{d}\theta} + \frac{\p g}{\p \theta} z\right) \nonumber \\
    &= \left[\dot{\lambda} + i\lambda \left(\frac{\p g}{\p z}z - \frac{\p g}{\p \bar{z}} \bar{z} + g\right)\right] \frac{\mathrm{d}z}{\mathrm{d}\theta} + i \lambda \frac{\p g}{\p \theta} z,
\end{align}
where we made use of Eq.~\eqref{eq:g_grad} and exploited the property in Eq.~\eqref{eq:der_relation} to transform the term with $\mathrm{d}z / \mathrm{d} \theta$ into a term with $\mathrm{d}\bar{z} / \mathrm{d} \theta$. This result can be substituted in Eq.~\eqref{eq:grad_calc}, obtaining
\begin{multline}
    \displaystyle \frac{\mathrm{d}}{\mathrm{d}\theta} \int_{t_0}^{T} \lambda G \, \mathrm{d}t + \text{c.c.} = \lambda(T)\frac{\mathrm{d}z(T)}{\mathrm{d}\theta} \\
    - \int_{t_0}^{T} \left[\dot{\lambda} + i\lambda\left(\frac{\p g}{\p z}z - \frac{\p g}{\p \bar{z}}\bar{z} + g\right)\right]\frac{\mathrm{d}z}{\mathrm{d}\theta}\,\mathrm{d}t - i\int_{t_0}^{T} \lambda \frac{\p g}{\p \theta} z \, \mathrm{d}t + \text{c.c.}
\end{multline}
The gradients in Eq.~\eqref{eq:U1_grads_lagr} become
\begin{align}
    \frac{\mathrm{d} \psi}{\mathrm{d}\theta} =  &\left(\frac{\p \mathcal{L}}{\mathrm{d} z(T)} - \lambda(T)\right) \frac{\mathrm{d}z(T)}{\mathrm{d}\theta} \nonumber \\
    &+ \int_{t_0}^{T} \left[\dot{\lambda} + i\lambda\left(\frac{\p g}{\p z}z - \frac{\p g}{\p \bar{z}}\bar{z} + g\right)\right]\frac{\mathrm{d}z}{\mathrm{d}\theta}\,\mathrm{d}t \nonumber \\
    &+ i\int_{t_0}^{T} \lambda \frac{\p g}{\p \theta} z \, \mathrm{d}t + \text{c.c.}
    \label{eq:U1_grad_psi_final}
\end{align}

Similarly to the real case, we get rid of the dependence on $\mathrm{d}z / \mathrm{d}\theta$ by imposing
\begin{empheq}[left=\empheqlbrace]{align}
    &\lambda(T) = 
    \displaystyle \frac{\p\mathcal{L}}{\p z(T)}, 
    \label{eq:U1_adj_bv} \\[5pt]
    &\dot{\lambda} + i\lambda g = -i\lambda\left(\frac{\p g}{\p z} z - \frac{\p g}{\p \bar{z}} \bar{z}\right) = 2\lambda \, \Im\left(\frac{\p g}{\p z} z\right). 
    \label{eq:U1_adj_ode}
\end{empheq}
The first equation sets the initial condition for the adjoint, while the second is the ODE for the adjoint. With these choices, the corresponding complex conjugate also vanish, hence no term involving $\mathrm{d} \bar{z} / \mathrm{d}\theta$ survives. The remaining term in the last line of Eq.~\eqref{eq:U1_grad_psi_final} and its complex conjugate corresponds to the gradients:
\begin{equation}
    \frac{\mathrm{d}\mathcal{L}(z(T), \bar{z}(T))}{\mathrm{d}\theta} = -i\int_{T}^{t_0}\frac{\p g}{\p \theta}(\lambda z-\bar{\lambda}\bar{z})\,\mathrm{d}t=2\int_{T}^{t_0}\frac{\p g}{\p \theta}\,\text{Im}(\lambda z)\,\mathrm{d}t.
    \label{eq:U1_final_grad}
\end{equation}
The gradients are manifestly real, as they are supposed to be. It is also possible to write a more compact solution by introducing the quantity $w = \lambda z$. Its time derivative is given by
\begin{align}
    \dot{w} = \dot{\lambda} z + \lambda \dot{z} = \dot{\lambda} z + \lambda (igz) = z (\dot{\lambda} + i \lambda g).
\end{align}
Substituting in Eq.~\eqref{eq:U1_adj_bv},~\eqref{eq:U1_adj_ode} and~\eqref{eq:U1_final_grad}, the initial condition and the equation of motion for this modified adjoint are
\begin{empheq}[left=\empheqlbrace]{align}
    &w(T) = \frac{\p\mathcal{L}}{\p z(T)}z(T),  \\[5pt]
    &\dot{w} = 2w\,\Im \left(\frac{\p g}{\p z}z\right),
\end{empheq}
and the gradients become
\begin{equation}
    \frac{\mathrm{d}\mathcal{L}(z(T), \bar{z}(T))}{\mathrm{d}\theta} = 2\int_{T}^{t_0} \frac{\p g}{\p \theta}\,\text{Im}\,w\,\mathrm{d}t.
\end{equation}

The implementation of the adjoint sensitivity method for $\text{U}(1)$ is almost identical to the case with real variables, except for the ODE solver. In order to stay in the group while integrating backwards, we use a solver that employs the exponential in a way similar to Eq.~\eqref{eq:discrete_flow}
\begin{align}
   z(t_{j})=\exp\left(-ig(z(t_{j+1}), \theta) \Delta t\right) z(t_{j+1}).
\end{align}
The exponent is purely imaginary, such that the exponential acts as a rotation of $z(t_j)$.

\begin{figure}
\minipage{0.5\textwidth}
\includegraphics[width=\textwidth]{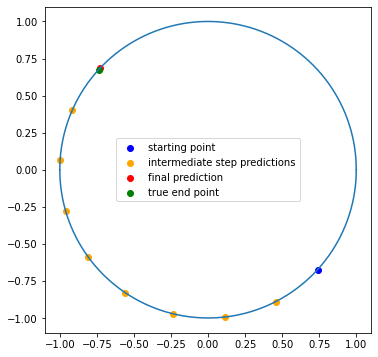}
\endminipage
\hfill
\minipage{0.5\textwidth}
\centering
\includegraphics[width=\textwidth]{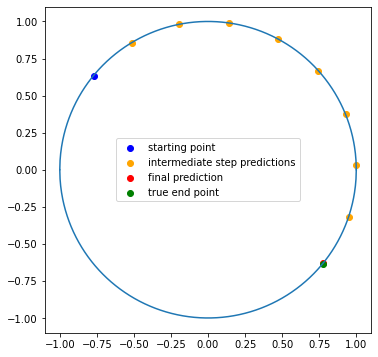}
\endminipage
\caption{Solution of an NODE of a $\text{U}(1)$ element with adjoint sensitivity method. We display the successful reconstruction of the time evolution of a complex variable constrained on the unit circle using the adjoint sensitivity method for two samples of our dataset. Since the starting point (blue) and the end point (green) are separated by a constant phase of $\pi$, the solution of the NODE is a complex variable rotating around the origin at fixed angular momentum. The intermediate flow times are equally spaced, which is why the predictions at such flow times (orange) are separated by constant rotations. The final prediction (red) lies close to the ground truth.}
\label{fig:U1_asm}
\end{figure}

A test of this derivation is run on a simple situation in which each input $z_0^i$ is separated by a constant phase $\Delta \varphi = \pi$ from the corresponding output $\tilde{z}_T^i$. In Fig.~\ref{fig:U1_asm}, we show the solution provided by a simple network. Specifically, the function $g$ is modeled with a dense network with two hidden layers each featuring 20 nodes and followed by a $\tanh$ activation function. The solution of the NODE is a constant motion, such that $z$ rotates at constant angular momentum around the origin. Given that the flow time steps $\Delta t$ are equal, the predictions at intermediate flow times have to be separated each by an approximately constant phase, as is the case in Fig.~\ref{fig:U1_asm}. We notice that the network is free to choose a clockwise or anticlockwise rotation for this problem. This ambiguity can be eliminated by providing intermediate values of the evolution and adding the MSE of the corresponding predictions to the loss function.

\subsection{Calculation for $\SU(N_c)$}

Let us now derive the adjoint sensitivity method in $\SU(N_c)$. The NODE we are dealing with is the neural gradient flow~\eqref{eq:ngf}, where the indices $x$ and $\mu$ will be dropped for simplicity:
\begin{align}
    \frac{\mathrm{d}U}{\mathrm{d}\tau} = iH(U(\tau), \bar{U}(\tau), \theta) \, U(\tau).
    \label{eq:asm_ngf}
\end{align}
Notice how we define $H$ as explicitly dependent on $\bar{U}$. This is because we will treat the elements of each matrix as independent complex numbers, in a way that resembles the $\text{U}(1)$ case. The initial condition of the dynamics are $U_0 = U(\tau=0)$, and the target final state $\tilde{U}_T$. We choose as loss function the Frobenius norm
\begin{align}
    \mathcal{L}(U(T), \bar{U}(T)) = \|U(T) - \tilde{U}_T\|^2,
\end{align}
defined by Eq.~\eqref{eq:frobenius}. The gradients depend on $U$ and $\bar{U}$:
\begin{align}
    \frac{\mathrm{d}\mathcal{L}}{\mathrm{d}\theta} = \frac{\p \mathcal{L}}{\p U_{ij}(T)} \frac{\mathrm{d} U_{ij}(T)}{\mathrm{d}\theta} + \frac{\p \mathcal{L}}{\p \bar{U}_{ij}(T)} \frac{\mathrm{d} \bar{U}_{ij}(T)}{\mathrm{d}\theta}.
\end{align}
The partial derivatives can be defined analogously to the Wirtinger derivatives in Eqs.~\eqref{eq:wirt_z} and~\eqref{eq:wirt_zbar}. We proceed as in the two cases previously examined, reworking the NODE in the form of the constraint
\begin{equation}
    \dot{U} - H(U(t), \theta) = \mathcal{H}(\dot{U}(t), U(t), \theta) = 0.
    \label{eq:sun_constr}
\end{equation}
In general, $\mathcal{H} \in \mathbb{C}^{N_c \times N_c}$. We introduce the Lagrange multiplier $\Lambda \in \mathbb{C}^{N_c\times N_c}$ and the modified loss function
\begin{align}
    \Psi &= \mathcal{L}(U(T), \bar{U}(T)) - \int_{t_0}^{T} \Tr \left(\Lambda \mathcal{H} + \overline{\Lambda \mathcal{H}}\right) \mathrm{d}\tau \nonumber \\
    &= \mathcal{L}(U(T), \bar{U}(T)) - \int_{t_0}^{T} \left(\Lambda_{ij} \mathcal{H}_{ji} + \bar{\Lambda}_{ij} \bar{\mathcal{H}}_{ji}\right) \mathrm{d}\tau.
    \label{eq:Psi}
\end{align}
The trace operation guarantees that the result of the integral is a scalar. The constraint~\eqref{eq:sun_constr} gives the following equalities for the gradients:
\begin{align}
    \frac{\mathrm{d}\mathcal{L}(U(T), \bar{U}(T))}{\mathrm{d}\theta} = \frac{\mathrm{d}\Psi}{\mathrm{d}\theta}
    = &\frac{\p \mathcal{L}}{\p U(T)}\frac{\mathrm{d}U(T)}{\mathrm{d}\theta} + \frac{\p \mathcal{L}}{\p \bar{U}(T)}\frac{\mathrm{d}\bar{U}(T)}{\mathrm{d}\theta} \nonumber \\
    & -\frac{\mathrm{d}}{\mathrm{d}\theta}\int_{t_0}^{T} \left(\Lambda \mathcal{H} + \bar{\Lambda}\bar{\mathcal{H}}\right)\mathrm{d}\tau.
    \label{eq:sun_grads_lagr}
\end{align}
Calculating the derivative of the integral in the last line of the above equation, we obtain
\begin{align}
    &\frac{\mathrm{d}}{\mathrm{d}\theta} \int_{t_0}^{T} \Tr \left(\Lambda \mathcal{H}\right) \mathrm{d}t = \frac{\mathrm{d}}{\mathrm{d}\theta} \int_{t_0}^{T} \Tr \left(\Lambda(\dot{U} - iHU)\right) \mathrm{d}\tau \nonumber \\
    &= \frac{\mathrm{d}}{\mathrm{d}\theta} \left[\Tr \left(\Lambda(T)U(T)\right) - \Tr \left(\Lambda(t_0)U(t_0)\right) - \int_{t_0}^{T} \Tr \left(\dot{\Lambda}U + i\Lambda HU\right)\mathrm{d}\tau\right] \nonumber \\
    &= \Lambda_{ji}(T)\frac{\mathrm{d}U_{ij}(T)}{\mathrm{d}\theta} - \frac{\mathrm{d}}{\mathrm{d}\theta} \int_{t_0}^{T} \Tr\left(\dot{\Lambda}U + i\Lambda HU\right)\mathrm{d}\tau,
    \label{eq:sun_grad_calc}
\end{align}
and a similar result for the complex conjugate. Noticing that a formula similar to Eq.~\eqref{eq:g_grad} holds also for $H$, we can write
\begin{align}
    &\frac{\mathrm{d}}{\mathrm{d}\theta} (\dot{\Lambda}_{ij}U_{ji} + i\Lambda_{ik} H_{kj}U_{ji} + \dot{\bar{\Lambda}}_{ij}\bar{U}_{ji} - i\bar{\Lambda}_{ik} \bar{H}_{kj}\bar{U}_{ji}) \nonumber \\
    &= \dot{\Lambda}_{ij} \frac{\mathrm{d}U_{ji}}{\mathrm{d}\theta} + i \Lambda_{ik} \left(\frac{\p H_{kj}}{\p U_{lm}} \frac{\mathrm{d}U_{lm}}{\mathrm{d}\theta} U_{ji} + \frac{\p H_{kj}}{\p \bar{U}_{lm}} \frac{\mathrm{d}\bar{U}_{lm}}{\mathrm{d}\theta} U_{ji} + \frac{\p H_{kj}}{\p \theta} U_{ji} + H_{kj} \frac{\mathrm{d}U_{ji}}{\mathrm{d}\theta}\right) \nonumber \\
    &+ \dot{\bar{\Lambda}}_{ij} \frac{\mathrm{d}\bar{U}_{ji}}{\mathrm{d}\theta} - i \bar{\Lambda}_{ik} \left(\frac{\p \bar{H}_{kj}}{\p U_{lm}} \frac{\mathrm{d}U_{lm}}{\mathrm{d}\theta} \bar{U}_{ji} + \frac{\p \bar{H}_{kj}}{\p \bar{U}_{lm}} \frac{\mathrm{d}\bar{U}_{lm}}{\mathrm{d}\theta} \bar{U}_{ji} + \frac{\p \bar{H}_{kj}}{\p \theta} \bar{U}_{ji} + \bar{H}_{kj} \frac{\mathrm{d}\bar{U}_{ji}}{\mathrm{d}\theta}\right) \nonumber \\
    &= \left(\dot{\Lambda}_{ij} + i \Lambda_{ml} \frac{\p H_{lk}}{\p U_{ji}} U_{km} - i \bar{\Lambda}_{ml} \frac{\p \bar{H}_{lk}}{\p U_{ji}} \bar{U}_{km} + i \Lambda_{ik} H_{kj} \right) \frac{\mathrm{d}U_{ji}}{\mathrm{d}\theta} + i \Lambda_{ik} \frac{\p H_{kj}}{\p \theta} U_{ji} \nonumber \\
    &+ \left(\dot{\bar{\Lambda}}_{ij} - i \bar{\Lambda}_{ml} \frac{\p \bar{H}_{lk}}{\p \bar{U}_{ji}} \bar{U}_{km} + i \Lambda_{ml} \frac{\p H_{lk}}{\p \bar{U}_{ji}} U_{km} - i \bar{\Lambda}_{ik} \bar{H}_{kj} \right) \frac{\mathrm{d}\bar{U}_{ji}}{\mathrm{d}\theta} - i \bar{\Lambda}_{ik} \frac{\p \bar{H}_{kj}}{\p \theta} \bar{U}_{ji}.
\end{align}
In the third equality, we have repositioned some terms and changed indices in order to be able to factor out the derivatives of $U$ and $\bar{U}$ with respect to the parameters $\theta$. We can substitute this result in Eq.~\eqref{eq:sun_grad_calc} and its complex conjugate, obtaining
\begin{align}
    \frac{\mathrm{d}}{\mathrm{d}\theta} \int_{t_0}^{T} &\Tr \left(\Lambda \mathcal{H}\right) \mathrm{d}\tau + \text{c.c.} = \Lambda_{ji}(T)\frac{\mathrm{d}U_{ij}(T)}{\mathrm{d}\theta} \nonumber \\
    &- \int_{t_0}^{T} \left[\dot{\Lambda}_{ij} + i \Lambda_{ik} H_{kj} - 2 \, \Im \left(\frac{\p H_{lk}}{\p U_{ji}} U_{km} \Lambda_{ml}\right) \right] \frac{\mathrm{d}U_{ji}}{\mathrm{d}\theta} \, \mathrm{d}\tau \nonumber \\
    &- i \int_{t_0}^{T} \Tr\left(\Lambda \frac{\p H}{\p \theta} U\right) \, \mathrm{d}\tau + \text{c.c.}
\end{align}
Finally, we can rewrite the gradients in Eq.~\eqref{eq:sun_grads_lagr} as
\begin{align}
    \frac{\mathrm{d} \Psi}{\mathrm{d}\theta} =  &\left(\frac{\p \mathcal{L}}{\mathrm{d} U_{ij}(T)} - \Lambda_{ji}(T)\right) \frac{\mathrm{d}U_{ij}(T)}{\mathrm{d}\theta} \nonumber \\
    &+ \int_{t_0}^{T} \left[\dot{\Lambda}_{ij} + i \Lambda_{ik} H_{kj} - 2 \, \Im \left(\frac{\p H_{lk}}{\p U_{ji}} U_{km} \Lambda_{ml} \right) \right] \frac{\mathrm{d}U_{ji}}{\mathrm{d}\theta} \, \mathrm{d}\tau \nonumber \\
    &+ i \int_{t_0}^{T} \Tr\left(\Lambda \frac{\p H}{\p \theta} U\right) \, \mathrm{d}\tau + \text{c.c.}
    \label{eq:sun_grad_psi_final}
\end{align}
We remove the dependence on $\mathrm{d}U / \mathrm{d}\theta$ and its complex conjugate imposing the following equations for the adjoint:
\begin{empheq}[left=\empheqlbrace]{align}
    &\Lambda_{ij}(T) = 
    \displaystyle \frac{\p\mathcal{L}}{\p U_{ji}(T)}, 
    \label{eq:sun_adj_bv} \\[5pt]
    &\dot{\Lambda}_{ij} + i \Lambda_{ik} H_{kj} = 2 \, \Im \left(\frac{\p H_{lk}}{\p U_{ji}} U_{km} \Lambda_{ml} \right),
    \label{eq:sun_adj_ode}
\end{empheq}
which implies that the contribution of the complex conjugates vanishes. With the terms left, we can compute the gradients:
\begin{equation}
    \frac{\mathrm{d}\mathcal{L}(U(T), \bar{U}(T))}{\mathrm{d}\theta} = 2 \int_{T}^{t_0} \Im \, \Tr \left( \frac{\p H}{\p \theta} U \Lambda \right) \mathrm{d}\tau.
\end{equation}
In analogy to U(1), we can introduce the quantity $W = U \Lambda$. The initial condition and the equation of motion of such a combination is
\begin{empheq}[left=\empheqlbrace]{align}
    &W_{ij}(T) = 
    \displaystyle U_{ik}(T)\frac{\p\mathcal{L}}{\p U_{jk}(T)}, \\[5pt]
    &\dot{W}_{ij} + i [W, H]_{ij} = 2 U_{ik} \Im \left(\frac{\p H_{lm}}{\p U_{jk}} W_{ml} \right),
\end{empheq}
and the gradients are easily expressed as
\begin{equation}
    \frac{\mathrm{d}\mathcal{L}(U(T), \bar{U}(T))}{\mathrm{d}\theta} = 2 \int_{T}^{t_0} \Im \, \Tr \left( \frac{\p H}{\p \theta} W \right) \mathrm{d}\tau.
\end{equation}
The implementation of the above equations is structured similarly to the real case, the main difference being the solver employed for the backward integration: in $\SU(N_c)$, an appropriate choice is represented by Eq.~\eqref{eq:discrete_flow}, adapted to evolve backwards in time. We have checked that this method provides approximately the right gradients for the $\SU(2)$ toy model by comparing the gradients obtained via standard backpropagation with the ones given by the adjoint sensitivity method, with the highest relative errors of the order of $10^{-3}$. Currently, the backward dynamics is integrated with a standard Runge-Kutta algorithm, but more precise results can be obtained with a Runge-Kutta method suited for Lie groups~\cite{Munthe-Kaas:1998}. We finally report that the adjoint sensitivity method in $\SU(N_c)$ has also been solved with a different approach in~\cite{Bacchio:2022vje}.

\chapter{Conclusions} \label{chap:conclusions}

In this dissertation, we have shown the benefits of imposing global and local symmetries on neural network architectures in applications to lattice field theory. In particular, we have initially dealt with the application of neural networks to a physical system characterized by translational symmetry. The discussion has been extended to gauge symmetries, which led to the formulation of neural networks that preserve such symmetries on the lattice. These networks have then been shown to be applicable to the generation of field configurations. In the following, we will recapitulate each of these points in more detail.

First, we have compared the performance of three different CNN architectures on a 1+1D translationally symmetric complex scalar field theory with quartic interaction and a non-zero chemical potential. One architecture type was translationally equivariant (\EQ), meaning that it respected the symmetry by construction, using convolutions and spatial pooling layer with a stride of one and a GAP after the convolutional part. Another type (\ST) was characterized by symmetry-breaking layers, specifically spatial pooling layers with a stride larger than one. The last type (\FLAT), in addition to the same non-equivariant layers of the \ST{} type, featured a flattening step instead of a GAP, which completely destroys the symmetry. The first two architecture types could be applied to different input sizes without transfer learning, thanks to the use of GAP, while the flattening step confined the \FLAT{} models to a fixed input size. \FLAT{} has been a popular choice in computer vision and has been employed in physics too, despite its shortcomings.

The complex scalar field was mapped into two two-dimensional integer fields by means of a duality transformation, and the datasets were generated with the worm algorithm. In the first task, the networks were required to regress the values of two observables, the particle density $n$ and the field average $|\phi|^2$, given the configurations of the two integer fields. In particular, for the training phase networks had access to data associated with a specific value of the chemical potential and only one lattice size, while testing was conducted on a wide range of the chemical potential and on various lattice sizes to investigate the generalization capabilities of the architectures. The most promising architectures for each type have been selected with \textit{optuna} to ensure a fair comparison. The \EQ{} architecture outperforms the architecture of the other two types. Increasing the number of training samples steadily improved the results of the \EQ{} type, whereas \ST{} and \FLAT{} remained almost unaffected. Remarkably, the results of the non-equivariant architectures were not ameliorated even by augmenting the data, in an attempt to prompt the networks to learn the symmetry during training. Equivariance proved to be relevant also when testing on configurations generated at chemical potentials higher than the one used for training, effectively exploring values of $n$ and $|\phi|^2$ absent from the training phase. This suggested that \EQ{} architectures are more resistant to generalization issues than the non-equivariant counterparts. The extension of the test to different lattice sizes between \EQ{} and \ST{} resulted again in favor of the former, additionally revealing that, for lattice sizes that are not a multiple of the stride used in the spatial pooling layers of \ST{}, part of the input information is lost, further worsening the score of the \ST{} architectures.

In the other two tasks, flux violations, which are associated with the presence of open worms, were artificially introduced in the configurations with a modification of the worm algorithm. The physical appeal of this study lies in its relationship with the calculation of $n$-point functions, and the interest from the machine learning point of view comes from the fact that networks containing only $1 \times 1$ convolutions cannot approximate the true function, since information from next neighbors is necessary to compute the flux. The generalization abilities included different values of the coupling constant and the mass, in addition to the chemical potential and the lattice size. The \EQ{} and the \ST{} architectures scored very similar results when asked to distinguish between configurations with and without open worms, while \FLAT{} performed much worse. In the third task, the networks were required to count the number of open worms from 0 to 10 having been trained on configurations containing either 0 or 5 open worms. Also in this case, the equivariant architectures proved to be highly reliable, whereas the non-equivariant ones struggle to make good predictions, especially for values of open worms between 1 and 3. In all tasks, we found that the networks suggested by \textit{optuna} featured a number of weights much smaller than in networks typically used in comparable studies in the literature.

As a result of these experiments, it is strongly advisable to employ translationally equivariant CNNs in problems symmetric under translations. A considerable performance boost is found when using GAP instead of a flattening step. An ideal scenario that is opened up by such a layer is to cheaply train a network on a small lattice size and apply it on larger ones without losing generalization abilities. The subsampling given by operations with a stride larger than one breaks translational symmetry and can be detrimental to the performance, although it seems to be problem-dependent.

The observation that equivariant networks achieve higher accuracy compared to non-equivariant ones in the context of translational symmetry served as a motivation for embarking on the construction of equivariant networks in the context of lattice gauge theories. Conventional CNNs break gauge symmetry because no parallel transporters are used when adding up contributions from different lattice sites in convolutional layers. Hence, novel layers had to be designed. The input of a gauge-equivariant network is the link configuration and a set of conveniently precomputed locally transforming objects, such as the Wilson loops evaluated at each lattice site. A convolutional layer with appropriate parallel transport was constructed, and its output could be combined in a bilinear layer forming larger Wilson loops without breaking gauge symmetry. A gauge-equivariant activation function was introduced to provide non-linearity. An exponential layer represented a gauge-equivariant operation for modifying the links. After a trace operation, the resulting objects can be further processed by a CNN without spoiling gauge symmetry. A lattice gauge equivariant convolutional neural network consists of stacks of these layers and can in principle learn an arbitrarily sized Wilson loop. More generally, it can be viewed as a universal approximator of gauge equivariant or invariant functions on the lattice.

We experimented the potential of this novel type of network testing it on regressions of Wilson loops of various size, from the trivial $1 \times 1$ to the challenging $4 \times 4$, in 1+1D comparing the performance with that of traditional CNNs. As the loop size grew, not even the best CNN was able to make acceptable predictions, while the \LGCNN{}s showed a robust performance, even when tested on lattice sizes larger than the one used for training. Additional experiments in 3+1D were conducted on $2 \times 2$ and $4 \times 4$ Wilson loops and on the topological charge, where \LGCNN{}s showed again a remarkably good behavior. They were also able to reconstruct Wilson flow of the topological charge by being trained on uncooled samples alone. These networks are very flexible, not only because of the freedom of choosing the most appropriate sequence of layers, but also for the possibility of picking any locally transforming object as input, e.g.~arbitrarily sized Wilson loops or Polyakov loops. Moreover, the code has been developed for $\SU(2)$ and extended to $\SU(3)$ in~\cite{Holland:2024muu}, and can easily be further extended to an arbitrary $\SU(N_c)$.

An application of these networks that we explored in this dissertation has been the generation of link configurations. The mechanism on which this proposal was based is the one of neural ordinary differential equations~\cite{Chen:2018}. An adaptation of the original method to the $\SU(N_c)$ case led us to a gradient flow type of equation (see also~\cite{Bacchio:2022vje}). The procedure is such that the network is provided with a dataset consisting of some initial configurations and their corresponding output ones. The starting configurations are chosen randomly, and are then evolved according to some action. After a successful training, the network is able to reconstruct the corresponding gradient flow.

We tested the neural gradient flow approach on an $\SU(2)$ toy model with a single link. In place of an \LGCNN{}, we employed a dense network whose output was projected onto $\SU(2)$, in order to constrain the evolution on the group. Encouraging results were observed, with networks showing a good generalization to gradient flow times much larger with respect to the interval used for training. The memory saturation issues that can originate from backpropagating through the whole evolution of the system can be overcome by the adjoint sensitivity method. We thoroughly reported our derivation of such a method using Lagrange multipliers for real variables, then in $\mathrm{U}(1)$ and $\SU(N_c)$. This method will enable the use of \LGCNN{}s without stumbling into memory saturation problems even for small architectures, and will allow us to extend the toy model onto the lattice.

A long-term goal and the most natural continuation of the research conducted so far is the generation of configurations for relevant problems in lattice gauge theories. This may include the addition of fermions in the construction of \LGCNN{}s, which still needs to be fully addressed. As discussed, \LGCNN{}s have also been applied to other physical problems such as the fixed-point action~\cite{Holland:2024muu}, they have been used as preconditioners for the Dirac operator~\cite{Lehner:2023bba, Lehner:2023prf}, and have been extended to incorporate global symmetries~\cite{Aronsson:2023rli}. Another exciting application is represented by the construction of a gauge-equivariant version of diffusion models~\cite{Wang:2023sry} using the layers that have been designed for \LGCNN{}s, again with the aim of generating gauge field configurations. The advantage of this approach is given by the inherent stochasticity of the diffusion models, which in principle allows the networks to explore topological sectors more efficiently than standard Monte Carlo algorithms. A very interesting possibility enabled by the gauge equivariance property of \LGCNN{}s is to train these networks to identify an observable in a specific gauge and investigate the properties of such a quantity in other gauges. This can be very useful, for example, in the case of center vortices, which are usually found in the maximal center gauge. Analyzing their characteristics in other gauges can lead to a deeper understanding of the problem of quark confinement. In general, the possibility of customizing novel layers can inspire future works and the versatility of \LGCNN{}s makes them a valuable tool that can find an application in several other lattice gauge theory problems.

\appendix

\numberwithin{equation}{chapter}

\chapter{Monte Carlo simulation of pure lattice gauge theory} \label{app:montecarlo}
\chaptermark{Monte Carlo simulation}

This is a short review of how to use MCMC for the generation of gauge link configurations in a pure $\SU(2)$ gauge theory on the lattice, largely based on~\cite{Gattringer:2010abc}).

The goal is the generation of a sequence of random configurations distributed according to the Wilson action $S_W[U]$ in Eq.~\eqref{eq:wilson_action}, whose corresponding probability functional is
\begin{align}
    \rho[U] \propto e^{-S_W[U]}.
\end{align}
We propose the following random link updates:
\begin{align}
    U'_{x,\mu} = V U_{x,\mu}, \label{eq:update}
\end{align}
where $V$ is a random $\SU(N_c)$ matrix that lies close to the identity. The Metropolis acceptance probability is
\begin{align}
    p[U, U'] = \mathrm{min}\left(1, e^{-S_W[U'] + S_W[U]} \right).
\end{align}
In order to generate the matrices $V$, we first generate random color vectors $X^a = A\, \eta^a$ with amplitude $A > 0$ and $a \in \{ 1, 2, \dots, N_c^2 - 1 \}$, where $\eta^a$ are sampled from a standard normal distribution. The color vectors are contracted with the generators of the group generators, which for $\SU(2)$ are the Pauli matrices~\eqref{eq:pauli_mat}, and the matrix exponential projects the result onto the gauge group, yielding
\begin{align}
    V = e^{i \sum_a T^a X^a} \in \SU(N_c).
\end{align}

In a single sweep, each link is updated according to~\eqref{eq:update} multiple times. In our implementation, we choose $A = 0.5$ and perform ten consecutive updates.

\chapter{\LGCNN{} architectures} \label{app:lcnns}

Details of the composition of the \LGCNN{} architectures employed Section~\ref{sec:comp_exp} are summarized in Table~\ref{tab:arch_lcnn_2d} and~\ref{tab:arch_lcnn_4d} respectively for the 1+1D and 3+1D case.

The \LCB{} operation emerges from the combination of \LConv{} and \LBL{} into a single layer, as explained in subsection~\ref{subsec:lbl}.
The notation \Conv{$K$}{$N_\mathrm{in}$}{$N_\mathrm{out}$} indicates that such a layer is characterized by a kernel size $K$ in every direction, $N_\mathrm{in}$ input and $N_\mathrm{out}$ output channels. Applying the \Trace{} to a configuration $(\mathcal{U}, \mathcal{W})$ with $N_\mathcal{W}$ channels for the $\mathcal{W}$ objects, we obtain $2 \, N_\mathcal{W}$ real numbers per site. The doubled number of channels is a consequence of treating real and imaginary parts separately in the linear layers, indicated by \Lin{$N_\mathrm{in}$}{$N_\mathrm{out}$}, where $N_\mathrm{in}$ is the number of input nodes and $N_\mathrm{out}$ the number of output nodes. Such linear layers act individually at every lattice site, leading to an output that is made up of observables defined over the whole lattice. Applying the same operation at each lattice site does not spoil the overall translational invariance of the networks. The output of the linear layers is not processed by an activation function. $N_\mathrm{param}$ indicates the number of trainable parameters.

\begin{table}
    \caption{\LGCNN{} architectures for the prediction of $W^{(1 \times 1)}$, $W^{(1 \times 2)}$, $W^{(2 \times 2)}$ and $W^{(4 \times 4)}$ in 1+1D. Table from~\cite{Favoni:2020reg}.}
    \label{tab:arch_lcnn_2d}
    \scriptsize
        \begin{tabular}{l | l | l | l}
            $W^{(1\times 1)}$ & & & \\
            \hline
            \hline
            & Small \hspace{5em} & \hspace{8em}  & \hspace{12em} \\
             \hline
            & \Conv{1}{1}{1} &  &  \\
            & \GTr &  &  \\
            & \Lin{2}{1} & & \\
             \hline
             $N_\mathrm{param}$ & 12 &  &  \\
             \hline
             \hline
            $W^{(1\times 2)}$ & & & \\
            \hline
            \hline
            & Small  & Medium & Large \\
             \hline
            & \Conv{2}{1}{2} & \Conv{3}{1}{4} & \Conv{4}{1}{8} \\
            & \GTr & \GTr & \GTr \\
            & \Lin{4}{1} & \Lin{8}{1} & \Lin{16}{1} \\
             \hline
             $N_\mathrm{param}$ & 35 & 117 & 329 \\
             \hline
             \hline
            $W^{(2\times 2)}$ & & & \\
            \hline
            \hline
            & Small & Medium & Large \\
             \hline
            & \Conv{2}{1}{2} & \Conv{3}{1}{4} & \Conv{4}{1}{8} \\
            & \Conv{2}{2}{2} & \Conv{3}{4}{4} & \Conv{4}{8}{8} \\
            & \GTr & \GTr & \GTr \\
            & \Lin{4}{1} & \Lin{8}{1} & \Lin{16}{1} \\
             \hline
             $N_\mathrm{param}$ & 125 & 1,305 & 13,521 \\
             \hline
             \hline
             $W^{(4 \times 4)}$ & & & \\
             \hline
             \hline
             & Small & Medium & Large \\
             \hline
            & \Conv{2}{1}{2} & \Conv{3}{1}{4} & \Conv{4}{1}{8} \\
            & \Conv{2}{2}{2} & \Conv{3}{4}{4} & \Conv{4}{8}{8} \\
            & \Conv{3}{2}{2} & \Conv{4}{4}{4} & \Conv{4}{8}{8} \\
            & \Conv{3}{2}{2} & \Conv{4}{4}{4} & \Conv{4}{8}{8} \\
            & \GTr & \GTr & \GTr \\
            & \Lin{4}{1} & \Lin{8}{1} & \Lin{16}{1} \\
             \hline
            $N_\mathrm{param}$ & 465 & 4,833 & 39,905 \\
        \end{tabular}
\end{table}

\begin{table}
    \caption{\LGCNN{} architectures for the prediction of $W^{(2 \times 2)}$,  $W^{(4 \times 4)}$ and $q^\mathrm{plaq}$ in 3+1D. Table from~\cite{Favoni:2020reg}.}
    \label{tab:arch_lcnn_4d}
        \scriptsize
        \begin{tabular}{l | l | l }
            $W^{(2 \times 2)}$ & & \\
            \hline
            \hline
            & Small \hspace{8em} & Medium \hspace{7em} \\
             \hline
            & \Conv{2}{6}{2} \hspace{.3cm} & \Conv{3}{6}{4} \hspace{.3cm} \\
            & \Conv{2}{2}{2} & \Conv{3}{4}{4} \\
            & \GTr & \GTr \\
            & \Lin{4}{1} & \Lin{8}{1} \\
             \hline
            $N_\mathrm{param} \hspace{.3cm}$ & 1,801 & 8,305  \\
            \hline
            \hline
            $W^{(4 \times 4)}$ & & \\
            \hline
            \hline
            & Small & Medium \\
             \hline
            & \Conv{2}{6}{2} & \Conv{3}{6}{4} \\
            & \Conv{2}{2}{2} & \Conv{3}{4}{4} \\
            & \Conv{3}{2}{2} & \Conv{4}{4}{4} \\
            & \Conv{3}{2}{2} & \Conv{4}{4}{4} \\
            & \GTr & \GTr \\
            & \Lin{4}{1} & \Lin{8}{1} \\
             \hline
             $N_\mathrm{param}$ & 2,109 & 14,377  \\
             \hline
            \hline
            $q^\mathrm{plaq}$ & & \\
            \hline
            \hline
            & Small &  \\
             \hline
            & \Conv{2}{6}{4} &  \\
            & \GTr &  \\
            & \Lin{8}{1} &  \\
             \hline
             $N_\mathrm{param}$ & 3,181 &   \\
        \end{tabular}
\end{table}

\newpage

\chapter{Baseline networks} \label{app:cnns}

Here are reported the tables containing the information about the traditional CNN architectures used in Section~\ref{sec:comp_exp} for the regressions of Wilson loops in 1+1D.

In order to find traditional architectures able to compete with \LGCNN{}s, we tried to cover different ranges of some hyperparameters, specifically the network depth and width. We categorize the architectures into small, medium, large and wide, and manually select three representative architectures of comparable size for each category. These architectures consist of two-dimensional convolutional layers, which in the tables are denoted by $\CCD{K}{N_\mathrm{in}}{N_\mathrm{out}}$, with $K$ indicating the size of the kernel \mbox{$(K \times K)$}, and $N_\mathrm{in}$ and $N_\mathrm{out}$ representing respectively the number of input and output channels. As previously introduced, the acronym GAP stands for global average pooling, which is used for all CNNs, in contrast to \LGCNN{}s where predictions are made at each lattice site. In the dense part, linear layers use the same notation as for \LGCNN{}s. We apply non-linear activation functions after every convolution and linear layer, except after the final output layer. Each architecture has been tried with four different activation functions, i.e.~\textit{LeakyReLU}, \textit{ReLU}, \textit{sigmoid} and \textit{tanh}. We also list the number of learnable weights $N_\mathrm{param}$.

Table~\ref{tab:arch_base_w1x2} lists the architectures employed for the prediction of $W^{(1 \times 1)}$ and $W^{(1 \times 2)}$, while Tables~\ref{tab:arch_base_w2x2} and~\ref{tab:arch_base_w4x4} report the architectures used for $W^{(2 \times 2)}$ and $W^{(4 \times 4)}$ respectively.

\begin{table}
    \caption{Baseline CNN architectures for the prediction of $W^{(1\times 1)}$ and $W^{(1\times 2)}$. Table adapted from~\cite{Favoni:2020reg}.}
    \label{tab:arch_base_w1x2}
    \scriptsize
        \begin{tabular}{l | l | l | l}
            \multicolumn{2}{l|}{$W^{(1 \times 1)}$, $W^{(1 \times 2)}$} & & \\
            \hline
            \hline
            Small & Architecture 1 & Architecture 2 & Architecture 3 \\
            \hline
            & \CCD{2}{$N_\mathrm{in}$}{4} & \CCD{2}{$N_\mathrm{in}$}{4} & \CCD{1}{$N_\mathrm{in}$}{8} \\
            & \CCD{1}{4}{8} & \CCD{2}{4}{4} & \CCD{2}{8}{4} \\
            & GAP & GAP & GAP \\
            & \Lin{8}{4} & \Lin{4}{4} & \Lin{4}{1} \\
            & \Lin{4}{1} & \Lin{4}{1} & - \\
            \hline
            $N_\mathrm{param}$ & 597 & 609 & 401 \\
            \hline
            \hline
            Medium & Architecture 1 & Architecture 2 & Architecture 3 \\
            \hline
            & \CCD{2}{$N_\mathrm{in}$}{8} & \CCD{2}{$N_\mathrm{in}$}{8} & \CCD{3}{$N_\mathrm{in}$}{4} \\
            & \CCD{2}{8}{8} & \CCD{2}{8}{8} & \CCD{2}{4}{8} \\
            & \CCD{2}{8}{8} & - & - \\
            & GAP & GAP & GAP \\
            & \Lin{8}{4} & \Lin{8}{4} & \Lin{8}{4} \\
            & \Lin{4}{1} & \Lin{4}{1} & \Lin{4}{1} \\
            \hline
            $N_\mathrm{param}$ & 1,601 & 1,337 & 1,333 \\
            \hline
            \hline
            Large & Architecture 1 & Architecture 2 & Architecture 3 \\
            \hline
            & \CCD{2}{$N_\mathrm{in}$}{16} & \CCD{3}{$N_\mathrm{in}$}{16} & \CCD{3}{$N_\mathrm{in}$}{16} \\
            & \CCD{2}{16}{16} & \CCD{3}{16}{8} & \CCD{1}{16}{8} \\
            & \CCD{2}{16}{16} & - & \CCD{3}{8}{16} \\
            & GAP & GAP & GAP \\
            & \Lin{16}{8} & \Lin{8}{8} & \Lin{16}{8} \\
            & \Lin{8}{1} & \Lin{8}{1} & \Lin{8}{1} \\
            \hline
            $N_\mathrm{param}$ & 4,289 & 5,865 & 6,073 \\
            \hline
            \hline
            Wide & Architecture 1 & Architecture 2 & Architecture 3 \\
            \hline
            & \CCD{2}{$N_\mathrm{in}$}{128} & \CCD{2}{$N_\mathrm{in}$}{256} & \CCD{2}{$N_\mathrm{in}$}{512} \\
            & - & \CCD{3}{256}{32} & - \\
            & GAP & GAP & GAP \\
            & \Lin{128}{1} & \Lin{32}{1} & \Lin{512}{64} \\
            & - & - & \Lin{64}{1} \\
           \hline
            $N_\mathrm{param}$ & 16,641 & 106,817 & 98,945 \\
        \end{tabular}
\end{table}

\begin{table}
    \caption{Baseline CNN architectures for the prediction of $W^{(2\times 2)}$. Table from~\cite{Favoni:2020reg}.} \label{tab:arch_base_w2x2}
    \scriptsize
        \begin{tabular}{l | l | l | l}
            $W^{(2 \times 2)}$ & & & \\
            \hline
            \hline
            Small  & Architecture 1 \hspace{2em} & Architecture 2 \hspace{2em} & Architecture 3 \hspace{2em}  \\
            \hline
            & \CCD{2}{32}{4} & \CCD{2}{32}{2} & \CCD{2}{32}{4}  \\
            & \CCD{2}{4}{4} & \CCD{1}{2}{4} & \CCD{2}{4}{2} \\
            & GAP & GAP & GAP \\
            & \Lin{4}{4} & \Lin{4}{1} & \Lin{2}{1} \\
            & \Lin{4}{1} & - &  \\
            \hline
            $N_\mathrm{param}$ & 609 & 275 & 553 \\
            \hline
            \hline
            Medium & Architecture 1 & Architecture 2 & Architecture 3 \\
            \hline
            & \CCD{2}{32}{4} & \CCD{2}{32}{8} & \CCD{3}{32}{4} \\
            & \CCD{2}{4}{8} & \CCD{2}{8}{8} & \CCD{2}{4}{8} \\
            & \CCD{2}{8}{8} & \CCD{2}{8}{8} & \CCD{3}{8}{8} \\
            & \CCD{2}{8}{8} & \CCD{2}{8}{8} & \CCD{2}{8}{8} \\
            & GAP & GAP & GAP \\
            & \Lin{8}{16} & \Lin{8}{8} & \Lin{8}{4} \\
            & \Lin{16}{1} & \Lin{8}{1} & \Lin{4}{1} \\
            \hline
            $N_\mathrm{param}$ & 1,341 & 1,905 & 2,181 \\
            \hline
            \hline
            Large & Architecture 1 & Architecture 2 & Architecture 3 \\
            \hline
            & \CCD{2}{32}{8} & \CCD{2}{32}{8} & \CCD{3}{32}{8} \\
            & \CCD{2}{8}{16} & \CCD{2}{8}{16} & \CCD{3}{8}{16} \\
            & \CCD{2}{16}{32} & \CCD{2}{16}{32} & \CCD{3}{16}{32} \\
            & \CCD{2}{32}{64} & \CCD{2}{32}{64} & \CCD{3}{32}{16} \\
            & - & \CCD{2}{64}{32} & - \\
            & GAP & GAP & GAP \\
            & \Lin{64}{16} & \Lin{32}{8} & \Lin{16}{8} \\
            & \Lin{16}{1} & \Lin{8}{1} & \Lin{8}{1} \\
            \hline
            $N_\mathrm{param}$ & 12,953 & 20,393 & 12,889
        \end{tabular}
\end{table}

\begin{table}
    \caption{Baseline CNN architectures for the prediction of $W^{(4\times 4)}$. Table from~\cite{Favoni:2020reg}.} \label{tab:arch_base_w4x4}
    \scriptsize
        \begin{tabular}{l | l | l | l}
            $W^{(4 \times 4)}$ & & & \\
            \hline
            \hline
            Small & Architecture 1 \hspace{2em} & Architecture 2 \hspace{2em} & Architecture 3 \hspace{2em} \\
            \hline
            & \CCD{2}{32}{4} & \CCD{2}{32}{4} & \CCD{2}{32}{4} \\
            & \CCD{2}{4}{4} & \CCD{1}{4}{8} & \CCD{2}{4}{2} \\
            & GAP & GAP & GAP \\
            & \Lin{4}{4} & \Lin{8}{4} & \Lin{2}{1} \\
            & \Lin{4}{1} & \Lin{4}{1} & - \\
            \hline
            $N_\mathrm{param}$ & 609 & 597 & 553 \\
            \hline
            \hline
            Medium & Architecture 1 & Architecture 2 & Architecture 3 \\
            \hline
            & \CCD{3}{32}{16} & \CCD{2}{32}{16} & \CCD{3}{32}{8} \\
            & \CCD{1}{16}{8} & \CCD{2}{16}{24} & \CCD{2}{8}{16} \\
            & \CCD{3}{8}{16} & \CCD{2}{24}{16} & \CCD{1}{16}{32} \\
            & - & - & \CCD{2}{32}{16} \\
            & - & - & \CCD{2}{16}{8} \\
            & GAP & GAP & GAP \\
            & \Lin{16}{8} & \Lin{16}{8} & \Lin{8}{8} \\
            & \Lin{8}{1} & \Lin{8}{1} & \Lin{8}{1} \\
            \hline
            $N_\mathrm{param}$ & 6,073 & 5,321 & 6,049 \\
            \hline
            \hline
            Large & Architecture 1 & Architecture 2 & Architecture 3 \\
            \hline
            & \CCD{3}{32}{16} & \CCD{2}{32}{16} & \CCD{4}{32}{16} \\
            & \CCD{3}{16}{32} & \CCD{2}{16}{32} & \CCD{4}{16}{32} \\
            & \CCD{3}{32}{64} & \CCD{2}{32}{64} & \CCD{4}{32}{32} \\
            & \CCD{3}{64}{32} & \CCD{2}{64}{64} & \CCD{4}{32}{16} \\
            & - & \CCD{2}{64}{32} & - \\
            & - & \CCD{2}{32}{16} & - \\
            & GAP & GAP & GAP \\
            & \Lin{32}{16} & \Lin{16}{16} & \Lin{16}{8} \\
            & \Lin{16}{1} & \Lin{16}{8} & \Lin{8}{8} \\
            & - & \Lin{8}{1} & \Lin{8}{1} \\
            \hline
            $N_\mathrm{param}$ & 46,769 & 39,553 & 41,273
        \end{tabular}
\end{table}

\newpage

\chapter{Test results} \label{app:results}

The detailed results of the experiments in Section~\ref{sec:comp_exp} are given in this Appendix. Tables~\ref{tab:results_w1x1},~\ref{tab:results_w1x2},~\ref{tab:results_w2x2} and~\ref{tab:results_w4x4} show the test results of both CNNs and \LGCNN{}s in 1+1D for $W^{(1\times 1)}$, $W^{(1\times 2)}$, $W^{(2\times 2)}$ and $W^{(4\times 4)}$ respectively. Table~\ref{tab:res_4d} displays the results of \LGCNN{} architectures for $W^{(2\times 2)}$, $W^{(4\times 4)}$ and the topological charge.

In each table, we report the median MSE of the model ensemble for each architecture type and each lattice size. The architectures employed for these experiments can be found in Appendix~\ref{app:lcnns} and~\ref{app:cnns}. The abbreviations are as follows: S stands for small, M for medium, L for large and W for wide. For CNN architectures, the number that follows such letters refers to whether architecture 1, 2 or 3 is used. For CNNs, we additionally specify the type of activation function used. In the first row, we also provide the variance of the observables in the test set. The values in boldface highlight which architecture type scored the lowest median MSE. The asterisk ($*$) signals the ensembles of CNNs and \LGCNN{}s featuring the best individual models according to the validation loss, which are the models used to produce Figs.~\ref{fig:d2_scatter} and~\ref{fig:d2_scatter_large}.

For $W^{(1\times 1)}$ in 1+1D, the \LGCNN{} architecture scores the best median MSE on $8 \times 8$ lattices, but interestingly the S1 type beats it when testing on larger lattices. For larger Wilson loops, the \LGCNN{}s consistently outperform the traditional architectures. For $W^{(1\times 2)}$, the best \LGCNN{} architecture beats every traditional architecture by at least four orders of magnitude. We notice that baseline models considerably improve their median MSE as the lattice size grows, which can be ascribed to the use of a GAP in the final layer of the baseline architectures, effectively reducing fluctuations in the predictions. The improvement for \LGCNN{}s is much milder. For $W^{(2 \times 2)}$ and $W^{(4 \times 4)}$, \LGCNN{}s maintain a low median MSE, with the medium and large-sized architectures being favored over the small one, which is instead the better solution for smaller loops. In 3+1D, we did not train traditional CNNs for a direct comparison, nevertheless we observe that the behavior of \LGCNN{}s is similar to the 1+1D case, with comparable values of MSE.

\begin{table}
    \caption{Test results of \LGCNN{} and CNN architectures (denoted as ``Base'') for the regression of $W^{(1\times 1)}$ in 1+1D. The architectures used are displayed in Tables~\ref{tab:arch_lcnn_2d} and~\ref{tab:arch_base_w1x2}. Table from~\cite{Favoni:2020reg}.}
    \label{tab:results_w1x1}
    \scriptsize
        \begin{tabular}{l | l | l | l | l}
            & $8 \times 8$ & $16 \times 16$ & $32 \times 32$  & $64 \times 64$   \\
            \hline
            {Variance} & {\nnum{5.97e-02}} & {\nnum{5.69e-02}} & {\nnum{5.64e-02}} & {\nnum{5.61e-02}} \\
            \hline
            \LGCNN{}$^*$& \bnum{2.19e-08} & \nnum{2.19e-08} & \nnum{2.19e-08} & \nnum{2.19e-08}  \\ 
            \hline 
            Base S1 (tanh)& \nnum{8.82e-06} & \nnum{3.43e-06} & \nnum{1.98e-06} & \nnum{1.53e-06}  \\ 
            Base S1 (sigm)& \nnum{5.07e-06} & \nnum{1.60e-06} & \nnum{7.89e-07} & \nnum{5.93e-07}  \\ 
            Base S1 (leaky)& \nnum{2.26e-08} & \nnum{8.23e-09} & \nnum{4.55e-09} & \nnum{3.42e-09}  \\ 
            Base S1 (relu)& \nnum{3.14e-08} & \bnum{6.99e-09} & \bnum{2.55e-09} & \bnum{1.20e-09}  \\ 
            \hline 
            Base S2 (tanh)& \nnum{1.57e-05} & \nnum{4.87e-06} & \nnum{2.19e-06} & \nnum{1.37e-06}  \\ 
            Base S2 (sigm)& \nnum{4.74e-06} & \nnum{1.84e-06} & \nnum{9.49e-07} & \nnum{7.83e-07}  \\ 
            Base S2 (leaky)$^*$& \nnum{2.67e-07} & \nnum{3.40e-08} & \nnum{1.47e-08} & \nnum{9.15e-09}  \\ 
            Base S2 (relu)& \nnum{4.82e-07} & \nnum{2.21e-07} & \nnum{6.37e-08} & \nnum{4.23e-08}  \\ 
            \hline 
            Base S3 (tanh)& \nnum{1.52e-06} & \nnum{5.78e-07} & \nnum{3.21e-07} & \nnum{2.48e-07}  \\ 
            Base S3 (sigm)& \nnum{1.40e-06} & \nnum{3.86e-07} & \nnum{1.27e-07} & \nnum{7.98e-08}  \\ 
            Base S3 (leaky)& \nnum{1.10e-07} & \nnum{3.78e-08} & \nnum{2.47e-08} & \nnum{1.83e-08}  \\ 
            Base S3 (relu)& \nnum{5.94e-07} & \nnum{2.25e-07} & \nnum{6.05e-08} & \nnum{4.95e-08}  \\ 
            \hline 
            Base M1 (tanh)& \nnum{1.39e-05} & \nnum{4.35e-06} & \nnum{1.62e-06} & \nnum{9.62e-07}  \\ 
            Base M1 (sigm)& \nnum{9.52e-06} & \nnum{3.76e-06} & \nnum{2.54e-06} & \nnum{2.28e-06}  \\ 
            Base M1 (leaky)& \nnum{2.41e-07} & \nnum{8.54e-08} & \nnum{4.93e-08} & \nnum{2.61e-08}  \\ 
            Base M1 (relu)& \nnum{5.22e-06} & \nnum{1.35e-06} & \nnum{6.63e-07} & \nnum{4.19e-07}  \\ 
            \hline 
            Base M2 (tanh)& \nnum{1.33e-05} & \nnum{4.27e-06} & \nnum{1.79e-06} & \nnum{1.08e-06}  \\ 
            Base M2 (sigm)& \nnum{3.86e-06} & \nnum{1.38e-06} & \nnum{7.22e-07} & \nnum{6.32e-07}  \\ 
            Base M2 (leaky)& \nnum{7.21e-08} & \nnum{1.83e-08} & \nnum{8.10e-09} & \nnum{4.81e-09}  \\ 
            Base M2 (relu)& \nnum{2.24e-06} & \nnum{7.41e-07} & \nnum{2.03e-07} & \nnum{1.25e-07}  \\ 
            \hline 
            Base M3 (tanh)& \nnum{7.21e-06} & \nnum{3.31e-06} & \nnum{2.46e-06} & \nnum{1.69e-06}  \\ 
            Base M3 (sigm)& \nnum{2.51e-06} & \nnum{9.12e-07} & \nnum{5.62e-07} & \nnum{5.02e-07}  \\ 
            Base M3 (leaky)& \nnum{1.17e-07} & \nnum{3.39e-08} & \nnum{1.23e-08} & \nnum{7.25e-09}  \\ 
            Base M3 (relu)& \nnum{6.51e-07} & \nnum{2.47e-07} & \nnum{4.64e-08} & \nnum{1.57e-08}  \\ 
            \hline 
            Base L1 (tanh)& \nnum{1.12e-05} & \nnum{3.71e-06} & \nnum{1.73e-06} & \nnum{1.04e-06}  \\ 
            Base L1 (sigm)& \nnum{5.97e-02} & \nnum{5.69e-02} & \nnum{5.64e-02} & \nnum{5.61e-02}  \\ 
            Base L1 (leaky)& \nnum{7.83e-07} & \nnum{1.89e-07} & \nnum{7.43e-08} & \nnum{5.22e-08}  \\ 
            Base L1 (relu)& \nnum{4.12e-06} & \nnum{9.68e-07} & \nnum{2.97e-07} & \nnum{1.37e-07}  \\ 
            \hline 
            Base L2 (tanh)& \nnum{1.91e-05} & \nnum{6.74e-06} & \nnum{2.88e-06} & \nnum{2.11e-06}  \\ 
            Base L2 (sigm)& \nnum{5.97e-02} & \nnum{5.69e-02} & \nnum{5.64e-02} & \nnum{5.61e-02}  \\ 
            Base L2 (leaky)& \nnum{3.00e-06} & \nnum{1.31e-06} & \nnum{9.96e-07} & \nnum{9.52e-07}  \\ 
            Base L2 (relu)& \nnum{3.21e-06} & \nnum{8.96e-07} & \nnum{2.86e-07} & \nnum{1.83e-07}  \\ 
            \hline 
            Base L3 (tanh)& \nnum{1.07e-05} & \nnum{3.38e-06} & \nnum{1.69e-06} & \nnum{1.35e-06}  \\ 
            Base L3 (sigm)& \nnum{2.34e-05} & \nnum{7.51e-06} & \nnum{2.90e-06} & \nnum{2.37e-06}  \\ 
            Base L3 (leaky)& \nnum{2.90e-06} & \nnum{7.75e-07} & \nnum{3.34e-07} & \nnum{1.73e-07}  \\ 
            Base L3 (relu)& \nnum{2.66e-06} & \nnum{8.01e-07} & \nnum{2.96e-07} & \nnum{1.41e-07}  \\ 
            \hline 
            Base W1 (tanh)& \nnum{7.26e-07} & \nnum{2.31e-07} & \nnum{7.88e-08} & \nnum{4.09e-08}  \\ 
            Base W1 (sigm)& \nnum{2.23e-06} & \nnum{6.94e-07} & \nnum{2.54e-07} & \nnum{1.58e-07}  \\ 
            Base W1 (leaky)& \nnum{8.99e-07} & \nnum{2.52e-07} & \nnum{8.42e-08} & \nnum{4.33e-08}  \\ 
            Base W1 (relu)& \nnum{1.68e-06} & \nnum{4.39e-07} & \nnum{1.50e-07} & \nnum{5.44e-08}  \\ 
            \hline 
            Base W2 (tanh)& \nnum{2.50e-04} & \nnum{6.50e-05} & \nnum{2.05e-05} & \nnum{1.05e-05}  \\ 
            Base W2 (sigm)& \nnum{5.97e-02} & \nnum{5.69e-02} & \nnum{5.64e-02} & \nnum{5.61e-02}  \\ 
            Base W2 (leaky)& \nnum{6.97e-05} & \nnum{1.77e-05} & \nnum{4.83e-06} & \nnum{1.59e-06}  \\ 
            Base W2 (relu)& \nnum{5.97e-02} & \nnum{5.69e-02} & \nnum{5.64e-02} & \nnum{5.61e-02}  \\ 
            \hline 
            Base W3 (tanh)& \nnum{2.70e-05} & \nnum{7.14e-06} & \nnum{3.41e-06} & \nnum{2.10e-06}  \\ 
            Base W3 (sigm)& \nnum{5.97e-02} & \nnum{5.69e-02} & \nnum{5.64e-02} & \nnum{5.62e-02}  \\ 
            Base W3 (leaky)& \nnum{5.03e-05} & \nnum{1.50e-05} & \nnum{7.16e-06} & \nnum{4.95e-06}  \\ 
            Base W3 (relu)& \nnum{3.25e-06} & \nnum{9.34e-07} & \nnum{3.44e-07} & \nnum{1.95e-07}  \\ 
        \end{tabular}
\end{table}

\begin{table}
    \caption{Test results of \LGCNN{} and CNN architectures (denoted as ``Base'') for the regression of $W^{(1\times 2)}$ in 1+1D. The architectures used are reported in Tables~\ref{tab:arch_lcnn_2d} and~\ref{tab:arch_base_w1x2}. Table from~\cite{Favoni:2020reg}.}
    \label{tab:results_w1x2}
    \scriptsize
        \begin{tabular}{l | l | l | l | l}
             & $8 \times 8$ & $16 \times 16$ & $32 \times 32$  & $64 \times 64$   \\
            \hline 
            {Variance} & {\nnum{4.50e-02}} & {\nnum{4.16e-02}} & {\nnum{4.08e-02}} & {\nnum{4.07e-02}} \\
            \hline
            \LGCNN{} S& \bnum{7.58e-09} & \bnum{7.15e-09} & \bnum{6.99e-09} & \bnum{6.97e-09}  \\ 
            \LGCNN{} M& \nnum{1.15e-08} & \nnum{1.10e-08} & \nnum{1.08e-08} & \nnum{1.08e-08}  \\ 
            \LGCNN{} L$^*$& \nnum{1.66e-08} & \nnum{1.60e-08} & \nnum{1.57e-08} & \nnum{1.57e-08}  \\ 
            \hline 
            Base S1 (tanh)& \nnum{2.34e-03} & \nnum{6.24e-04} & \nnum{1.63e-04} & \nnum{6.52e-05}  \\ 
            Base S1 (sigm)& \nnum{2.25e-03} & \nnum{5.96e-04} & \nnum{1.62e-04} & \nnum{6.29e-05}  \\ 
            Base S1 (leaky)& \nnum{2.20e-03} & \nnum{5.59e-04} & \nnum{1.45e-04} & \nnum{4.59e-05}  \\ 
            Base S1 (relu)& \nnum{2.17e-03} & \nnum{5.59e-04} & \nnum{1.50e-04} & \nnum{5.32e-05}  \\ 
            \hline 
            Base S2 (tanh)& \nnum{2.32e-03} & \nnum{6.05e-04} & \nnum{1.65e-04} & \nnum{7.00e-05}  \\ 
            Base S2 (sigm)& \nnum{2.23e-03} & \nnum{5.85e-04} & \nnum{1.52e-04} & \nnum{5.22e-05}  \\ 
            Base S2 (leaky)& \nnum{2.14e-03} & \nnum{5.55e-04} & \nnum{1.44e-04} & \nnum{5.33e-05}  \\ 
            Base S2 (relu)& \nnum{2.09e-03} & \nnum{5.31e-04} & \nnum{1.48e-04} & \nnum{5.38e-05}  \\ 
            \hline 
            Base S3 (tanh)& \nnum{2.19e-03} & \nnum{5.57e-04} & \nnum{1.54e-04} & \nnum{5.80e-05}  \\ 
            Base S3 (sigm)$^*$& \nnum{2.07e-03} & \nnum{5.26e-04} & \nnum{1.32e-04} & \nnum{4.29e-05}  \\ 
            Base S3 (leaky)& \nnum{2.01e-03} & \nnum{5.08e-04} & \nnum{1.39e-04} & \nnum{4.73e-05}  \\ 
            Base S3 (relu)& \nnum{2.03e-03} & \nnum{5.14e-04} & \nnum{1.38e-04} & \nnum{4.81e-05}  \\ 
            \hline 
            Base M1 (tanh)& \nnum{2.51e-03} & \nnum{6.70e-04} & \nnum{1.78e-04} & \nnum{7.41e-05}  \\ 
            Base M1 (sigm)& \nnum{2.35e-03} & \nnum{6.21e-04} & \nnum{1.61e-04} & \nnum{6.22e-05}  \\ 
            Base M1 (leaky)& \nnum{2.18e-03} & \nnum{5.63e-04} & \nnum{1.46e-04} & \nnum{5.39e-05}  \\ 
            Base M1 (relu)& \nnum{2.29e-03} & \nnum{6.04e-04} & \nnum{1.77e-04} & \nnum{7.70e-05}  \\ 
            \hline 
            Base M2 (tanh)& \nnum{2.51e-03} & \nnum{6.84e-04} & \nnum{1.83e-04} & \nnum{7.66e-05}  \\ 
            Base M2 (sigm)& \nnum{2.31e-03} & \nnum{6.04e-04} & \nnum{1.53e-04} & \nnum{5.46e-05}  \\ 
            Base M2 (leaky)& \nnum{2.11e-03} & \nnum{5.33e-04} & \nnum{1.38e-04} & \nnum{4.87e-05}  \\ 
            Base M2 (relu)& \nnum{2.16e-03} & \nnum{5.52e-04} & \nnum{1.42e-04} & \nnum{5.36e-05}  \\ 
            \hline 
            Base M3 (tanh)& \nnum{2.89e-03} & \nnum{7.80e-04} & \nnum{2.24e-04} & \nnum{1.10e-04}  \\ 
            Base M3 (sigm)& \nnum{2.43e-03} & \nnum{6.39e-04} & \nnum{1.76e-04} & \nnum{6.92e-05}  \\ 
            Base M3 (leaky)& \nnum{2.31e-03} & \nnum{6.02e-04} & \nnum{1.53e-04} & \nnum{5.35e-05}  \\ 
            Base M3 (relu)& \nnum{2.43e-03} & \nnum{6.39e-04} & \nnum{1.77e-04} & \nnum{6.40e-05}  \\ 
            \hline 
            Base L1 (tanh)& \nnum{2.63e-03} & \nnum{7.16e-04} & \nnum{2.10e-04} & \nnum{9.13e-05}  \\ 
            Base L1 (sigm)& \nnum{2.38e-03} & \nnum{6.33e-04} & \nnum{1.76e-04} & \nnum{7.41e-05}  \\ 
            Base L1 (leaky)& \nnum{2.30e-03} & \nnum{5.93e-04} & \nnum{1.55e-04} & \nnum{5.49e-05}  \\ 
            Base L1 (relu)& \nnum{2.33e-03} & \nnum{6.23e-04} & \nnum{2.10e-04} & \nnum{1.14e-04}  \\ 
            \hline 
            Base L2 (tanh)& \nnum{2.83e-03} & \nnum{8.03e-04} & \nnum{2.60e-04} & \nnum{1.42e-04}  \\ 
            Base L2 (sigm)& \nnum{2.87e-03} & \nnum{7.49e-04} & \nnum{2.25e-04} & \nnum{1.12e-04}  \\ 
            Base L2 (leaky)& \nnum{2.41e-03} & \nnum{6.24e-04} & \nnum{1.72e-04} & \nnum{7.11e-05}  \\ 
            Base L2 (relu)& \nnum{2.51e-03} & \nnum{6.89e-04} & \nnum{2.32e-04} & \nnum{1.29e-04}  \\ 
            \hline 
            Base L3 (tanh)& \nnum{2.70e-03} & \nnum{7.38e-04} & \nnum{2.08e-04} & \nnum{9.14e-05}  \\ 
            Base L3 (sigm)& \nnum{2.53e-03} & \nnum{6.70e-04} & \nnum{1.87e-04} & \nnum{8.39e-05}  \\ 
            Base L3 (leaky)& \nnum{2.38e-03} & \nnum{6.55e-04} & \nnum{1.91e-04} & \nnum{8.43e-05}  \\ 
            Base L3 (relu)& \nnum{2.36e-03} & \nnum{6.38e-04} & \nnum{1.81e-04} & \nnum{7.93e-05}  \\ 
            \hline 
            Base W1 (tanh)& \nnum{3.73e-03} & \nnum{1.49e-03} & \nnum{8.38e-04} & \nnum{6.60e-04}  \\ 
            Base W1 (sigm)& \nnum{2.39e-03} & \nnum{6.35e-04} & \nnum{1.72e-04} & \nnum{6.35e-05}  \\ 
            Base W1 (leaky)& \nnum{2.14e-03} & \nnum{5.69e-04} & \nnum{1.78e-04} & \nnum{9.01e-05}  \\ 
            Base W1 (relu)& \nnum{2.14e-03} & \nnum{5.41e-04} & \nnum{1.55e-04} & \nnum{6.43e-05}  \\ 
            \hline 
            Base W2 (tanh)& \nnum{2.76e-03} & \nnum{7.59e-04} & \nnum{2.09e-04} & \nnum{9.09e-05}  \\ 
            Base W2 (sigm)& \nnum{4.50e-02} & \nnum{4.16e-02} & \nnum{4.09e-02} & \nnum{4.07e-02}  \\ 
            Base W2 (leaky)& \nnum{2.37e-03} & \nnum{6.75e-04} & \nnum{1.81e-04} & \nnum{7.37e-05}  \\ 
            Base W2 (relu)& \nnum{4.50e-02} & \nnum{4.16e-02} & \nnum{4.09e-02} & \nnum{4.07e-02}  \\ 
            \hline 
            Base W3 (tanh)& \nnum{2.81e-03} & \nnum{9.02e-04} & \nnum{4.35e-04} & \nnum{3.15e-04}  \\ 
            Base W3 (sigm)& \nnum{4.50e-02} & \nnum{4.16e-02} & \nnum{4.09e-02} & \nnum{4.07e-02}  \\ 
            Base W3 (leaky)& \nnum{2.51e-03} & \nnum{6.94e-04} & \nnum{2.41e-04} & \nnum{1.35e-04}  \\ 
            Base W3 (relu)& \nnum{2.12e-03} & \nnum{5.55e-04} & \nnum{1.78e-04} & \nnum{8.62e-05}  \\ 
        \end{tabular}
\end{table}

\begin{table}
    \caption{Test results of \LGCNN{} and CNN architectures (denoted as ``Base'') for the regression of $W^{(2\times 2)}$ in 1+1D. The architectures used are reported in Tables~\ref{tab:arch_lcnn_2d} and~\ref{tab:arch_base_w2x2}. Table from~\cite{Favoni:2020reg}.}
    \label{tab:results_w2x2}
    \scriptsize
        \begin{tabular}{l | l | l | l | l}
             & $8 \times 8$ & $16 \times 16$ & $32 \times 32$  & $64 \times 64$   \\
            \hline
            {Variance} & {\nnum{1.96e-02}} & {\nnum{1.55e-02}} & {\nnum{1.47e-02}} & {\nnum{1.45e-02}} \\
            \hline
            \LGCNN{} S& \nnum{1.17e-07} & \nnum{6.91e-08} & \nnum{6.79e-08} & \nnum{6.77e-08}  \\ 
            \LGCNN{} M$^*$& \bnum{3.24e-08} & \bnum{1.96e-08} & \bnum{1.68e-08} & \bnum{1.64e-08}  \\ 
            \LGCNN{} L& \nnum{6.67e-08} & \nnum{3.89e-08} & \nnum{3.18e-08} & \nnum{3.02e-08}  \\ 
            \hline 
            Base S1 (tanh)& \nnum{4.15e-03} & \nnum{1.10e-03} & \nnum{3.15e-04} & \nnum{1.27e-04}  \\ 
            Base S1 (sigm)& \nnum{3.88e-03} & \nnum{9.98e-04} & \nnum{2.81e-04} & \nnum{9.60e-05}  \\ 
            Base S1 (leaky)& \nnum{3.88e-03} & \nnum{1.01e-03} & \nnum{2.91e-04} & \nnum{1.11e-04}  \\ 
            Base S1 (relu)& \nnum{3.93e-03} & \nnum{1.01e-03} & \nnum{2.96e-04} & \nnum{1.09e-04}  \\ 
            \hline 
            Base S2 (tanh)& \nnum{3.80e-03} & \nnum{9.75e-04} & \nnum{2.82e-04} & \nnum{1.04e-04}  \\ 
            Base S2 (sigm)& \nnum{3.82e-03} & \nnum{1.00e-03} & \nnum{2.95e-04} & \nnum{1.12e-04}  \\ 
            Base S2 (leaky)$^*$& \nnum{3.71e-03} & \nnum{9.54e-04} & \nnum{2.61e-04} & \nnum{8.63e-05}  \\ 
            Base S2 (relu)& \nnum{3.86e-03} & \nnum{9.89e-04} & \nnum{2.77e-04} & \nnum{1.00e-04}  \\ 
            \hline 
            Base S3 (tanh)& \nnum{4.15e-03} & \nnum{1.11e-03} & \nnum{3.20e-04} & \nnum{1.26e-04}  \\ 
            Base S3 (sigm)& \nnum{3.85e-03} & \nnum{9.74e-04} & \nnum{2.60e-04} & \nnum{8.31e-05}  \\ 
            Base S3 (leaky)& \nnum{3.89e-03} & \nnum{1.02e-03} & \nnum{2.93e-04} & \nnum{1.14e-04}  \\ 
            Base S3 (relu)& \nnum{3.86e-03} & \nnum{1.01e-03} & \nnum{2.86e-04} & \nnum{1.06e-04}  \\ 
            \hline 
            Base M1 (tanh)& \nnum{4.19e-03} & \nnum{1.08e-03} & \nnum{3.08e-04} & \nnum{1.21e-04}  \\ 
            Base M1 (sigm)& \nnum{3.98e-03} & \nnum{1.04e-03} & \nnum{2.93e-04} & \nnum{1.04e-04}  \\ 
            Base M1 (leaky)& \nnum{3.87e-03} & \nnum{9.96e-04} & \nnum{2.77e-04} & \nnum{1.02e-04}  \\ 
            Base M1 (relu)& \nnum{4.13e-03} & \nnum{1.11e-03} & \nnum{3.42e-04} & \nnum{1.66e-04}  \\ 
            \hline 
            Base M2 (tanh)& \nnum{4.20e-03} & \nnum{1.14e-03} & \nnum{3.61e-04} & \nnum{1.77e-04}  \\ 
            Base M2 (sigm)& \nnum{3.99e-03} & \nnum{1.06e-03} & \nnum{2.95e-04} & \nnum{1.21e-04}  \\ 
            Base M2 (leaky)& \nnum{3.94e-03} & \nnum{1.02e-03} & \nnum{2.83e-04} & \nnum{1.01e-04}  \\ 
            Base M2 (relu)& \nnum{4.18e-03} & \nnum{1.11e-03} & \nnum{3.51e-04} & \nnum{1.68e-04}  \\ 
            \hline 
            Base M3 (tanh)& \nnum{4.57e-03} & \nnum{1.19e-03} & \nnum{3.72e-04} & \nnum{1.82e-04}  \\ 
            Base M3 (sigm)& \nnum{4.26e-03} & \nnum{1.15e-03} & \nnum{3.62e-04} & \nnum{1.71e-04}  \\ 
            Base M3 (leaky)& \nnum{4.04e-03} & \nnum{1.05e-03} & \nnum{2.95e-04} & \nnum{1.12e-04}  \\ 
            Base M3 (relu)& \nnum{4.47e-03} & \nnum{1.21e-03} & \nnum{4.29e-04} & \nnum{2.30e-04}  \\ 
            \hline 
            Base L1 (tanh)& \nnum{4.26e-03} & \nnum{1.13e-03} & \nnum{3.21e-04} & \nnum{1.26e-04}  \\ 
            Base L1 (sigm)& \nnum{1.96e-02} & \nnum{1.55e-02} & \nnum{1.47e-02} & \nnum{1.45e-02}  \\ 
            Base L1 (leaky)& \nnum{4.01e-03} & \nnum{1.07e-03} & \nnum{3.14e-04} & \nnum{1.25e-04}  \\ 
            Base L1 (relu)& \nnum{1.96e-02} & \nnum{1.55e-02} & \nnum{1.47e-02} & \nnum{1.45e-02}  \\ 
            \hline 
            Base L2 (tanh)& \nnum{4.19e-03} & \nnum{1.12e-03} & \nnum{3.43e-04} & \nnum{1.54e-04}  \\ 
            Base L2 (sigm)& \nnum{1.96e-02} & \nnum{1.55e-02} & \nnum{1.47e-02} & \nnum{1.45e-02}  \\ 
            Base L2 (leaky)& \nnum{4.07e-03} & \nnum{1.07e-03} & \nnum{3.32e-04} & \nnum{1.54e-04}  \\ 
            Base L2 (relu)& \nnum{4.26e-03} & \nnum{1.14e-03} & \nnum{3.50e-04} & \nnum{1.69e-04}  \\ 
            \hline 
            Base L3 (tanh)& \nnum{4.34e-03} & \nnum{1.17e-03} & \nnum{3.60e-04} & \nnum{1.70e-04}  \\ 
            Base L3 (sigm)& \nnum{1.96e-02} & \nnum{1.55e-02} & \nnum{1.47e-02} & \nnum{1.45e-02}  \\ 
            Base L3 (leaky)& \nnum{4.07e-03} & \nnum{1.08e-03} & \nnum{3.18e-04} & \nnum{1.38e-04}  \\ 
            Base L3 (relu)& \nnum{1.96e-02} & \nnum{1.55e-02} & \nnum{1.47e-02} & \nnum{1.45e-02}  \\ 
        \end{tabular}
\end{table}

\begin{table}
    \caption{Test results of \LGCNN{} and CNN architectures (denoted as ``Base'') for the regression of $W^{(4 \times 4)}$ in 1+1D. The notation is the same as in Table~\ref{tab:results_w1x1}. The architectures used are reported in Tables~\ref{tab:arch_lcnn_2d} and~\ref{tab:arch_base_w4x4}. Table from~\cite{Favoni:2020reg}.}
    \label{tab:results_w4x4}
    \scriptsize
        \begin{tabular}{l | l | l | l | l}
            & $8 \times 8$ & $16 \times 16$ & $32 \times 32$  & $64 \times 64$   \\
            \hline
            {Variance} & {\nnum{4.79e-03}} & {\nnum{1.14e-03}} & {\nnum{2.97e-04}} & {\nnum{8.53e-05}} \\
            \hline
            \LGCNN{} S$^*$& \nnum{3.34e-07} & \nnum{1.51e-07} & \nnum{1.17e-07} & \nnum{1.06e-07}  \\ 
            \LGCNN{} M& \bnum{2.06e-07} & \bnum{7.15e-08} & \bnum{4.00e-08} & \bnum{3.10e-08}  \\ 
            \LGCNN{} L& \nnum{2.82e-07} & \nnum{1.09e-07} & \nnum{6.11e-08} & \nnum{5.26e-08}  \\ 
            \hline 
            Base S1 (tanh)& \nnum{4.80e-03} & \nnum{1.15e-03} & \nnum{2.95e-04} & \nnum{8.52e-05}  \\ 
            Base S1 (sigm)& \nnum{4.79e-03} & \nnum{1.14e-03} & \nnum{2.88e-04} & \nnum{7.88e-05}  \\ 
            Base S1 (leaky)& \nnum{4.79e-03} & \nnum{1.13e-03} & \nnum{2.89e-04} & \nnum{7.88e-05}  \\ 
            Base S1 (relu)$^*$& \nnum{4.79e-03} & \nnum{1.14e-03} & \nnum{2.97e-04} & \nnum{8.53e-05}  \\ 
            \hline 
            Base S2 (tanh)& \nnum{4.80e-03} & \nnum{1.14e-03} & \nnum{2.95e-04} & \nnum{8.35e-05}  \\ 
            Base S2 (sigm)& \nnum{4.80e-03} & \nnum{1.13e-03} & \nnum{2.89e-04} & \nnum{7.97e-05}  \\ 
            Base S2 (leaky)& \nnum{4.79e-03} & \nnum{1.14e-03} & \nnum{2.92e-04} & \nnum{8.16e-05}  \\ 
            Base S2 (relu)& \nnum{4.79e-03} & \nnum{1.14e-03} & \nnum{2.92e-04} & \nnum{8.09e-05}  \\ 
            \hline 
            Base S3 (tanh)& \nnum{4.80e-03} & \nnum{1.14e-03} & \nnum{2.92e-04} & \nnum{8.13e-05}  \\ 
            Base S3 (sigm)& \nnum{4.80e-03} & \nnum{1.14e-03} & \nnum{2.91e-04} & \nnum{8.03e-05}  \\ 
            Base S3 (leaky)& \nnum{4.80e-03} & \nnum{1.14e-03} & \nnum{2.94e-04} & \nnum{8.28e-05}  \\ 
            Base S3 (relu)& \nnum{4.80e-03} & \nnum{1.14e-03} & \nnum{2.92e-04} & \nnum{8.20e-05}  \\ 
            \hline 
            Base M1 (tanh)& \nnum{4.81e-03} & \nnum{1.15e-03} & \nnum{2.93e-04} & \nnum{8.22e-05}  \\ 
            Base M1 (sigm)& \nnum{4.79e-03} & \nnum{1.14e-03} & \nnum{2.95e-04} & \nnum{8.30e-05}  \\ 
            Base M1 (leaky)& \nnum{4.80e-03} & \nnum{1.14e-03} & \nnum{2.90e-04} & \nnum{8.02e-05}  \\ 
            Base M1 (relu)& \nnum{4.79e-03} & \nnum{1.14e-03} & \nnum{2.94e-04} & \nnum{8.40e-05}  \\ 
            \hline 
            Base M2 (tanh)& \nnum{4.80e-03} & \nnum{1.14e-03} & \nnum{2.93e-04} & \nnum{8.15e-05}  \\ 
            Base M2 (sigm)& \nnum{4.79e-03} & \nnum{1.14e-03} & \nnum{2.96e-04} & \nnum{8.33e-05}  \\ 
            Base M2 (leaky)& \nnum{4.80e-03} & \nnum{1.14e-03} & \nnum{2.99e-04} & \nnum{8.76e-05}  \\ 
            Base M2 (relu)& \nnum{4.79e-03} & \nnum{1.14e-03} & \nnum{2.97e-04} & \nnum{8.54e-05}  \\ 
            \hline 
            Base M3 (tanh)& \nnum{4.79e-03} & \nnum{1.14e-03} & \nnum{2.94e-04} & \nnum{8.45e-05}  \\ 
            Base M3 (sigm)& \nnum{4.79e-03} & \nnum{1.14e-03} & \nnum{2.97e-04} & \nnum{8.55e-05}  \\ 
            Base M3 (leaky)& \nnum{4.79e-03} & \nnum{1.14e-03} & \nnum{2.99e-04} & \nnum{8.68e-05}  \\ 
            Base M3 (relu)& \nnum{4.79e-03} & \nnum{1.14e-03} & \nnum{2.97e-04} & \nnum{8.53e-05}  \\ 
            \hline 
            Base L1 (tanh)& \nnum{4.83e-03} & \nnum{1.14e-03} & \nnum{2.99e-04} & \nnum{8.74e-05}  \\ 
            Base L1 (sigm)& \nnum{4.79e-03} & \nnum{1.14e-03} & \nnum{2.97e-04} & \nnum{8.55e-05}  \\ 
            Base L1 (leaky)& \nnum{4.79e-03} & \nnum{1.13e-03} & \nnum{2.92e-04} & \nnum{8.16e-05}  \\ 
            Base L1 (relu)& \nnum{4.79e-03} & \nnum{1.14e-03} & \nnum{2.97e-04} & \nnum{8.54e-05}  \\ 
            \hline 
            Base L2 (tanh)& \nnum{4.80e-03} & \nnum{1.14e-03} & \nnum{2.95e-04} & \nnum{8.41e-05}  \\ 
            Base L2 (sigm)& \nnum{4.79e-03} & \nnum{1.14e-03} & \nnum{2.97e-04} & \nnum{8.54e-05}  \\ 
            Base L2 (leaky)& \nnum{4.79e-03} & \nnum{1.14e-03} & \nnum{2.97e-04} & \nnum{8.54e-05}  \\ 
            Base L2 (relu)& \nnum{4.79e-03} & \nnum{1.14e-03} & \nnum{2.97e-04} & \nnum{8.55e-05}  \\ 
            \hline 
            Base L3 (tanh)& \nnum{4.81e-03} & \nnum{1.15e-03} & \nnum{3.00e-04} & \nnum{8.86e-05}  \\ 
            Base L3 (sigm)& \nnum{4.79e-03} & \nnum{1.14e-03} & \nnum{2.97e-04} & \nnum{8.55e-05}  \\ 
            Base L3 (leaky)& \nnum{4.79e-03} & \nnum{1.14e-03} & \nnum{2.89e-04} & \nnum{8.20e-05}  \\ 
            Base L3 (relu)& \nnum{4.79e-03} & \nnum{1.14e-03} & \nnum{2.97e-04} & \nnum{8.54e-05}  \\ 
        \end{tabular}
\end{table}

\begin{table}[htbp]
    \caption{Test results of \LGCNN{} architectures for all regression tasks in 3+1D. Architecture details are provided in table~\ref{tab:arch_lcnn_4d}. Image from~\cite{Favoni:2020reg}.}
    \label{tab:res_4d}
    \scriptsize
        \begin{tabular}{l | l | l | l | l}
             & $4 \times 8^3$ & $6 \times 8^3$ & $6 \times 12^3$  & $8 \times 16^3$   \\
            \hline
            $W^{(2 \times 2)}$ & & & & \\
            {Variance} & {\nnum{7.03e-02}} & {\nnum{7.08e-02}} & {\nnum{7.05e-02}} & {\nnum{7.05e-02}} \\
            \LGCNN{} S& \bnum{1.64e-07} & \bnum{1.63e-07} & \bnum{1.63e-07} & \nnum{1.63e-07}  \\ 
            \LGCNN{} M& \nnum{9.16e-07} & \nnum{6.18e-07} & \nnum{2.17e-07} & \bnum{1.30e-07}  \\ 
            \hline
            $W^{(4 \times 4)}$ & & & & \\
            {Variance} & {\nnum{2.00e-02}} & {\nnum{2.08e-02}} & {\nnum{2.04e-02}} & {\nnum{2.03e-02}} \\
            \LGCNN{} S& \bnum{3.77e-07} & \bnum{3.79e-07} & \bnum{3.74e-07} & \bnum{3.74e-07}  \\ 
            \LGCNN{} M& \nnum{8.26e-07} & \nnum{8.16e-07} & \nnum{7.99e-07} & \nnum{7.99e-07}  \\ 
            \hline
            $Q_P$ & & & & \\
            {Variance} & {\nnum{2.91e-07}} & {\nnum{1.91e-07}} & {\nnum{6.27e-08}} & {\nnum{1.87e-08}} \\
            \LGCNN{} S& \nnum{3.18e-09} & \nnum{3.17e-09} & \nnum{3.17e-09} & \nnum{3.17e-09}  \\ 
        \end{tabular}
\end{table}

\newpage

\bibliographystyle{JHEP.bst}
\bibliography{bibliography}

\end{document}